\documentclass[useAMS,usenatbib]{mn2e}
\usepackage[toc,page]{appendix}
\usepackage{natbib}
\usepackage{amssymb}
\usepackage{epsf}
\usepackage{pstricks}
\usepackage{color}
\usepackage{graphicx}
\usepackage{slashed}
\usepackage{relsize}
\usepackage{morefloats}
\usepackage{multirow}
\usepackage {array}
\usepackage{hyperref}
\usepackage{url}
\usepackage[justification=centering]{caption}

\begin{document}

\title[A Study of Multi-frequency Polarization Pulse Profiles of Millisecond Pulsars]{A Study of Multi-frequency Polarization Pulse Profiles of Millisecond Pulsars}
\author[S. Dai et al.]{S. Dai$^{1,2}$, G. Hobbs$^2$, R. N. Manchester$^2$, M. Kerr$^2$, R. M. Shannon$^2$, W. van Straten$^3$,   
\newauthor A. Mata$^4$, M. Bailes$^3$, N. D. R. Bhat$^5$, S. Burke-Spolaor$^6$, W. A. Coles$^7$, S. Johnston$^2$, 
\newauthor M. J. Keith$^8$, Y. Levin$^9$, S. Os\l owski$^{10,11}$, D. Reardon$^{9,2}$, V. Ravi$^{12}$, J. M. Sarkissian$^{13}$, 
\newauthor C. Tiburzi$^{14,15,2}$, L. Toomey$^{2}$, H. G. Wang$^{16,2}$, J.-B. Wang$^{17}$, L. Wen$^{18}$, R. X. Xu$^{1,19}$, 
\newauthor W. M. Yan$^{17}$, X.-J. Zhu$^{18}$\\
$^1$School of Physics and State Key Laboratory of Nuclear Physics and Technology, Peking University, Beijing 100871, China\\
$^2$CSIRO Astronomy and Space Science, Australia Telescope National Facility, Box 76 Epping NSW 1710, Australia\\
$^3$Centre for Astrophysics and Supercomputing, Swinburne University of Technology, PO Box 218, Hawthorn, VIC 3122, Australia\\
$^4$Center for Advanced Radio Astronomy, University of Texas, Rio Grande Valley, Brownsville, TX 78520, USA\\
$^5$International Centre for Radio Astronomy Research, Curtin University, Bentley, WA 6102, Australia\\
$^6$Department of Astronomy, California Institute of Technology, Pasadena, CA 91125, USA\\
$^7$Department of Electrical and Computer Engineering, University of California, San Diego, La Jolla, CA 92093, USA\\
$^8$Jodrell Bank Centre for Astrophysics, University of Manchester, M13 9PL, UK\\
$^9$School of Physics and Astronomy, Monash University, Victoria 3800, Australia\\
$^{10}$Max-Planck-Institut f\"{u}r Radioastronomie, Auf dem H\"{u}gel 69, D-53121 Bonn, Germany\\
$^{11}$Department of Physics, Universit\"{a}t Bielefeld Universit\"{a}tsstr. 25 D-33615 Bielefeld, Germany\\
$^{12}$School of Physics, University of Melbourne, Parkville, VIC 3010, Australia\\
$^{13}$CSIRO Astronomy and Space Science, Parkes Observatory, Box 276, Parkes NSW 2870, Australia\\
$^{14}$INAF–Osservatorio Astronomico di Cagliari, Via della Scienza, I-09047 Selargius (CA), Italy\\
$^{15}$Dipartimento di Fisica, Universit\`a di Cagliari, Cittadella Universitaria, I-09042 Monserrato (CA), Italy\\
$^{16}$School of Physics and Electronic Engineering, Guangzhou University, 510006 Guangzhou, China\\
$^{17}$Xinjiang Astronomical Observatory, Chinese Academy of Sciences, 150 Science 1-Street, Urumqi, Xinjiang 830011, China\\
$^{18}$School of Physics, University of Western Australia, Crawley, WA 6009, Australia\\
$^{19}$Kavli Institute for Astronomy and Astrophysics, Peking University, Beijing 100871, China\\
}

\maketitle

\begin{abstract}

We present high signal-to-noise ratio, multi-frequency polarization pulse profiles 
for $24$ millisecond pulsars that are being observed as part of the Parkes 
Pulsar Timing Array (PPTA) project. 
The pulsars are observed in three bands, centred close to $730$, $1400$
and $3100$ MHz, using a dual-band 10\,cm/50\,cm receiver and the 
central beam of the 20\,cm multibeam receiver. 
Observations spanning approximately six years have been carefully calibrated and summed to 
produce high S/N profiles. This allows us to study the individual profile 
components and in particular how they evolve with frequency. We also identify previously undetected 
profile features.   
For many pulsars we show that pulsed emission extends across almost the entire 
pulse profile. The pulse component widths and component separations follow a complex 
evolution with frequency; in some cases these parameters increase and in other cases 
they decrease with increasing frequency. 
The evolution with frequency of the polarization properties of the profile is also 
non-trivial. We provide evidence that the pre- and post-cursors generally have higher 
fractional linear polarization than the main pulse.  
We have obtained the spectral index and rotation measure for each pulsar by 
fitting across all three observing bands. For the majority of pulsars, the spectra follow 
a single power-law and the position angles follow a $\lambda^2$ relation, as expected.  
However, clear deviations are seen for some pulsars.  
We also present phase-resolved measurements of the spectral index, fractional linear 
polarization and rotation measure. All these properties are shown to vary systematically 
over the pulse profile.

\end{abstract}


\section{Introduction}

Millisecond pulsars (MSPs) are a special subgroup of radio pulsars. 
Compared with `normal' pulsars, they have shorter spin periods 
and much smaller spin-down rates, and therefore have larger characteristic 
ages and weaker implied dipole magnetic fields.
The short spin periods and highly stable average pulse shapes of MSPs 
make them powerful tools to investigate a large variety of astrophysical phenomena.
In particular, much recent work has been devoted to a search for a gravitational-wave 
background using observations of a large sample of MSPs in a ``Pulsar Timing Array"~\citep[e.g.,][]{Foster90}.
The Parkes Pulsar Timing Array (PPTA) project (Manchester et al. 2013) 
regularly observes $24$ MSPs. 
The PPTA search for gravitational waves has been described in other papers 
including~\citet{Shannon13b},~\citet{Zhu14} and~\citet{Wang15}. 

We have not yet detected gravitational waves. In order to do so we will need 
to observe a larger set of pulsars, increase the span of the observations and/or 
to increase the timing precision achieved for each observation~\citep[e.g.,][]{Cordes12}. 
Determining whether it is possible to improve the timing precision and, if so, 
by how much relies on our understanding of the stability of pulse profiles~\citep[e.g.,][]{Shannon14} 
and also on the profile frequency evolution and polarization properties.
For our work we study the large number of well calibrated, high signal-to-noise 
ratio (S/N) multi-frequency polarization pulse profiles that have been obtained as 
part of the PPTA project. 

An earlier analysis of the 20\,cm pulse profiles from the PPTA sample was published by~\citet{Yan11}. 
This earlier work is extended in this paper as:
(1) we include four new pulsars that have recently been added to the PPTA sample;
(2) we utilise more modern pulsar backend instrumentation than was available to~\citet{Yan11};
(3) we use longer data sets enabling higher S/N profiles; and
(4) we provide polarization pulse profiles in three independent bands 
(at 10, 20 and 50\,cm).
We note that, even though we have mainly the same sample of pulsars as was described by \citet{Yan11}, 
our data sets are independent (i.e., no data is in common between this and the earlier 
publication).

It has been shown that, compared with normal pulsars, the pulse profiles of MSPs usually cover a much larger fraction of the pulse period and, for measurements with the same S/N, often exhibit a larger number of components~\citep{Yan11}. 
However, the spectra of MSPs and normal pulsars are similar~\citep{Toscano98,Kramer98,Kramer99}.
Both MSPs and normal pulsars often have a high degree of linear polarization and orthogonal-mode 
position angle (PA) jumps~\citep[see e.g.,][]{Thorsett90,Navarro97,Stairs99,Manchester04,Ord04}.
For MSPs the PAs often vary significantly with pulse phase and, in most cases, they do not fit 
the `rotating vector model'~\citep[RVM,][]{Radhakrishnan69}.

Various models exist to explain complex pulse profiles.  Multiple emission cones have been proposed 
and discussed by several authors~\citep{Rankin83,Kramer94b,Gupta03}. In  another 
model the emission beam contains randomly 
distributed emission patches~\citep{Lyne88,Manchester95_2,Han01}. 
It has also been suggested that the emission from at least some young pulsars arises from the outermost 
open field lines at relatively high altitudes \citep{Johnston06}.
Similarly, \citet{Kara07} proposed that radio emission is confined to a region close 
to the last open field lines and arises from a wide range of altitudes above the surface of 
the star at a particular frequency.
Based on investigations of the radio and gamma-ray beaming properties of both normal 
pulsars and MSPs,~\citet{Manchester05b} and~\citet{Ravi10} proposed that the radio emission 
of young and MSPs originates in wide beams from regions high in the pulsar magnetosphere (up to 
or even beyond the null-charge surface) and that features in the radio profile represent 
caustics in the emission beam pattern.

To date, no single model can describe the observations.  
This paper is an observationally-based publication that we hope will shed new 
light on the MSP emission mechanism.  
We present the new profiles in three widely separated observing bands and describe how they were 
created. We determine various observationally-derived properties of the profiles (such as 
spectral indices, polarization fractions, etc.) and study how such parameters vary 
between pulsars and with frequency. 
Using these high S/N profiles, we also carry out phase-resolved studies of 
the spectral index~\citep[e.g.,][]{Lyne88,Kramer94a,Manchester04,Chen07}, linear polarization 
fraction, and rotation measures (RMs)~\citep[e.g.,][]{Ramach04,Han06,Noutsos09}.
The data described here will be used in a subsequent paper to study the stability 
of the pulse profiles as a function of time, which is relevant for high-precision pulsar 
timing experiments. In a further paper, we will apply new methods~\citep[e.g.,][]{Pennucci14,Liu14} 
to improve our timing precision using frequency-dependent pulse templates.  
Our data sets are publically available, enabling anyone to compare the actual 
observations with their models of the pulse profiles.

Details of the observation, data processing, and data access are given in 
Section $2$. 
In Section $3$, we present the multi-frequency polarization pulse profiles.
In Section $4$, the pulse widths, flux densities and spectral 
indices, polarization parameters and rotational measures are discussed. 
A summary of our results and conclusions are given in Section $5$.

\section{Observations and Analysis}

\subsection{Observations}

We selected observations from the PPTA project of $24$ MSPs. 
The pulsars are observed regularly, with an approximate observing cadence of 
three weeks, in three bands centred close to $730$ MHz (50\,cm), $1400$ MHz (20\,cm) and $3100$ MHz (10\,cm), 
using a dual-band 10\,cm/50\,cm receiver and the central beam 
of the 20\,cm multibeam receiver. The observing 
bandwidth was $64$, $256$ and $1024$ MHz respectively for the 50\,cm, 20\,cm 
and 10\,cm bands. 
We used both digital polyphase filterbank spectrometers (PDFB4 at 10\,cm 
and PDFB3 at 20\,cm) and a coherent dedispersion machine (CASPSR at 50\,cm). 
In Table \ref{obs}, we summarize the observational parameters for the $24$ PPTA MSPs. 
For each band, we give the number of frequency channels across 
the band, the number of bins across the pulse period, the total number 
of observations and the total integration time.
In Table \ref{psr}, we give the basic pulsar parameters from the ATNF Pulsar 
Catalogue~\citep{Manchester05}.
For each observing band, we also give the dispersion smearing and the pulse 
broadening time caused by scattering (in units of profile bins).
The dispersion smearing across each frequency channel is calculated according to
\begin{equation}
\Delta t_{\rm{DM}}\approx 8.30 \times 10^{6}\ \rm{DM}\ \Delta \nu\ \nu^{-3}\ \rm{ms},
\label{dm}
\end{equation}
where $\Delta \nu$ is the channel width in MHz, $\nu$ is the band central frequency in MHz, 
and DM is the dispersion measure in units of $\rm{cm^{-3}\ pc}$.
The pulse broadening time caused by scattering is estimated according to
\begin{equation}
\tau_{\rm{d}} = \frac{1}{2\pi\nu_0},
\end{equation}
where $\nu_0$ is the scintillation bandwidth. We calculate the broadening time 
in the 20\,cm band using scintillation bandwidths measured by~\citet{Keith13}, and 
then scale it to the 10\,cm and 50\,cm bands according to $\tau_{\rm{d}}\propto\nu^{-4}$.
For MSPs not in the sample of~\citet{Keith13}, we measure the scintillation 
bandwidths using the autocorrelation function (ACF) of the dynamic spectrum~\citep[e.g.,][]{Wang05}.
We note that in Table \ref{psr}, we only list $\tau_{\rm{d}}$ values that are 
$\ge 0.0001\ \rm{bin}$ and set others as zero.

\begin{table*}
\caption{Observational parameters for the $24$ PPTA MSPs.}
\label{obs}
\begin{center}
\begin{tabular}{p{1.5cm}p{0.9cm}<{\centering}p{0.9cm}<{\centering}p{0.9cm}<{\centering}p{0.9cm}<{\centering}p{0.9cm}<{\centering}p{0.9cm}<{\centering}p{0.9cm}<{\centering}p{0.9cm}<{\centering}p{0.9cm}<{\centering}p{0.9cm}<{\centering}p{0.9cm}<{\centering}p{0.9cm}<{\centering}}
\hline
PSR         &     \multicolumn{3}{c}{No. of channels}   &   \multicolumn{3}{c}{No. of phase bins}  &    \multicolumn{3}{c}{No. of observation epochs}   &    \multicolumn{3}{c}{Integration time}      \\
            &         &                 &          &         &             &          &         &             &          &         &   (h)            &       \\
            & 50\,cm &    20\,cm     & 10\,cm &  50\,cm &    20\,cm     & 10\,cm &  50\,cm &    20\,cm     & 10\,cm &  50\,cm &    20\,cm     & 10\,cm     \\
\hline
J0437$-$4715&  256    &    1024         &   1024   &  1024   &  1024       &  2048    &  177    &  669        & 281      &  142.9  &    502.2         &  248.8   \\
J0613$-$0200&  256    &    1024         &   1024   &  1024   &  512        &  512     &  64     &  160        & 111      &  66.0   &    159.3         &  113.9   \\
J0711$-$6830&  256    &    1024         &   1024   &  1024   &  1024       &  1024    &  72     &  161        & 102      &  65.9   &    161.1         &  102.2   \\
J1017$-$7156&  256    &    2048         &   2048   &  1024   &  256        &  512     &  85     &  135        & 73       &  86.5   &    130.4         &  76.3    \\
J1022$+$1001&  256    &    1024         &   1024   &  1024   &  2048       &  2048    &  65     &  148        & 117      &  58.4   &    138.3         &  110.5    \\
						&         &                 &          &         &             &          &         &             &          &         &                  &           \\
J1024$-$0719&  256    &    1024         &   1024   &  1024   &  1024       &  1024    &  34     &  112        & 59       &  36.1   &    111.0         &  61.5     \\
J1045$-$4509&  256    &    2048         &   1024   &  1024   &  512        &  1024    &  63     &  137        & 103      &  42.7   &    138.9         &  104.5    \\ 
J1446$-$4701&  256    &    512          &   1024   &  1024   &  512        &  1024    &  19     &  50         & 9        &  15.2   &    39.4          &  8.8    \\ 
J1545$-$4550&  256    &    1024         &   1024   &  1024   &  512        &  1024    &  15     &  21         & 15       &  13.2   &    20.6          &  12.2   \\ 
J1600$-$3053&  256    &    1024         &   1024   &  1024   &  512        &  512     &  53     &  139        & 106      &  56.6   &    129.9         &  108.0   \\ 
						&         &                 &          &         &             &          &         &             &          &         &                  &          \\
J1603$-$7202&  256    &    2048         &   1024   &  1024   &  1024       &  1024    &  52     &  131        & 49       &  44.4   &    127.4         &  50.6    \\ 
J1643$-$1224&  256    &    2048         &   1024   &  1024   &  512        &  1024    &  53     &  116        & 93       &  53.7   &    117.0         &  93.4     \\ 
J1713$+$0747&  256    &    1024         &   1024   &  1024   &  1024       &  1024    &  66     &  155        & 110      &  67.8   &    132.0         &  107.9    \\ 
J1730$-$2304&  256    &    1024         &   1024   &  1024   &  1024       &  2048    &  57     &  104        & 62       &  51.0   &    105.8         &  62.2    \\ 
J1744$-$1134&  256    &    512          &   1024   &  1024   &  1024       &  1024    &  65     &  129        & 96       &  66.0   &    126.7         &  99.5    \\ 
						&         &                 &          &         &             &          &         &             &          &         &                  &          \\
J1824$-$2452A&  256    &    2048         &   1024   &  1024   &  256        &  512     &  33     &  88         & 54       &  33.0   &    82.9          &  53.6    \\ 
J1832$-$0836&  256    &    1024         &   1024   &  1024   &  512        &  1024    &  12     &  19         & 11       &  9.0    &    16.9          &  10.1    \\ 
J1857$+$0943&  256    &    1024         &   1024   &  1024   &  1024       &  1024    &  54     &  99         & 68       &  27.8   &    50.9          &  35.5   \\ 
J1909$-$3744&  256    &    1024         &   1024   &  1024   &  512        &  1024    &  95     &  218        & 138      &  91.3   &    191.1         &  129.4   \\ 
J1939$+$2134&  256    &    1024         &   1024   &  512    &  256        &  256     &  58     &  102        & 91       &  26.4   &    49.4          &  46.0    \\ 
						&         &                 &          &         &             &          &         &             &          &         &                  &     \\
J2124$-$3358&  256    &    1024         &   1024   &  1024   &  1024       &  1024    &  40     &  134        & 78       &  20.3   &    68.5          &  40.5    \\ 
J2129$-$5721&  256    &    1024         &   1024   &  1024   &  512        &  512     &  59     &  116        & 17       &  31.1   &    112.6         &  9.0     \\ 
J2145$-$0750&  256    &    1024         &   1024   &  1024   &  2048       &  2048    &  70     &  134        & 117      &  65.1   &    129.3         &  111.2   \\ 
J2241$-$5236&  256    &    1024         &   1024   &  1024   &  512        &  1024    &  75     &  188        & 93       &  69.8   &    152.3         &  92.9   \\ 
\hline
\end{tabular}
\end{center}
\end{table*}

\begin{table*}
\caption{Pulsar parameters for the $24$ PPTA MSPs.}
\label{psr}
\begin{center}
\begin{tabular}{lcccccccccc}
\hline
PSR                  &  RAJ      &  DECJ    &   P     &       DM               &     \multicolumn{3}{c}{DM smear}   &   \multicolumn{3}{c}{$\tau_{\rm{d}}$}       \\
                     &  (hms)    &  (dms)   &  (ms)   &  ($\rm{cm^{-3}\ pc}$)  &     \multicolumn{3}{c}{(bin)}      &   \multicolumn{3}{c}{(bins)}                \\
			               &           &          &         &                        &  50\,cm  & 20\,cm  & 10\,cm        & 50\,cm  &     20\,cm      & 10\,cm          \\
\hline
J0437$-$4715$^{a,b}$ & 04:37:15.9  &  $-$47:15:09.0 &  5.757  &  2.64     & 7.9      & 0.4       & 0.3    &  0.0004  &  0.0000  &  0.0000  \\ 
J0613$-$0200$^{a,b}$ & 06:13:44.0  &  $-$02:00:47.2 &  3.062  &  38.78    & 218.0    & 5.2       & 1.8    &  0.4058  &  0.0162  &  0.0006  \\ 
J0711$-$6830         & 07:11:54.2  &  $-$68:30:47.6 &  5.491  &  18.41    & 57.7     & 2.8       & 1.0    &  0.0103  &  0.0008  &  0.0000  \\ 
J1017$-$7156         & 10:17:51.3  &  $-$71:56:41.6 &  2.339  &  94.22    & 693.4    & 4.2       & 2.9    &  0.7923  &  0.0158  &  0.0012  \\ 
J1022$+$1001         & 10:22:58.0  &  $+$10:01:52.8 &  16.453 &  10.25    & 10.7     & 1.0       & 0.4    &  0.0019  &  0.0003  &  0.0000  \\ 
                     &             &                &         &           &          &           &        &          &          &          \\ 
J1024$-$0719$^{f}$   & 10:24:38.7  &  $-$07:19:19.2 &  5.162  &  6.49     & 21.6     & 1.0       & 0.4    &  0.0015  &  0.0001  &  0.0000  \\ 
J1045$-$4509         & 10:45:50.2  &  $-$45:09:54.1 &  7.474  &  58.17    & 133.9    & 1.6       & 2.2    &  2.9005  &  0.1160  &  0.0088  \\ 
J1446$-$4701$^{d}$   & 14:46:35.7  &  $-$47:01:26.8 &  2.195  &  55.83    & 437.8    & 21.1      & 7.3    &  0.3439  &  0.0138  &  0.0010  \\ 
J1545$-$4550         & 15:45:55.9  &  $-$45:50:37.5 &  3.575  &  68.39    & 329.2    & 7.9       & 5.5    &  0.5182  &  0.0207  &  0.0016  \\ 
J1600$-$3053$^{f}$   & 16:00:51.9  &  $-$30:53:49.3 &  3.598  &  52.33    & 250.3    & 6.0       & 2.1    &  6.2935  &  0.2516  &  0.0096  \\ 
                     &             &                &         &           &          &           &        &          &          &          \\ 
J1603$-$7202         & 16:03:35.7  &  $-$72:02:32.7 &  14.842 &  38.05    & 44.1     & 1.1       & 0.7    &  0.0275  &  0.0022  &  0.0001  \\ 
J1643$-$1224         & 16:43:38.2  &  $-$12:24:58.7 &  4.622  &  62.41    & 232.4    & 2.8       & 3.9    &  20.0424 &  0.8014  &  0.0610  \\
J1713$+$0747$^{f}$   & 17:13:49.5  &  $+$07:47:37.5 &  4.570  &  15.99    & 60.2     & 2.9       & 1.0    &  0.0186  &  0.0015  &  0.0001  \\ 
J1730$-$2304         & 17:30:21.7  &  $-$23:04:31.3 &  8.123  &  9.62     & 20.4     & 1.0       & 0.7    &  0.0202  &  0.0016  &  0.0001  \\ 
J1744$-$1134$^{a,b}$ & 17:44:29.4  &  $-$11:34:54.7 &  4.075  &  3.14     & 13.3     & 1.3       & 0.2    &  0.0083  &  0.0007  &  0.0000  \\ 
                     &             &                &         &           &          &           &        &          &          &          \\ 
J1824$-$2452A$^{g,h}$& 18:24:32.0  &  $-$24:52:10.8 &  3.054  &  120.50   & 675.5    & 4.1       & 5.6    & 26.6882  &  0.5335  &  0.0406  \\ 
J1832$-$0836         & 18:32:27.6  &  $-$08:36:55.0 &  2.719  &  28.18    & 178.3    & 4.3       & 3.0    &  0.6245  &  0.0250  &  0.0019  \\ 
J1857$+$0943         & 18:57:36.4  &  $+$09:43:17.3 &  5.362  &  13.30    & 42.7     & 2.1       & 0.7    &  0.0691  &  0.0055  &  0.0002  \\ 
J1909$-$3744         & 19:09:47.4  &  $-$37:44:14.4 &  2.947  &  10.39    & 60.7     & 1.5       & 1.0    &  0.0187  &  0.0007  &  0.0001  \\ 
J1939$+$2134$^{e}$   & 19:39:38.6  &  $+$21:34:59.1 &  1.558  &  71.04    & 392.3    & 9.4       & 3.3    &  0.5451  &  0.0218  &  0.0008  \\ 
                     &             &                &         &           &          &           &        &          &          &          \\ 
J2124$-$3358$^{a,b}$ & 21:24:43.9  &  $-$33:58:44.7 &  4.931  &  4.60     & 16.0     & 0.8       & 0.3    &  0.0004  &  0.0000  &  0.0000  \\ 
J2129$-$5721         & 21:29:22.8  &  $-$57:21:14.2 &  3.726  &  31.85    & 147.1    & 3.5       & 1.2    &  0.0320  &  0.0013  &  0.0000  \\ 
J2145$-$0750         & 21:45:50.5  &  $-$07:50:18.4 &  16.052 &  9.00     & 9.7      & 0.9       & 0.3    &  0.0007  &  0.0001  &  0.0000  \\ 
J2241$-$5236$^{c}$   & 22:41:42.0  &  $-$52:36:36.2 &  2.187  &  11.41    & 89.8     & 2.1       & 1.5    &  0.0661  &  0.0026  &  0.0002  \\ 
\hline
\end{tabular}
\end{center}
~\\
Gamma-ray loud: $^a$~\citet{Abdo09}; $^b$~\citet{Abdo10}; $^c$~\citet{Keith11}; $^d$~\citet{Keith12}; 

$^e$~\citet{Guillemot12}; $^f$~\citet{Espinoza13}; $^g$~\citet{Abdo13}; $^h$~\citet{Barr13}.
\end{table*}

To calibrate the gain and phase of the receiver system, a linearly polarized 
broad-band and pulsed calibration signal is injected into the two orthogonal 
channels through a calibration probe at $45^{\circ}$ to the 
signal probes. The pulsed calibration signal was recorded for $2\sim3$ min prior to 
each pulsar observation.
Signal amplitudes were placed on a flux density scale using observations of 
Hydra A, assuming a flux density of 43.1\,Jy at 1400\,MHz and a spectral 
index of $-0.91$ over the PPTA frequency range.
All data were recorded using the PSRFITS data format~\citep{Hotan04} with 
$1$-min subintegrations and the full spectral resolution
\citep[for further details see][and references therein]{Manchester13}. 

\subsection{Analysis}

The data were processed using the PSRCHIVE software package~\citep{Hotan04}. 
We removed 5 per cent of the bandpass at each edge and excised data 
affected by narrow band and impulsive radio-frequency interference for each 
subintegration.
The polarization was then calibrated by correcting for differential gain and 
phase between the receptors using the associated calibration files.
For 20\,cm observations with the multibeam receiver, we corrected for 
cross coupling between the feeds through a model derived from observations of 
PSR J0437$-$4715 that covered a wide range of parallactic angles~\citep{VanStraten04}.

The Stokes parameters are in accordance with the astronomical conventions described 
by~\citet{vanStraten10}. Stokes $V$ is defined as $I_{\rm{LH}}-I_{\rm{RH}}$, 
using the IEEE definition for sense of circular polarization. 
The baseline region was determined with the Stokes $I$ profile. The baseline duty 
cycle used for each MSP are presented in Table~\ref{ref}. Baselines for 
the Stokes $I$, $Q$, $U$ and $V$ profiles were set to zero mean.
The linear polarization $L$ was calculated as $L=(Q^2+U^2)^{1/2}$, and the 
noise bias in $L$ was corrected according to Equation 11 in~\citet{Everett01}. 
The similar bias in $|V|$ was corrected as described in~\citet{Yan11}.
The position angles (PAs) of the linear polarization refer to the band central 
frequency and were calculated as $\psi=0.5\tan^{-1}(U/Q)$ when the linear 
polarization exceeds four times of the baseline root mean square (rms) noise. 
They are absolute and measured from celestial north towards east, i.e. 
counterclockwise on the sky.
Errors on the PA values were estimated according to Equation 12 in~\citet{Everett01}.

In order to add the data in time to form a final mean profile, pulse times of arrival 
were obtained for each observation using an analytic template based on an existing 
high S/N pulse profile. The TEMPO2 pulsar timing software package~\citep{Hobbs06}  
was then used to fit pulsar spin, astrometric, and binary parameters, and also to fit 
harmonic waves as necessary to give ‘white’ timing residuals for each pulsar. Finally, 
the separate observations were summed using this timing model to determine relative 
phases and form the final Stokes-parameter profiles. 

To give the best possible S/N in the polarization pulse profiles, the individual 
observation profiles were weighted by their $(\rm{S/N})^2$ when forming the average profile. 
As many of the pulsars scintillate strongly, this weighting implies that, for a few pulsars, 
the average profiles are dominated by a few individual observations with a high S/N. As discussed in 
Section $4$ this can affect measurements of the spectral index, fractional polarizations, and 
RMs. Also, if the pulse profile varies with flux density (for instance, 
as seen for PSR~J0437$-$4715 by Os{\l}owski et al. 2014) then this weighted profile will be 
biased towards the profile shape at high flux density. We therefore have also produced 
average profiles using only the observation time for weighting. 

Since the PA of the linear polarization suffers Faraday rotation in the interstellar medium
and in the Earth's ionosphere, this Faraday rotation must be removed to form the mean polarization 
profiles.
According to~\citet{Yan11b}, the interstellar RMs of PPTA MSPs are stable, and for our initial analysis we 
used the best-available interstellar RM values for our sample~\citep{Keith11,Yan11,Keith12,Burgay13}.
To account for the contribution of the Earth's ionosphere, we used the International 
Reference Ionosphere (IRI) model~\footnote{See \url{http://iri.gsfc.nasa.gov} for a general description
of the IRI.}. 

For each MSP, we aligned the average pulse profile in the 10 and 50\,cm bands with 
respect to that in the 20\,cm band.
The technique we used is described in detail in~\citet{Taylor92}, which 
was originally developed for the measurement of pulse arrival times.
We derived the phase shift between profiles and the profile in the 20\,cm band 
in the frequency domain, rotated the profiles and then transformed them back to 
the time domain.
With these aligned three-band profiles, we calculated the phase-resolved 
spectral indices, fractional linear polarizations and RMs for each MSP. 
The spectral index was fitted using a power-law of the form $S=S_{0}\nu^{\alpha}$ 
and the fractional linear polarization was defined as $\langle L \rangle/S$, 
where $S=\langle I\rangle$ is the total intensity and $L$ is the linear polarization.
The RM was obtained by fitting the PA across bands according to $\psi=\rm{RM}\ \lambda^{2}$,
where $\lambda=c/\nu$ is the radio wavelength corresponding to radio 
frequency $\nu$.
As many of the MSP profiles have multiple components which vary significantly with frequency, 
it is difficult to determine an absolute profile alignment. Our cross-correlation 
method is a straight-forward and reproducible technique. However, the reader should 
note that, when studying the phase-resolved parameters, other alignment methods 
may produce slightly different results.

\subsection{Data access}

The raw data and calibration files used in this paper are available from the 
Parkes Observatory Pulsar Data Archive~\citep{Hobbs11}.  
The scripts used to create the results given in this paper and the resulting averaged 
(weighted by their $(\rm{S/N})^2$ and by the observing time) profiles are available for 
public access\footnote{\url{http://dx.doi.org/10.4225/08/54F3990BDF3F1}}.

\section{Multi-frequency Polarization Profiles}

Our main results are the polarization pulse profiles for the PPTA pulsars in the three bands.  
These are shown, for each of the 24 pulsars, in Figures~\ref{0437} to~\ref{2241}. 
The left-hand panels show the pulse profile in the 10\,cm (top), 20\,cm (second panel) and 
50\,cm (third panel) observing bands. 
The bottom panel on the left-hand side presents the phase-resolved spectral index.   
In order to obtain the phase-resolved spectral index, we divided the 10\,cm and 20\,cm band 
into four subbands and the 50\,cm band into three subbands (details are given in Section 3 for the few cases in which we used a different number of subbands).
We rebinned the profile in each subband into $256$ phase bins to gain higher S/N. Only 
phase bins whose signal exceeds three times the baseline rms noise in all subbands 
are used and we only plot spectral indices whose uncertainty is smaller than one. 

In the right-hand panels we have two panels for each of the 10, 20 and 50\,cm bands. The upper 
panel shows the PA of the linear polarization (in degrees) determined when the 
linear polarization exceeds four times the baseline rms noise.  
The lower panels shows a zoom-in around profile baseline to show weaker profile features. 
The bottom two panels on the right-hand side show the phase-resolved fractional linear polarization 
for the three observing bands and the phase-resolved apparent RM.
In order to obtain the phase-resolved fractional linear polarization, we rebinned the profile 
in each band into $128$ phase bins to gain higher S/N and only phase bins whose linear polarization 
exceeds three times the baseline rms noise were used. 
The phase-resolved RMs were obtained with the frequency-averaged profile in each band 
without any rebinning. In order to avoid low S/N regions and obtain smaller uncertainties of 
the PA, only phase bins whose linear polarization exceeds five times the baseline rms noise 
were used. We only plot RMs whose uncertainty is smaller than $3\ \rm{rad\ m^{-2}}$.
Further details on the figures are given in the appendix.

In almost all cases our results are consistent with earlier measurements~\citep[such as][]{Ord04,Yan11}
where these exist.
Specific comments for each individual pulsar and on the comparison with 
previous work are given in the caption of each figure.   
In particular, we have discovered weak components for PSRs J1603$-$7202, J1713$+$0747, 
J1730$-$2304, J2145$-$0750 and J2241$-$5236. We also show new details of the PA curves, 
including new orthogonal transitions for PSRs J0437$-$4715, J1643$-$1224, J2124$-$3358, 
J2129$-$5721 and J2241$-$5236; and new non-orthogonal transitions for PSRs J1045$-$4509, 
J1857$+$0943 and J2124$-$3358.

\section{Discussion}

\subsection{Pulse widths}

One of the most fundamental properties of the pulse profile is the pulse width.    
The frequency dependence of the pulse width has been extensively studied for normal pulsars~\citep[e.g.,][]{Cordes78,Thorsett91}.
A recent study of 150 normal pulsars~\citep{Chen14} shows that 81 pulsars in their sample exhibit 
considerable profile narrowing at high frequencies, 29 pulsars exhibit profile broadening at high 
frequencies, and the remaining 40 pulsars only have a marginal change in pulse width. 
Studies of the pulse width as a function of frequency for MSPs have also been carried 
out~\citep[e.g.,][]{Kramer99}.

However, the pulse width is difficult to interpret, particularly for profiles 
that contain multiple components. Comparing pulse widths across wide frequency 
bands is even more challenging as the components often differ in spectral index or new 
components appear in the profile. Traditionally pulse widths are published as the 
width of the profile at 10 and 50 per cent of the peak flux density ($W_{10}$ and $W_{50}$ 
respectively).  
For comparison with previous work, $W_{10}$ and $W_{50}$ are given in Table~\ref{tableWidth} 
for the three observing bands of each pulsar (PSRs J1545$-$4550 and J1832$-$0836 
have very low S/N profiles in the 50\,cm band, therefore we do not present their 
pulse widths in the 50\,cm band). 
However, these results have limited value. For instance, the $W_{10}$ measurement for 
PSR J1939$+$2134 in all three bands provides a measure of the width between the 
two distinct components. The $W_{50}$ measurement does the same for the 20\,cm and the 
50\,cm observing bands, but in the 10\,cm band one of the components does not reach 
the 50 per cent height of the peak component. The meaning of the $W_{50}$ measurement 
is therefore different in the 10\,cm band. 

Following~\citet{Yan11} we also present the ``overall pulse width" for the three 
bands of each pulsar. This is measured to give the pulse width in which the pulse 
intensity significantly exceeds the baseline noise (3$\sigma$). This value is presented 
in the first three columns of Table~\ref{tableWidth}.   
The overall widths have, in most cases, increased from the results published by~\citet{Yan11} 
as our higher S/N profiles have allowed us to identify new low-level emission over 
more of the pulse profile. With the S/N currently achievable (approximately 33,500 for 
PSR J0437$-$4715 at 20\,cm) we find that 18 of the 24 pulsars exhibit emission 
over more than half of the pulse period. 
Even though the individual pulse components vary with observing frequency, the 
overall pulse width is relatively constant for pulsars that have high S/N profiles in all 
three bands. This suggests that, even though the properties of individual components 
vary across observing bands, the absolute width of the emission beam is more constant.
To understand the wide profiles of MSPs, \citet{Ravi10} suggested that the MSP radio 
emission is emitted from the outer magnetosphere and that caustic effects may account 
for the broad frequency-independent pulse profiles~\citep{Dyks03,Watters09}.

In terms of pulsar timing, the ``sharpness" of the profile provides a measure of 
how precisely pulse times-of-arrival can be measured. We measure the sharpness of 
profiles with the effective pulse width defined as 
\begin{equation}
W_{\rm{s}}=\frac{\Delta \phi}{\sum_{i}[I(\phi_{i+1})-I(\phi_{i})]^2},
\end{equation}
where $\Delta \phi$ is the phase resolution of the pulse profile (measured in units 
of time), and the profile is normalized to have a maximum intensity of unity \citep{Cordes10,Shannon14}.
This parameter for each of the observing bands is presented in the last three columns 
of Table~\ref{tableWidth}.  

For some MSPs in our sample it is possible to identify a well-defined pulse component 
over multiple observing bands. This allows us to investigate the frequency 
evolution of the component width and separation. 
Such components have been identified in Fig.~\ref{0437} to~\ref{2241} with component 
numbers (C1 to C28).  
The width of each component is shown in Table~\ref{tableWidth2}.  
In order to mitigate the effects of surrounding components and low-level features, for
each component we provide a measure of its width at 50 and 80 per cent of its 
peak flux density ($W_{50}$ and $W_{80}$ respectively) as a function of observing frequency.
We estimated the uncertainties on these measurements by determining how the width changes when 
the 50 and 80 per cent flux density cuts across the profile move up or down by the baseline rms 
noise level.
In most cases the pulse component widths decrease with increasing frequency despite the 
relatively large uncertainty. 
For PSRs~J1939$+$2134 and J2241$-$5236, we see small increases of the pulse component
widths with increasing observing frequency compared with their uncertainties. 
This is likely because of substructure in the components.

The component separations are shown in Table~\ref{separation}. We estimated the 
uncertainties of measurements as the variation of component separations when we 
adjust the peak flux density by the amount of the baseline rms noise.
Within the uncertainty, most cases show no significent frequency evolution of 
component separations, consistent with the caustic interpretation of profile components. 
For PSR~J0711$-$6830, we see an increase of the component separation with 
decreasing observing frequency, which is likely because of the steep spectrum 
at the trailing edge of the main pulse.

Under the conventional radius-frequency-mapping scenario~\citep{Cordes78}, which assumes that the 
emission is narrowband at a given altitude and the emission frequency increases 
with decreasing altitude, our results suggest that the radio emission happens over 
a very narrow height range at least between 730\,MHz to 3100\,MHz.
However, according to a fan beam model developed by~\citet{Wang14}, the spectral 
variation across the emission region is responsible for the frequency dependence 
of the pulse width.

\begin{table*}
\begin{center}
\caption{Pulse widths for PPTA MSPs.}
\label{tableWidth}
\begin{tabular}{lcccccccccccc}
\hline
PSR              & \multicolumn{3}{c}{Overall width} &        & $W_{10}$&        &       &  $W_{50}$ &      &       &  $W_{\rm{s}}$&       \\
								 &  50\,cm & 20\,cm & 10\,cm         & 50\,cm & 20\,cm  & 10\,cm & 50\,cm& 20\,cm    &10\,cm&50\,cm &  20\,cm      &10\,cm \\
								 &  (deg) &  (deg) & (deg)           & (deg)  & (deg)   & (deg)  & (deg) &   (deg)   & (deg)& ($\rm{\mu s}$) & ($\rm{\mu s}$) & ($\rm{\mu s}$)  \\
\hline
J0437$-$4715     & 321.3 &  300.2 &  350.5  &   130.5     & 63.4   & 18.6   & 15.4  & 8.9   & 5.6     &  127.5  &  77.3   & 45.3    \\
J0613$-$0200     & 143.0 &  145.1 &  126.1  &   105.9     & 109.1  & 105.4  & 10.5  & 54.9  & 30.4    &  19.7   &  42.0   & 49.5    \\
J0711$-$6830     & 272.7 &  284.7 &  238.9  &   180.9     & 168.2  & 167.8  & 131.4 & 124.3 & 108.7   &  92.8   &  74.3   & 93.6    \\
J1017$-$7156     & 46.6  &  69.2  &  46.6   &   22.2      & 21.7   & 34.4   & 16.1  & 10.7  & 11.0    &  28.0   &  37.2   & 43.4    \\
J1022$+$1001     & 66.9  &  71.8  &  61.9   &   41.9      & 43.0   & 35.8   & 16.5  & 21.1  & 8.2     &  171.3  &  124.5  & 171.8   \\
	               &       &        &         &             &        &        &       &       &         &         &         &         \\     
J1024$-$0719     & 153.4 &  271.0 &  124.6  &   123.6     & 109.6  & 113.7  & 67.3  & 35.7  & 32.0    &  54.3   &  66.8   & 62.9    \\
J1045$-$4509     & 236.0 &  250.1 &  229.7  &   70.3      & 69.7   & 66.6   & 33.5  & 36.6  & 35.7    &  328.7  &  278.3  & 297.8   \\
J1446$-$4701     & 53.5  &  91.6  &  23.2   &   49.3      & 45.2   & 37.7   & 12.4  & 12.2  & 11.5    &  36.7   &  45.0   & 39.4    \\
J1545$-$4550     &       &  189.5 &  49.3   &             & 56.8   & 43.9   &       & 12.8  & 9.2     &         &  55.4   & 39.1    \\
J1600$-$3053     & 55.7  &  76.8  &  63.4   &   48.6      & 41.3   & 42.1   & 11.2  & 9.3   & 22.7    &  70.5   &  62.5   & 46.2    \\
	               &       &        &         &             &        &        &       &       &         &         &         &         \\     
J1603$-$7202     & 76.4  &  230.1 &  222.4  &   48.3      & 41.8   & 38.5   & 32.4  & 29.4  & 7.0     &  203.8  &  143.4  & 147.7   \\
J1643$-$1224     & 164.1 &  221.9 &  192.3  &   83.8      & 72.6   & 65.7   & 32.8  & 24.9  & 20.5    &  245.2  &  209.1  & 159.7   \\
J1713$+$0747     & 98.9  &  198.5 &  99.6   &   42.3      & 30.3   & 29.6   & 16.4  & 8.8   & 8.3     &  120.4  &  64.6   & 58.6    \\
J1730$-$2304     & 188.3 &  252.3 &  198.5  &   68.9      & 76.0   & 73.0   & 34.2  & 43.2  & 43.8    &  164.9  &  99.1   & 90.2    \\
J1744$-$1134     & 167.9 &  200.6 &  160.8  &   24.0      & 21.9   & 20.1   & 13.1  & 12.3  & 8.8     &  65.2   &  64.8   & 57.1    \\
	               &       &        &         &             &        &        &       &       &         &         &         &         \\     
J1824$-$2452A    & 288.0 &  283.8 &  190.6  &   219.1     & 191.0  & 170.0  & 113.4 &115.4  &7.7      &  47.2   &  30.1   & 40.9    \\
J1832$-$0836     &       &  285.3 &  253.6  &             & 244.1  & 213.7  &       & 113.2 &  6.9    &         &  13.5   & 22.9    \\
J1857$+$0943     & 223.8 &  242.5 &  232.6  &   219.0     & 202.4  & 203.4  & 42.4  & 35.2  & 31.2    &  101.2  &  106.7  & 59.4    \\
J1909$-$3744     & 178.2 &  190.2 &  19.0   &   13.1      & 11.0   & 9.2    & 6.9   &  5.3  & 4.3     &  27.7   &  22.8   & 19.4    \\
J1939$+$2134     & 306.4 &  337.4 &  306.4  &   207.0     & 199.3  & 204.5  & 195.1 &182.1  & 10.5    &  16.3   &  25.0   & 21.5    \\
	               &       &        &         &             &        &        &       &       &         &         &         &         \\     
J2124$-$3358     & 320.6 &  332.2 &  281.2  &   255.1     & 269.7  & 282.9  & 168.3 & 37.5  & 31.8    &  96.2   &  153.4  & 121.9   \\
J2129$-$5721     & 72.6  &  157.8 &  67.6   &   37.5      & 60.0   & 88.4   & 22.9  & 25.5  & 53.8    &  74.7   &  78.8   & 50.9    \\
J2145$-$0750     & 256.5 &  267.4 &  180.9  &   94.1      & 93.6   & 91.1   & 9.1   & 7.6   & 7.8     &  206.8  &  206.6  & 196.0   \\
J2241$-$5236     & 74.7  &  209.9 &  43.7   &   18.8      & 20.3   & 21.0   & 10.3  & 10.6  & 9.8     &  26.3   &  28.7   & 26.8    \\
\hline
\end{tabular}
\end{center}
\end{table*}

\begin{table*}
\begin{center}
\caption{Widths of pulse components for PPTA MSPs whose mean pulse profiles have multiple, well defined components.}
\label{tableWidth2}
\begin{tabular}{lccccccc}
\hline
PSR                           & Component   &   \multicolumn{3}{c}{$W_{50}$}&  \multicolumn{3}{c}{$W_{80}$}     \\
								              &             & 50\,cm  & 20\,cm   & 10\,cm   & 50\,cm  & 20\,cm   & 10\,cm   \\
								              &             & (deg)   & (deg)    &  (deg)   &  (deg)  & (deg)    &   (deg)  \\
\hline
J0711$-$6830                  & C1   &15   $\pm$ 4   & 10   $\pm$ 1   & 8.8  $\pm$ 0.7  & 5.6  $\pm$ 0.7 & 5.3 $\pm$ 0.7 & 4.2 $\pm$ 0.7 \\ 
J1017$-$7156                  & C3   &16   $\pm$ 1   & 10   $\pm$ 3   & 10   $\pm$ 1    & 9    $\pm$ 3   & 6   $\pm$ 3   & 6   $\pm$ 3   \\ 
J1600$-$3053                  & C9   &11   $\pm$ 1   & 9.2  $\pm$ 0.7 & 7    $\pm$ 3    & 6    $\pm$ 1   & 5   $\pm$ 1   & 3   $\pm$ 1   \\ 
                              &      &               &                &                 &                &               &               \\
\multirow{2}{*}{J1603$-$7202} & C10  &12.7 $\pm$ 0.7 & 7.4  $\pm$ 0.7 & 7.0  $\pm$ 0.7  & 6.7  $\pm$ 0.7 & 3.9 $\pm$ 0.7 & 3.9 $\pm$ 0.4 \\  
                              & C11  &14.1 $\pm$ 0.7 & 11.3 $\pm$ 0.7 & 9.9  $\pm$ 0.7  & 7.4  $\pm$ 0.4 & 6.3 $\pm$ 0.7 & 5   $\pm$ 2   \\ 
                              &      &               &                &                 &                &               &               \\
J1643$-$1224                  & C12  &33   $\pm$ 1   & 25   $\pm$ 1   & 20   $\pm$ 1    & 17   $\pm$ 1   & 11  $\pm$ 1   & 8   $\pm$ 1   \\ 
J1713$+$0747                  & C13  &17   $\pm$ 1   & 8.8  $\pm$ 0.7 & 8    $\pm$ 2    & 6.3  $\pm$ 0.7 & 4.2 $\pm$ 0.7 & 3.9 $\pm$ 0.7 \\ 
J1744$-$1134                  & C16  &13.0 $\pm$ 0.7 & 12.3 $\pm$ 0.7 & 8.8  $\pm$ 0.7  & 7    $\pm$ 1   & 4.9 $\pm$ 0.7 & 3.9 $\pm$ 0.7 \\ 
J1824$-$2452A                 & C18  &13   $\pm$ 3   & 9    $\pm$ 3   & 9    $\pm$ 3    & 7    $\pm$ 3   & 4   $\pm$ 3   & 6   $\pm$ 2   \\ 
J1909$-$3744                  & C21  &6    $\pm$ 1   & 5    $\pm$ 1   & 4    $\pm$ 1    & 2    $\pm$ 1   & 2   $\pm$ 1   & 2   $\pm$ 1   \\ 
                              &      &               &                &                 &                &               &               \\
\multirow{2}{*}{J1939$+$2134} & C22  &11   $\pm$ 3   & 14   $\pm$ 3   & 11   $\pm$ 3    & 7    $\pm$ 3   & 9   $\pm$ 3   & 7   $\pm$ 3   \\ 
                              & C23  &13   $\pm$ 3   & 16   $\pm$ 3   & 11   $\pm$ 3    & 7    $\pm$ 3   & 9   $\pm$ 3   & 6   $\pm$ 3   \\ 
                              &      &               &                &                 &                &               &               \\
J2145$-$0750                  & C26  &8.8  $\pm$ 0.7 & 7.7  $\pm$ 0.7 & 8.1  $\pm$ 0.7  & 4.6  $\pm$ 0.7 & 3.5 $\pm$ 0.7 & 3.2 $\pm$ 0.7 \\ 
J2241$-$5236                  & C28  &11   $\pm$ 1   & 11   $\pm$ 1   & 10   $\pm$ 1    & 4    $\pm$ 1   & 6   $\pm$ 1   & 6   $\pm$ 1   \\ 
\hline                                                                                                         
\end{tabular}
\end{center}
\end{table*}

\begin{table*}
\begin{center}
\caption{Component separation for PPTA MSPs whose mean pulse profiles have multiple, well defined components.}
\label{separation}
\begin{tabular}{lcccc}
\hline
PSR              & Component        &         & Component separation &      \\
								 &                  & 730 MHz & 1400 MHz             & 3100 MHz \\
								 &                  &  (deg)  & (deg)                &   (deg)  \\
\hline
J0711$-$6830     &  C1, C2          &99.6  $\pm$ 0.6  & 97.1  $\pm$ 0.4 & 91.1 $\pm$ 0.6  \\
J1022$+$1001     &  C4, C5          &12.0  $\pm$ 0.4  & 12.0  $\pm$ 0.5 & 12.0 $\pm$ 0.5  \\
J1024$-$0719     &  C6, C7          &16.2  $\pm$ 0.5  & 16.5  $\pm$ 0.4 & 17.6 $\pm$ 0.3  \\
J1600$-$3053     &  C8, C9          & 9    $\pm$ 3    & 12.0  $\pm$ 0.8 &  14  $\pm$ 1    \\
	               &                  &                 &                 &                 \\
J1603$-$7202     &  C10, C11        &21.1  $\pm$ 0.7  & 21.8  $\pm$ 0.5 & 21.8 $\pm$ 0.9  \\
J1730$-$2304     &  C14, C15        &17.9  $\pm$ 0.4  & 17.2  $\pm$ 0.5 & 12.7 $\pm$ 0.5  \\
J1824$-$2452A    &  C17, C18        &  106 $\pm$ 2    &  107  $\pm$ 2   & 110  $\pm$ 2    \\
J1857$+$0943     &  C19, C20        & 165.4$\pm$ 0.6  & 164.0 $\pm$ 0.4 &163.3 $\pm$ 0.7  \\
	               &                  &                 &                 &                 \\
J1939$+$2134     &  C22, C23        & 172  $\pm$ 2    & 174   $\pm$ 2   &  172 $\pm$ 2    \\
J2129$-$5721     &  C24, C25        & 10.6 $\pm$ 0.8  &   11  $\pm$ 1   &   8  $\pm$ 3    \\
J2145$-$0750     &  C26, C27        & 78.1 $\pm$ 0.4  & 79.2  $\pm$ 0.4 &79.5  $\pm$ 0.8  \\ 
\hline
\end{tabular}
\end{center}
\end{table*}

\subsection{Flux Densities and Spectral indices}

\begin{figure*}
\begin{center}
\includegraphics[width=6 in,trim=0 3cm 0 3cm]{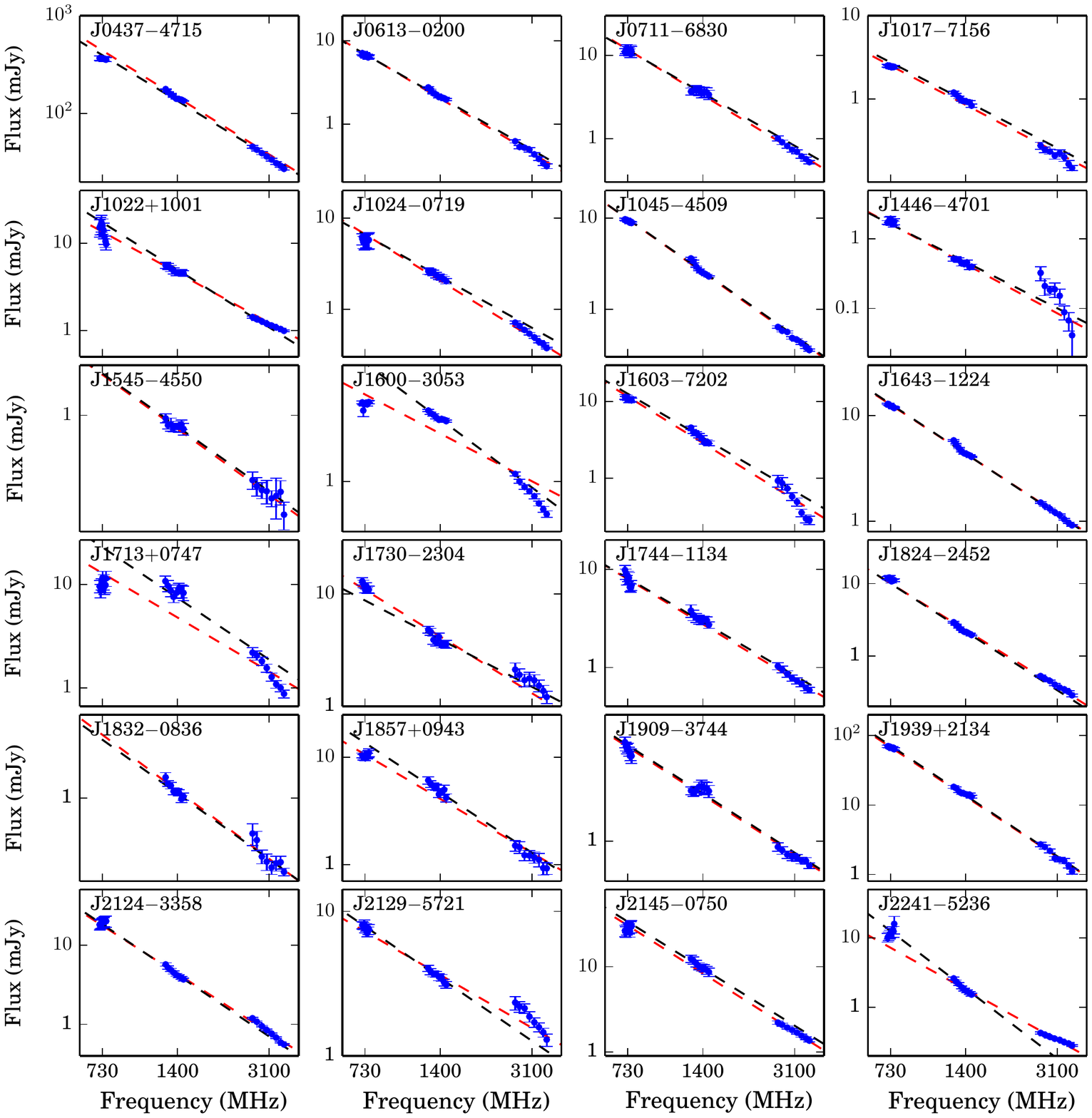}
\caption{Flux density spectra for $24$ MSPs. Red and black dashed lines show the power-law 
spectra with spectral indices $\alpha_1$ and $\alpha_2$ respectively.} 
\label{index}
\end{center}
\end{figure*}

In Table~\ref{tableFlux}, we present the flux densities and spectral indices 
for all the MSPs in our sample. As described in Section $2.2$, measuring flux 
densities is not trivial as each pulsar's flux density varies because 
of diffractive and refractive scintillation. Using summed profiles weighted by 
(S/N)$^2$ leads to results that are biased high. For the analysis presented 
here we therefore make use of the individual profiles. 

For each individual observation of each pulsar we calculate the mean flux 
density by averaging over the entire Stokes I profile. 
The $S_{730}$, $S_{1400}$ and $S_{3100}$ measurements given in Table~\ref{tableFlux} 
are calculated by averaging all the mean flux densities for a given pulsar 
in an observing band.  
The variance of the individual measurements in the three bands are tabulated 
as $S^{\rm{RMS}}_{730}$, $S^{\rm{RMS}}_{1400}$ and $S^{\rm{RMS}}_{3100}$ respectively. 
The uncertainty of the mean flux density is estimated as, $S^{\rm{RMS}}/(N-1)^{1/2}$, 
and $N$ is the number of observations.
The mean flux densities of several pulsars (e.g., PSRs J0711$-$6830, J1022$+$1001) 
are significantly different from~\citet{Yan11}. For these pulsars we found that 
they have relatively large flux variances compared with their mean flux densities, 
indicating that the flux discrepancies with previous work are caused by interstellar 
scintillation effects.

The good S/N that we measure in individual observations for most of the pulsars allows us 
to obtain measurements of the variation in the flux density within each observing band. 
We therefore divided each band into eight subbands (for PSRs J1545$-$4550 and J1832$-$0836, 
we only have a few observations in the 50\,cm band and the S/N are low, therefore we 
did not present their flux densities in the 50\,cm band).
Flux densities are obtained in each subband and are plotted in Fig.~\ref{index}. 
In this case, the best fit power-law spectra are indicated with red dashed lines and the corresponding 
spectral indices, $\alpha_1$, are given in Table~\ref{tableFlux}.
For several pulsars (e.g., PSRs J0437$-$4715, J1022$+$1001, J2241$-$5236), the flux density 
fluctuations caused by insterstellar scintillation result in large uncertainties in 
mean flux densities and affect the fitting for spectral indices, especially when the 
spectra deviate from a single power-law.
Therefore, for comparison, we also calculated flux densities using the summed profiles 
only weighted by the observing time. The uncertainty of flux density is estimated 
as the baseline rms noise of the profile. 
The best fit power-law spectra are indicated with black dashed lines in Fig.~\ref{index} 
and the corresponding spectral indices, $\alpha_2$, are given in the last column of Table~\ref{tableFlux}.

As shown in Fig.~\ref{index}, the spectrum of some MSPs can be generally modelled as 
a single power-law across a wide range of frequency (e.g., PSRs J0613$-$0200, J0711$-$6830, 
J1017$-$7156, J1643$-$1224, J1824$-$2452A, J1939$+$2134). 
For most pulsars whose spectra deviate from a single power-law, their spectra 
become steeper at high frequencies (e.g., PSRs J0437$-$4715, J1024$+$0719, 
J1603$-$7202) as also reported in normal pulsars~\citep[e.g.,][]{Maron00}.
Exceptions are PSRs J1022$+$1001 and J2241$-$5236 whose spectra become flatter at high 
frequencies.
For PSRs J1600$-$3053, J1713$+$0747, J2124$-$3358, J2145$-$0750 and J2241$-$5236, we 
observed positive spectral indices within the 50\,cm band. Such spectral features have 
been observed in normal pulsars~\citep[e.g.,][]{Kijak11}, but not in MSPs.
For pulsars whose spectra significantly deviate from a single power-law and have large 
flux density fluctuations, for instance PSRs J1022$+$1001, J1024$-$0719  
and J2241$-$5236, the spectral indices, $\alpha_1$ and $\alpha_2$, show large 
differences.
We note that in order to produce high S/N polarization profiles, in the data processing 
we have abandoned observations that are either too weak to see any profile, or have bad 
calibration files or are affected by radio-frequency interference. 
Therefore, for MSPs that have relatively steep spectra and have only a few available 
observations, the flux densities in the 10\,cm band are likely to be biased by several 
bright observations. Two examples are PSRs J1446$-$4701 and J2129$-$5721.  

The spectral indices are consistent with the results presented in~\citet{Toscano98}, 
but our measurements have significantly smaller uncertainties. However, compared with 
~\citet{Kramer99}, the spectral indices do show discrepancies for some pulsars. 
For instance, \citet{Kramer99} published a spectral index of $-1.17\pm0.06$ for
PSR J0437$-$4715, and we obtained a much steeper spectrum with a spectral index of 
$-1.69\pm0.03$. Fig.~\ref{index} shows that our fitting is dominated by the shape of 
the spectra in the 10\,cm and 20\,cm bands, and the spectrum becomes flatter in the 
50\,cm band. Therefore the discrepancy is likely because \citet{Kramer99} used a 
very wide frequency range without any information within bands. 
We derived a mean spectral index of $-1.76\pm0.01$ for $\alpha_1$ and $-1.81\pm0.01$ 
for $\alpha_2$. This is consistent with previous results of MSPs~\citep{Toscano98,Kramer99} 
and close to the observed spectral index of normal pulsars~\citep{Lorimer95,Maron00}. 

The bottom part of the left-side panels of Fig. \ref{0437} to 
\ref{2241} shows the phase-resolved spectral index for each MSP. 
As the phase-resolved spectral index is derived from the summed profiles weighted 
by the observing time, we compare them with the mean spectral index, $\alpha_2$, 
which is shown with a red dashed line in each figure and its uncertainty shown as 
yellow highlighted region.
In many cases the spectral indices vary significantly at different profile phases. 
For instance, in PSR~J0437$-$4715 the spectral index varies from approximately $-1$ to $-2$ in 
different parts of the profile. For PSR~J1022$+$1001 one component has a spectral 
index of approximately $-1.5$ and the other $-2.5$. 

In most, but not all cases, the variations in the spectral index as a function of 
pulse phase follow the components in the total intensity profile. 
Although we do not find strong correlations between the phase-resolved spectral 
index and the pulse profile, we clearly see that different pulse profile components usually 
have different spectral indices and they overlap with each other. In some cases, 
the peaks of pulse profile components coincide with the local maximum or minimum 
of the phase-resolved spectral index, which can naturally explain the frequency 
evolution of the width of pulse component presented in Table \ref{tableWidth2}. 
For models assuming that the emission from a single subregion of pulsar magnetosphere, 
e.g., a flux tube of plasma flow, is broadband~\citep[e.g.,][]{Michel87,Dyks10,Wang14},
such features imply a spatial spectral distribution within each subregion.

\begin{figure}
\begin{center}
\includegraphics[width=3 in,trim=0 3cm 0 3cm]{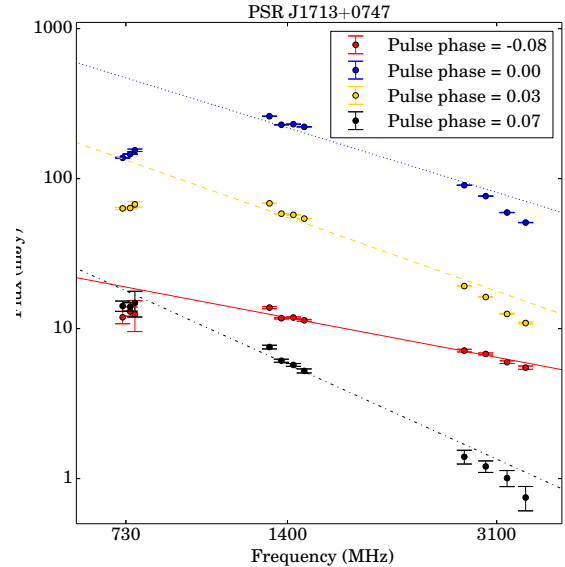}
\caption{Flux density spectra for PSR J1713$+$0747 at different pulse phases. The best fit power-law 
spectra are indicated with different types of lines.} 
\label{1713SI}
\end{center}
\end{figure}

The uncertainties placed on the phase-resolved spectral indices are determined from 
the errors in determining the flux density in the different observing bands and also 
from the goodness-of-fit for the single power-law model.  
Regions with high uncertainties but high S/N profiles are therefore regions in 
which the spectra do not fit a single power-law. 
Fig.~\ref{1713SI} shows the flux density spectra for PSR J1713$+$0747 at several 
different pulse phases. Close to phase zero, the turnover of the spectrum at around 
1400\,MHz becomes significant and therefore the uncertainty of the phase-resolved 
spectral index is much larger than those at other pulse phases.
For almost all pulsars in our sample, the uncertainty of the phase-resolved spectral 
index varies across the profile, indicating that different profile components  
can have quite different spectral shapes. For some pulsars (e.g., PSRs J0613$-$0200, 
J1643$-$1224 and J1939$+$2134), even though their mean flux density follows a 
single power-law very well across bands, the spectrum of individual profile 
components significantly deviate from a single power-law.

\begin{table*}
\centering
\caption{Flux densities and spectral indices for PPTA MSPs.}
\label{tableFlux}
\begin{tabular}{lcccccccc}
\hline
PSR              & $S_{730}$&$S^{\rm{RMS}}_{730}$&$S_{1400}$&$S^{\rm{RMS}}_{1400}$&$S_{3100}$&$S^{\rm{RMS}}_{3100}$& \multicolumn{2}{c}{Spectral index} \\
								 &  (mJy)   &    (mJy)           & (mJy)    &    (mJy)            &  (mJy)   &    (mJy)            &   $\alpha_{1}$ & $\alpha_{2}$ \\
\hline
 J0437$-$4715  &  364.3 $\pm$ 19.2 &  255.2 &  150.2 $\pm$ 1.6  &  42.2 &  35.6 $\pm$ 1.2  &  20.5  &  $-$1.69 $\pm$ 0.03 &  $-$1.65 $\pm$ 0.02 \\ 
 J0613$-$0200  &  6.7   $\pm$ 0.3  &  2.3   &  2.25  $\pm$ 0.03 &  0.4  &  0.45 $\pm$ 0.01 &  0.1   &  $-$1.90 $\pm$ 0.03 &  $-$1.83 $\pm$ 0.03 \\ 
 J0711$-$6830  &  11.4  $\pm$ 1.0  &  8.5   &  3.7   $\pm$ 0.4  &  5.7  &  0.72 $\pm$ 0.04 &  0.4   &  $-$1.94 $\pm$ 0.03 &  $-$1.83 $\pm$ 0.05 \\ 
 J1017$-$7156  &  2.5   $\pm$ 0.1  &  0.8   &  0.99  $\pm$ 0.04 &  0.4  &  0.21 $\pm$ 0.01 &  0.1   &  $-$1.67 $\pm$ 0.04 &  $-$1.64 $\pm$ 0.04 \\ 
 J1022$+$1001  &  14.2  $\pm$ 2.8  &  22.9  &  4.9   $\pm$ 0.4  &  4.6  &  1.18 $\pm$ 0.03 &  0.4   &  $-$1.66 $\pm$ 0.03 &  $-$1.91 $\pm$ 0.06 \\ 
               &	                 &        &                   &       &                  &        &                     &                     \\ 
 J1024$-$0719  &  5.6   $\pm$ 0.8  &  4.9   &  2.3   $\pm$ 0.2  &  1.7  &  0.52 $\pm$ 0.01 &  0.1   &  $-$1.80 $\pm$ 0.03 &  $-$1.62 $\pm$ 0.05 \\ 
 J1045$-$4509  &  9.2   $\pm$ 0.2  &  1.8   &  2.74  $\pm$ 0.04 &  0.5  &  0.48 $\pm$ 0.01 &  0.1   &  $-$2.06 $\pm$ 0.02 &  $-$2.04 $\pm$ 0.03 \\ 
 J1446$-$4701  &  1.8   $\pm$ 0.1  &  0.5   &  0.46  $\pm$ 0.02 &  0.2  &  0.15 $\pm$ 0.02 &  0.07  &  $-$2.05 $\pm$ 0.07 &  $-$1.93 $\pm$ 0.09 \\ 
 J1545$-$4550  &                   &        &  0.87  $\pm$ 0.05 &  0.2  &  0.34 $\pm$ 0.04 &  0.1   &  $-$1.15 $\pm$ 0.07 &  $-$1.13 $\pm$ 0.06 \\ 
 J1600$-$3053  &  2.9   $\pm$ 0.1  &  0.4   &  2.44  $\pm$ 0.04 &  0.4  &  0.84 $\pm$ 0.02 &  0.2   &  $-$0.83 $\pm$ 0.07 &  $-$1.19 $\pm$ 0.05 \\ 
               &	                 &        &                   &       &                  &        &                     &                     \\   
 J1603$-$7202  &  10.9  $\pm$ 0.7  &  4.9   &  3.5   $\pm$ 0.2  &  1.7  &  0.55 $\pm$ 0.06 &  0.4   &  $-$2.15 $\pm$ 0.06 &  $-$2.03 $\pm$ 0.05 \\ 
 J1643$-$1224  &  12.4  $\pm$ 0.2  &  1.4   &  4.68  $\pm$ 0.06 &  0.7  &  1.18 $\pm$ 0.02 &  0.2   &  $-$1.64 $\pm$ 0.01 &  $-$1.66 $\pm$ 0.02 \\ 
 J1713$+$0747  &  10.1  $\pm$ 0.8  &  6.2   &  9.1   $\pm$ 0.7  &  8.4  &  2.6  $\pm$ 0.2  &  1.6   &  $-$1.06 $\pm$ 0.07 &  $-$1.2  $\pm$ 0.1 \\ 
 J1730$-$2304  &  11.5  $\pm$ 0.5  &  3.9   &  4.0   $\pm$ 0.2  &  2.0  &  1.7  $\pm$ 0.2  &  1.5   &  $-$1.46 $\pm$ 0.06 &  $-$1.22 $\pm$ 0.07 \\ 
 J1744$-$1134  &  8.0   $\pm$ 0.7  &  5.7   &  3.2   $\pm$ 0.3  &  3.2  &  0.77 $\pm$ 0.05 &  0.5   &  $-$1.63 $\pm$ 0.03 &  $-$1.58 $\pm$ 0.05 \\ 
               &	                 &        &                   &       &                  &        &                     &                     \\  
 J1824$-$2452A &  11.4  $\pm$ 0.5  &  2.9   &  2.30  $\pm$ 0.05 &  0.4  &  0.39 $\pm$ 0.01 &  0.1   &  $-$2.28 $\pm$ 0.03 &  $-$2.35 $\pm$ 0.03 \\ 
 J1832$-$0836  &	                 &        &  1.18  $\pm$ 0.07 &  0.3  &  0.32 $\pm$ 0.03 &  0.1   &  $-$1.66 $\pm$ 0.06 &  $-$1.60 $\pm$ 0.07 \\ 
 J1857$+$0943  &  10.4  $\pm$ 0.4  &  3.0   &  5.1   $\pm$ 0.3  &  2.9  &  1.2  $\pm$ 0.1  &  0.9   &  $-$1.46 $\pm$ 0.04 &  $-$1.63 $\pm$ 0.07 \\ 
 J1909$-$3744  &  4.9   $\pm$ 0.3  &  3.1   &  2.5   $\pm$ 0.2  &  3.2  &  0.76 $\pm$ 0.04 &  0.5   &  $-$1.29 $\pm$ 0.02 &  $-$1.29 $\pm$ 0.03 \\ 
 J1939$+$2134  &  67.8  $\pm$ 2.7  &  20.9  &  15.2  $\pm$ 0.6  &  6.2  &  1.82 $\pm$ 0.09 &  0.9   &  $-$2.52 $\pm$ 0.02 &  $-$2.54 $\pm$ 0.02 \\ 
               &	                 &        &                   &       &                  &        &                     &                     \\   
 J2124$-$3358  &  19.3  $\pm$ 2.7  &  17.2  &  4.5   $\pm$ 0.2  &  2.2  &  0.82 $\pm$ 0.01 &  0.1   &  $-$2.15 $\pm$ 0.03 &  $-$2.25 $\pm$ 0.03 \\ 
 J2129$-$5721  &  5.9   $\pm$ 0.5  &  3.9   &  1.28  $\pm$ 0.09 &  1.0  &  0.34 $\pm$ 0.05 &  0.2   &  $-$2.12 $\pm$ 0.07 &  $-$2.52 $\pm$ 0.05 \\ 
 J2145$-$0750  &  27.4  $\pm$ 3.4  &  28.5  &  10.3  $\pm$ 1.0  &  11.2 &  1.75 $\pm$ 0.07 &  0.8   &  $-$1.98 $\pm$ 0.03 &  $-$1.94 $\pm$ 0.04 \\ 
 J2241$-$5236  &  11.9  $\pm$ 1.8  &  16.2  &  1.95  $\pm$ 0.09 &  1.2  &  0.35 $\pm$ 0.01 &  0.1   &  $-$2.12 $\pm$ 0.04 &  $-$2.93 $\pm$ 0.07 \\ 
\hline
\end{tabular}
\end{table*}

\subsection{Polarization properties}

In Table \ref{tablePol}, the fractional linear polarization $\langle L \rangle/S$, 
the fractional net circular polarization $\langle V \rangle/S$ and the fractional absolute 
circular polarization $\langle|V|\rangle/S$ at different frequencies are presented. 
The means are taken across the pulse profile where the total intensity exceeds 
three times the baseline rms noise.
All the polarization parameters are calculated from the average polarization 
profiles and the uncertainties are estimated using the baseline rms noise 
(PSRs J1545$-$4550 and J1832$-$0836 have very low S/N profiles in the 50\,cm band, 
therefore we did not present these results in the 50\,cm band). 

For nine pulsars, we see a clear decrease in the mean fractional linear polarization 
with increasing frequency. In contrast, for PSRs J1045$-$4509, J1603$-$7202 and 
J1730$-$2304 and J1824$-$2452A, the mean fractional linear polarization significantly 
increases with frequency. 
Different profile components of a pulsar can show different frequency evolution 
of the fractional linear polarization. For instance, for PSR J1643$-$1224, the fractional 
linear polarization of the leading edge of the main pulse increases with decreasing 
frequency while that of the trailing edge decreases with decreasing frequency.
There is no evidence that highly polarized sources depolarize rapidly 
with increasing frequency as reported previously~\citep{Kramer99}.

Circular polarization also has complicated variations with both frequency and pulse 
phase, with different components often having different signs of circular polarization 
and/or opposite frequency dependence in the degree of circular polarization. For example, 
for PSR J1603$-$7202, the two main components have the same sign of circular polarization, 
but for the leading component, the circular polarization is much stronger at high frequency, 
whereas for the trailing component the opposite frequency dependence is seen. For J1017$-$7156, 
the main peak of the profile has overlapping components, one with negative $V$ and the 
other with positive $V$. These two components have very different spectral indices, so 
that at high frequencies the negative $V$ component dominates, whereas at low frequencies, 
the positive $V$ component, which is slightly narrower, is dominant.

We note that the high fractional linear and circular polarization of pulsars has been 
suggested as a way to distinguish pulsars from other point radio sources in a 
continuum survey~\citep[e.g.,][]{Crawford00}. However, in continuum surveys the signal 
is averaged over pulse phases. The Stokes parameters $Q$ and $U$ are initially
averaged separately in time and then the average is combined to form the linear polarization.
Therefore, the linear polarization of a continuum survey, $L_{\rm{C}}$, is calculated 
as $\langle L_{\rm{C}}\rangle = \sqrt{\langle Q\rangle^{2}+\langle U\rangle^{2}}$, which 
is often much less than $\langle L\rangle$ since $Q$ and $U$ can change sign across the profile. 
In order to aid predictions of the measured linear polarization for MSPs in future continuum 
surveys we therefore present, in Table~\ref{tablePol2}, the fractional amount of Stokes 
$Q$ and $U$ and the fractional linear polarization $\langle L_{\rm{C}} \rangle/S$.
As expected, these results show that the fractional linear polarization of a pulsar will 
be reduced in a continuum survey and therefore any predictions for the discovery of 
pulsars in future continuum surveys should make use of the results in Table~\ref{tablePol2}.
We note that the RM and DM for a particular source may not be known at or 
shortly after the time of a continuum survey, and therefore the fractional 
linear polarization could be further reduced. However, the circularly polarised 
flux component should remain unaffected.

The bottom parts of the right-side panels of Fig.~\ref{0437} to \ref{2241} show the 
phase-resolved fractional linear polarization for each MSP. 
For most of our MSPs, the phase-resolved fractional linear polarization is remarkably 
similar at different observing bands (examples include PSR~J0437$-$4715 and J1857$+$0943). 
However, for a few pulsars (such as PSR~J1022$+$1001) the fractional linear polarization 
differs between bands. We find no strong correlation between the phase-resolved spectral index 
and the fractional linear polarization. 
In pulsars such as PSRs J1603$-$7202, J1730$-$2304, J1939$+$2134, J2145$-$0750 and J2241$-$5236 
we see evidence that the main component has a lower fractional linear polarization 
than leading or trailing components~\citep[e.g.,][]{Basu15}. However, for PSR J1744$-$1134, 
we do not see high fractional linear polarizations in the precursor pulse.

At phase ranges where a PA transition occurs, the fractional linear 
polarization is significantly lower than other phase ranges, which can be explained as 
the overlap of orthogonal modes. However, we do not see significantly lower or higher 
fractional net circular polarization close to PA transitions.
We do not find strong relations between the size of the PA transition and 
the fractional linear polarization. Orthogonal mode transitions normally correspond 
to lower fractional linear polarization, but we also see low fractional linear 
polarizations for non-orthogonal transitions, for instance in PSRs J1045$-$4509 
and J1730$-$2304. 

\begin{figure*}
\begin{center}
\includegraphics[width=5.5 in,trim=0 6cm 0 6cm]{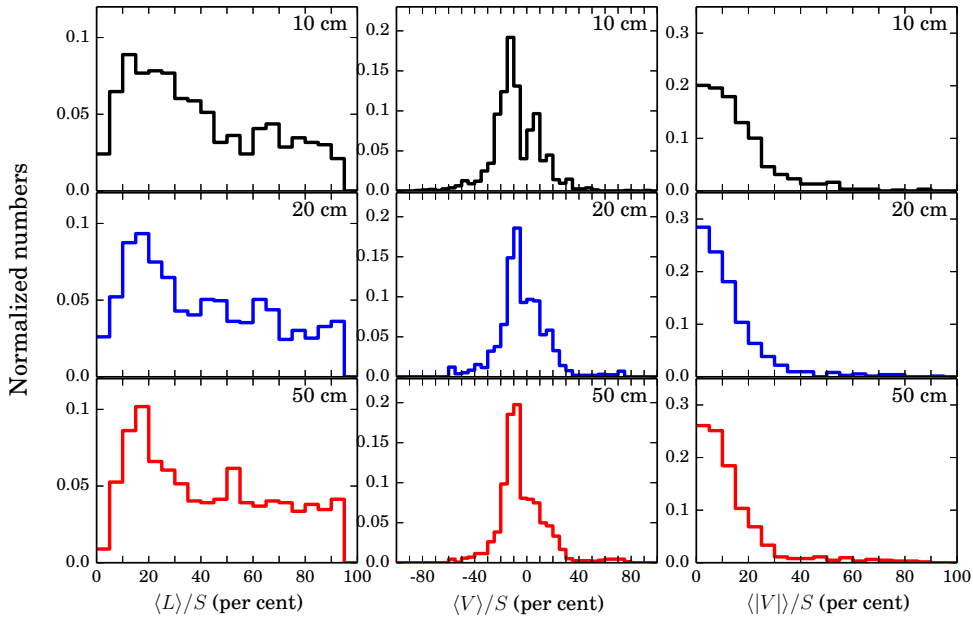}
\caption{Histograms of the phase-resolved fractional linear, net circular and absolute circular 
polarization for $24$ MSPs in three bands.}
\label{polHist}
\end{center}
\end{figure*}

In Fig.~\ref{polHist}, the distribution of phase-resolved fractional linear, 
circular and net circular polarization for 24 MSPs in three bands are shown.
To obtain the phase-resolved values, we rebinned the profile in each band into 
$128$ phase bins and only phase bins whose linear or circular polarization 
exceeds three times their baseline rms noise were used. 
While the distributions of the fractional linear polarization are similar 
across three bands, we see that both the distribution of fractional net circular 
and absolute circular polarization becomes narrower at lower 
frequencies. 
This indicates that the fractional circular and net circular polarization 
decrease with decreasing frequency.

\begin{table*}
\begin{center}
\caption{The fractional linear polarization $\langle L \rangle/S$, 
the fractional net circular polarization $\langle V \rangle/S$ and the fractional absolute 
circular polarization $\langle|V|\rangle/S$ for 24 PPTA MSPs.}
\label{tablePol}
\begin{tabular}{lccccccccc}
\hline
PSR              &                  &    $\langle L \rangle/S$    &                  &               & $\langle V \rangle/S$       &                  &      &      $\langle|V|\rangle/S$       &                      \\
								 &    50\,cm      &   20\,cm       &    10\,cm &    50\,cm      &   20\,cm       &    10\,cm &    50\,cm      &   20\,cm       &    10\,cm              \\
								 &     (per cent)   &         (per cent)          &     (per cent)   &    (per cent)   &         (per cent)          &     (per cent)   &   (per cent)   &         (per cent)          &     (per cent)  \\
\hline
J0437$-$4715& 26.6 $\pm$ 0.0& 25.1 $\pm $ 0.0& 20.4 $\pm$ 0.0&$ -4.2$ $\pm$ 0.0 &$ -2.9$ $\pm$ 0.0 &$ -8.0$ $\pm$ 0.0 & 15.4 $\pm$ 0.0 & 11.3 $\pm$ 0.0 & 12.4 $\pm$ 0.0 \\
J0613$-$0200& 28.9 $\pm$ 0.3& 21.0 $\pm $ 0.1& 14.7 $\pm$ 0.5&$ -6.5$ $\pm$ 0.3 &$ 5.2 $ $\pm$ 0.1 &$ 10.7$ $\pm$ 0.6 &  8.9 $\pm$ 0.3 &  5.6 $\pm$ 0.1 & 11.2 $\pm$ 0.6 \\
J0711$-$6830& 24.6 $\pm$ 0.2& 14.1 $\pm $ 0.1& 17   $\pm$ 2  &$-12.7$ $\pm$ 0.2 &$-12.9$ $\pm$ 0.1 &$ -24 $ $\pm$ 2   & 12.7 $\pm$ 0.2 & 13.1 $\pm$ 0.1 & 24   $\pm$ 2 \\
J1017$-$7156& 44.5 $\pm$ 0.7& 35.4 $\pm $ 0.3& 42   $\pm$ 1  &$  6.9$ $\pm$ 0.8 &$-28.9$ $\pm$ 0.2 &$ -38 $ $\pm$ 2   & 18.5 $\pm$ 0.8 & 29.5 $\pm$ 0.2 & 42   $\pm$ 2 \\
J1022$+$1001& 67.9 $\pm$ 0.1& 56.3 $\pm $ 0.0& 23.5 $\pm$ 0.2&$-13.4$ $\pm$ 0.1 &$-11.6$ $\pm$ 0.0 &$ -2.7$ $\pm$ 0.2 & 13.4 $\pm$ 0.1 & 12.6 $\pm$ 0.0 & 5.6  $\pm$ 0.2 \\
            &               &                &               &                &                &                &                &                &                \\
J1024$-$0719& 69.0 $\pm$ 0.6& 67.9 $\pm $ 0.1& 61.7 $\pm$ 0.8&$ 1.1 $ $\pm$ 0.6 &$  5.5$ $\pm$ 0.2 &$ 6.1 $ $\pm$ 0.7 &  3.7 $\pm$ 0.6 &  6.3 $\pm$ 0.2 & 6.7  $\pm$ 0.7 \\
J1045$-$4509& 18.7 $\pm$ 0.3& 22.5 $\pm $ 0.1& 30.2 $\pm$ 0.5&$ 8.2 $ $\pm$ 0.3 &$ 14.7$ $\pm$ 0.1 &$ 16.4$ $\pm$ 0.6 & 10.6 $\pm$ 0.3 & 16.6 $\pm$ 0.1 & 16.5 $\pm$ 0.6 \\
J1446$-$4701& 60.4 $\pm$ 2.8& 38   $\pm $ 1  & 0.0  $\pm$ 3.5&$ -13 $ $\pm$ 2   &$ -9  $ $\pm$ 1   &$ 0.0 $ $\pm$ 3.1 &  15  $\pm$ 3   &   11 $\pm$ 1   & 0.0  $\pm$ 3.1 \\
J1545$-$4550&               & 58   $\pm $ 1  & 59   $\pm$ 2  &                  &$-13.2$ $\pm$ 0.9 &$ -10 $ $\pm$ 2   &                & 17.1 $\pm$ 0.9 & 11   $\pm$ 2 \\
J1600$-$3053& 33   $\pm$ 2  & 31.3 $\pm $ 0.1& 36.8 $\pm$ 0.3&$ 0.4 $ $\pm$ 2   &$  3.8$ $\pm$ 0.1 &$ -2.3$ $\pm$ 0.3 &    3 $\pm$ 2   &  4.0 $\pm$ 0.1 & 4.7  $\pm$ 0.3 \\
            &               &                &               &                  &                  &                  &                &                &       \\
J1603$-$7202& 16.6 $\pm$ 0.2& 18.6 $\pm $ 0.1& 31.6 $\pm$ 0.7&$33.6 $ $\pm$ 0.3 &$ 29.0$ $\pm$ 0.1 &$ 15.3$ $\pm$ 0.8 & 34.2 $\pm$ 0.3 & 32.4 $\pm$ 0.1 & 22.3 $\pm$ 0.8 \\
J1643$-$1224& 20.0 $\pm$ 0.3& 17.4 $\pm $ 0.1& 19.9 $\pm$ 0.2&$ 6.8 $ $\pm$ 0.2 &$  0.4$ $\pm$ 0.1 &$ -6.6$ $\pm$ 0.2 & 13.9 $\pm$ 0.2 & 13.8 $\pm$ 0.1 & 10.4 $\pm$ 0.2 \\
J1713$+$0747& 33.3 $\pm$ 0.3& 31.5 $\pm $ 0.0& 27.0 $\pm$ 0.1&$-2.8 $ $\pm$ 0.2 &$  1.1$ $\pm$ 0.0 &$ -1.1$ $\pm$ 0.1 &  3.9 $\pm$ 0.2 &  3.8 $\pm$ 0.0 & 3.8  $\pm$ 0.1 \\
J1730$-$2304& 26.2 $\pm$ 0.3& 29.2 $\pm $ 0.1& 44.9 $\pm$ 0.2&$-19.1$ $\pm$ 0.3 &$-19.4$ $\pm$ 0.1 &$-11.9$ $\pm$ 0.2 & 19.2 $\pm$ 0.3 & 20.6 $\pm$ 0.1 & 15.9 $\pm$ 0.2 \\
J1744$-$1134& 88.9 $\pm$ 0.4& 91.8 $\pm $ 0.1& 88.0 $\pm$ 0.4&$ 0.2 $ $\pm$ 0.4 &$  2.9$ $\pm$ 0.1 &$  1.5$ $\pm$ 0.3 &  0.7 $\pm$ 0.4 &  2.9 $\pm$ 0.1 & 1.6  $\pm$ 0.3 \\
            &               &                &               &                  &                  &                  &                &                &               \\
J1824$-$2452A& 70.9 $\pm$ 0.5& 77.8 $\pm $ 0.2& 84.2 $\pm$ 1.0&$ 0.1 $ $\pm$ 0.3 &$  3.5$ $\pm$ 0.2 &$ -0.8$ $\pm$ 0.8 &  3.8 $\pm$ 0.3 &  4.4 $\pm$ 0.2 & 5.5  $\pm$ 0.8 \\
J1832$-$0836&               & 36   $\pm $ 2  & 43   $\pm$ 11 &                  &$   3 $ $\pm$ 1   &$   -4$ $\pm$ 10  &                &   10 $\pm$ 1   & 11   $\pm$ 10   \\
J1857$+$0943& 20.9 $\pm$ 0.9& 14.5 $\pm $ 0.1& 14.1 $\pm$ 0.4&$ -1.2$ $\pm$ 0.7 &$  2.5$ $\pm$ 0.1 &$  0.3$ $\pm$ 0.4 &  4.7 $\pm$ 0.7 &  5.8 $\pm$ 0.1 & 7.3  $\pm$ 0.4 \\
J1909$-$3744& 61.2 $\pm$ 0.4& 48.7 $\pm $ 0.1& 26.3 $\pm$ 0.2&$ 13.1$ $\pm$ 0.4 &$ 14.9$ $\pm$ 0.1 &$  5.0$ $\pm$ 0.2 & 15.4 $\pm$ 0.4 & 16.1 $\pm$ 0.1 & 6.6  $\pm$ 0.2 \\
J1939$+$2134& 38.1 $\pm$ 0.1& 30.0 $\pm $ 0.0& 24.3 $\pm$ 0.2&$ 0.9 $ $\pm$ 0.1 &$  3.3$ $\pm$ 0.0 &$ -0.2$ $\pm$ 0.2 &  1.1 $\pm$ 0.1 &  3.3 $\pm$ 0.0 & 1.2  $\pm$ 0.2 \\
            &               &                &               &                  &                  &                  &                &                &               \\
J2124$-$3358& 46.2 $\pm$ 0.2& 33.1 $\pm $ 0.1& 49   $\pm$ 1  &$ -2.5$ $\pm$ 0.2 &$  0.4$ $\pm$ 0.1 &$ -3.9$ $\pm$ 1.0 &  3.8 $\pm$ 0.2 &  5.5 $\pm$ 0.1 & 7    $\pm$ 1   \\
J2129$-$5721& 66.8 $\pm$ 0.6& 47.3 $\pm $ 0.2& 39   $\pm$ 8  &$-27.0$ $\pm$ 0.6 &$-24.8$ $\pm$ 0.2 &$ -16 $ $\pm$ 8   & 35.5 $\pm$ 0.6 & 26.6 $\pm$ 0.2 & 17   $\pm$ 8  \\
J2145$-$0750& 19.2 $\pm$ 0.1& 15.9 $\pm $ 0.0& 10.9 $\pm$ 0.1&$  5.9$ $\pm$ 0.1 &$  9.2$ $\pm$ 0.0 &$  0.9$ $\pm$ 0.1 &  9.5 $\pm$ 0.1 & 10.0 $\pm$ 0.0 & 8.1  $\pm$ 0.1 \\
J2241$-$5236& 20.0 $\pm$ 0.2& 12.6 $\pm $ 0.1& 12.5 $\pm$ 0.7&$ -2.9$ $\pm$ 0.2 &$ -0.7$ $\pm$ 0.1 &$ -4.2$ $\pm$ 0.7 &  4.7 $\pm$ 0.2 &  6.2 $\pm$ 0.1 & 8.9  $\pm$ 0.7 \\
\hline
\end{tabular}
\end{center}
\end{table*}

\begin{table*}
\begin{center}
\caption{The fractional amount of Stokes $Q$ and $U$ and the 
fractional linear polarization $\langle L_{\rm{C}} \rangle/S$ for 24 PPTA MSPs.}
\label{tablePol2}
\begin{tabular}{lccccccccc}
\hline
PSR              &                  &    $\langle Q \rangle/S$    &                  &               & $\langle U \rangle/S$       &                  &      &      $\langle L_{\rm{C}}\rangle/S$       &                      \\
								 &    50\,cm      &   20\,cm       &    10\,cm &    50\,cm      &   20\,cm       &    10\,cm &    50\,cm      &   20\,cm       &    10\,cm              \\
								 &     (per cent)   &         (per cent)          &     (per cent)   &    (per cent)   &         (per cent)          &     (per cent)   &   (per cent)   &         (per cent)          &     (per cent)  \\
\hline
J0437$-$4715 &$-3.0 $ $\pm$ 0.0 & $-1.8 $ $\pm$ 0.0 & $0.2  $ $\pm$ 0.0 & $0.7  $ $\pm$ 0.0 & $-4.3 $ $\pm$ 0.0 & $0.9  $ $\pm$ 0.0 & 3.0  $\pm$ 0.0 & 4.7  $\pm$ 0.0 & 0.9  $\pm$ 0.0 \\
J0613$-$0200 &$9.9  $ $\pm$ 0.3 & $8.9  $ $\pm$ 0.1 & $6.1  $ $\pm$ 0.4 & $-6.9 $ $\pm$ 0.3 & $8.3  $ $\pm$ 0.1 & $-3.1 $ $\pm$ 0.4 & 12.1 $\pm$ 0.3 & 12.1 $\pm$ 0.1 & 6.8  $\pm$ 0.4 \\
J0711$-$6830 &$-14.1$ $\pm$ 0.2 & $-6.2 $ $\pm$ 0.1 & $-1.0 $ $\pm$ 0.3 & $7.0  $ $\pm$ 0.1 & $-1.0 $ $\pm$ 0.1 & $0.6  $ $\pm$ 0.3 & 15.7 $\pm$ 0.2 & 6.3  $\pm$ 0.1 & 1.2  $\pm$ 0.3 \\
J1017$-$7156 &$23.6 $ $\pm$ 0.7 & $-17.9$ $\pm$ 0.2 & $-22.2$ $\pm$ 1.2 & $-22.2$ $\pm$ 0.7 & $22.3 $ $\pm$ 0.3 & $31.2 $ $\pm$ 1.3 & 32.4 $\pm$ 0.7 & 28.6 $\pm$ 0.3 & 38.4 $\pm$ 1.3 \\
J1022$+$1001 &$14.6 $ $\pm$ 0.1 & $28.5 $ $\pm$ 0.0 & $18.1 $ $\pm$ 0.2 & $22.7 $ $\pm$ 0.1 & $8.4  $ $\pm$ 0.1 & $-0.5 $ $\pm$ 0.2 & 27.0 $\pm$ 0.1 & 29.7 $\pm$ 0.1 & 18.1 $\pm$ 0.2 \\
             &                &                 &                &                   &                   &                   &                &                &                 \\
J1024$-$0719 &$-20.8$ $\pm$ 0.5 & $-49.3$ $\pm$ 0.1 & $-30.6$ $\pm$ 0.4 & $42.1 $ $\pm$ 0.5 & $27.4 $ $\pm$ 0.1 & $14.9 $ $\pm$ 0.5 & 47.0 $\pm$ 0.5 & 56.4 $\pm$ 0.1 & 34.1 $\pm$ 0.5 \\
J1045$-$4509 &$-6.9 $ $\pm$ 0.2 & $8.6  $ $\pm$ 0.1 & $0.2  $ $\pm$ 0.4 & $-9.6 $ $\pm$ 0.3 & $-8.7 $ $\pm$ 0.1 & $-14.6$ $\pm$ 0.4 & 11.8 $\pm$ 0.3 & 12.2 $\pm$ 0.1 & 14.6 $\pm$ 0.4 \\
J1446$-$4701 &$34.6 $ $\pm$ 2.1 & $1.7  $ $\pm$ 1.0 & $6.6  $ $\pm$ 4.2 & $-36.2$ $\pm$ 2.2 & $28.8 $ $\pm$ 0.9 & $12.3 $ $\pm$ 4.1 & 50.1 $\pm$ 2.2 & 28.8 $\pm$ 0.9 & 14.0 $\pm$ 4.1 \\
J1545$-$4550 &                  & $-14.9$ $\pm$ 0.8 & $-40.3$ $\pm$ 1.7 &                   & $-32.5$ $\pm$ 0.6 & $-12.4$ $\pm$ 1.5 &                & 35.7 $\pm$ 0.7 & 42.2 $\pm$ 1.7 \\
J1600$-$3053 &$-4.4 $ $\pm$ 0.9 & $-5.7 $ $\pm$ 0.1 & $2.1  $ $\pm$ 0.3 & $-18.0$ $\pm$ 1.0 & $-5.3 $ $\pm$ 0.1 & $-16.0$ $\pm$ 0.3 & 18.5 $\pm$ 1.0 & 7.8  $\pm$ 0.1 & 16.1 $\pm$ 0.3 \\
             &                &                 &                &                   &                   &                   &                &                &                 \\
J1603$-$7202 &$-3.5 $ $\pm$ 0.2 & $-2.5 $ $\pm$ 0.1 & $-10.0$ $\pm$ 0.5 & $-3.1 $ $\pm$ 0.3 & $0.4  $ $\pm$ 0.1 & $0.5  $ $\pm$ 0.5 & 4.7  $\pm$ 0.2 & 2.5  $\pm$ 0.1 & 10.0 $\pm$ 0.5 \\
J1643$-$1224 &$12.9 $ $\pm$ 0.2 & $2.3  $ $\pm$ 0.1 & $4.5  $ $\pm$ 0.2 & $-5.3 $ $\pm$ 0.2 & $-2.7 $ $\pm$ 0.1 & $0.6  $ $\pm$ 0.2 & 14.0 $\pm$ 0.2 & 3.5  $\pm$ 0.1 & 4.5  $\pm$ 0.2 \\
J1713$+$0747 &$-9.4 $ $\pm$ 0.3 & $-2.9 $ $\pm$ 0.0 & $-2.0 $ $\pm$ 0.1 & $11.9 $ $\pm$ 0.3 & $4.2  $ $\pm$ 0.0 & $6.0  $ $\pm$ 0.1 & 15.2 $\pm$ 0.3 & 5.1  $\pm$ 0.0 & 6.4  $\pm$ 0.1 \\
J1730$-$2304 &$9.4  $ $\pm$ 0.2 & $-11.6$ $\pm$ 0.1 & $-19.0$ $\pm$ 0.2 & $13.3 $ $\pm$ 0.3 & $19.3 $ $\pm$ 0.1 & $18.2 $ $\pm$ 0.2 & 16.3 $\pm$ 0.3 & 22.5 $\pm$ 0.1 & 26.3 $\pm$ 0.2 \\
J1744$-$1134 &$-70.5$ $\pm$ 0.4 & $-48.7$ $\pm$ 0.1 & $-28.7$ $\pm$ 0.3 & $34.8 $ $\pm$ 0.4 & $67.7 $ $\pm$ 0.1 & $69.3 $ $\pm$ 0.4 & 78.6 $\pm$ 0.4 & 83.4 $\pm$ 0.1 & 75.0 $\pm$ 0.4 \\
             &                &                 &                &                   &                   &                   &                &                &                 \\
J1824$-$2452A&$-34.7$ $\pm$ 0.5 & $-19.0$ $\pm$ 0.2 & $-33.9$ $\pm$ 0.9 & $26.8 $ $\pm$ 0.4 & $43.2 $ $\pm$ 0.1 & $35.3 $ $\pm$ 0.6 & 43.8 $\pm$ 0.5 & 47.2 $\pm$ 0.1 & 48.9 $\pm$ 0.8 \\
J1832$-$0836 &                  & $-1.2 $ $\pm$ 0.9 & $-15.4$ $\pm$ 2.1 &                   & $2.8  $ $\pm$ 0.8 & $-8.8 $ $\pm$ 2.2 &                & 3.1  $\pm$ 0.8 & 17.7 $\pm$ 2.1 \\
J1857$+$0943 &$2.0  $ $\pm$ 0.5 & $0.3  $ $\pm$ 0.1 & $0.3  $ $\pm$ 0.3 & $0.9  $ $\pm$ 0.4 & $-0.6 $ $\pm$ 0.1 & $4.8  $ $\pm$ 0.3 & 2.2  $\pm$ 0.5 & 0.6  $\pm$ 0.1 & 4.8  $\pm$ 0.3 \\
J1909$-$3744 &$-54.0$ $\pm$ 0.4 & $-42.9$ $\pm$ 0.1 & $-23.5$ $\pm$ 0.3 & $-12.9$ $\pm$ 0.4 & $-15.1$ $\pm$ 0.1 & $-9.9 $ $\pm$ 0.2 & 55.5 $\pm$ 0.4 & 45.4 $\pm$ 0.1 & 25.5 $\pm$ 0.3 \\
J1939$+$2134 &$-30.5$ $\pm$ 0.1 & $2.0  $ $\pm$ 0.0 & $0.1  $ $\pm$ 0.2 & $-11.8$ $\pm$ 0.1 & $16.2 $ $\pm$ 0.0 & $-6.9 $ $\pm$ 0.2 & 32.7 $\pm$ 0.1 & 16.3 $\pm$ 0.0 & 6.9  $\pm$ 0.2 \\
             &                &                 &                &                   &                   &                   &                &                &                 \\
J2124$-$3358 &$19.4 $ $\pm$ 0.1 & $9.0  $ $\pm$ 0.1 & $-2.0 $ $\pm$ 0.4 & $8.2  $ $\pm$ 0.2 & $10.6 $ $\pm$ 0.1 & $5.4  $ $\pm$ 0.5 & 21.1 $\pm$ 0.2 & 13.9 $\pm$ 0.1 & 5.8  $\pm$ 0.5 \\
J2129$-$5721 &$22.2 $ $\pm$ 0.5 & $-24.4$ $\pm$ 0.2 & $-3.1 $ $\pm$ 2.1 & $45.1 $ $\pm$ 0.5 & $29.6 $ $\pm$ 0.2 & $16.4 $ $\pm$ 2.2 & 50.2 $\pm$ 0.5 & 38.3 $\pm$ 0.2 & 16.7 $\pm$ 2.2 \\
J2145$-$0750 &$-5.9 $ $\pm$ 0.1 & $-2.0 $ $\pm$ 0.0 & $2.5  $ $\pm$ 0.1 & $6.2  $ $\pm$ 0.1 & $3.2  $ $\pm$ 0.0 & $2.3  $ $\pm$ 0.1 & 8.6  $\pm$ 0.1 & 3.8  $\pm$ 0.0 & 3.3  $\pm$ 0.1 \\
J2241$-$5236 &$7.8  $ $\pm$ 0.2 & $0.4  $ $\pm$ 0.1 & $8.5  $ $\pm$ 0.7 & $-14.2$ $\pm$ 0.2 & $5.0  $ $\pm$ 0.1 & $-3.8 $ $\pm$ 0.7 & 16.2 $\pm$ 0.2 & 5.0  $\pm$ 0.1 & 9.3  $\pm$ 0.7 \\
\hline
\end{tabular}
\end{center}
\end{table*}

\subsection{Rotation measures}


With the aligned, three-band profiles, we can not only determine new  
RM values, but also investigate whether the polarization PAs obey the 
expected $\lambda^2$ law.
To gain enough S/N, we typically split the 10\,cm and 20\,cm bands into four subbands 
and the 50\,cm band into three subbands. For pulsars whose linear polarization is weak 
and has low S/N, we split the bands into fewer subbands or fully average them in frequency 
(specific comments are given in the footnotes of Table~\ref{rm}). For PSR J1832$-$0836, 
the S/N of profile is low and the linear polarization is weak in both 10\,cm and 50\,cm bands, 
therefore we excluded it from our RM measurements.

As the PAs vary significantly with pulse phase and also with observing frequency, 
we have selected small regions in pulse phase in which the PAs are generally stable across 
the three bands. 
Phase ranges we used for each pulsar are listed in the third column of Table~\ref{rm}.
In order to avoid low S/N regions and obtain smaller uncertainties of the PA, only 
phase bins whose linear polarization exceeds five times the baseline rms noise 
were used.

\begin{figure*}
\begin{center}
\includegraphics[width=6 in,trim=0 3cm 0 3cm]{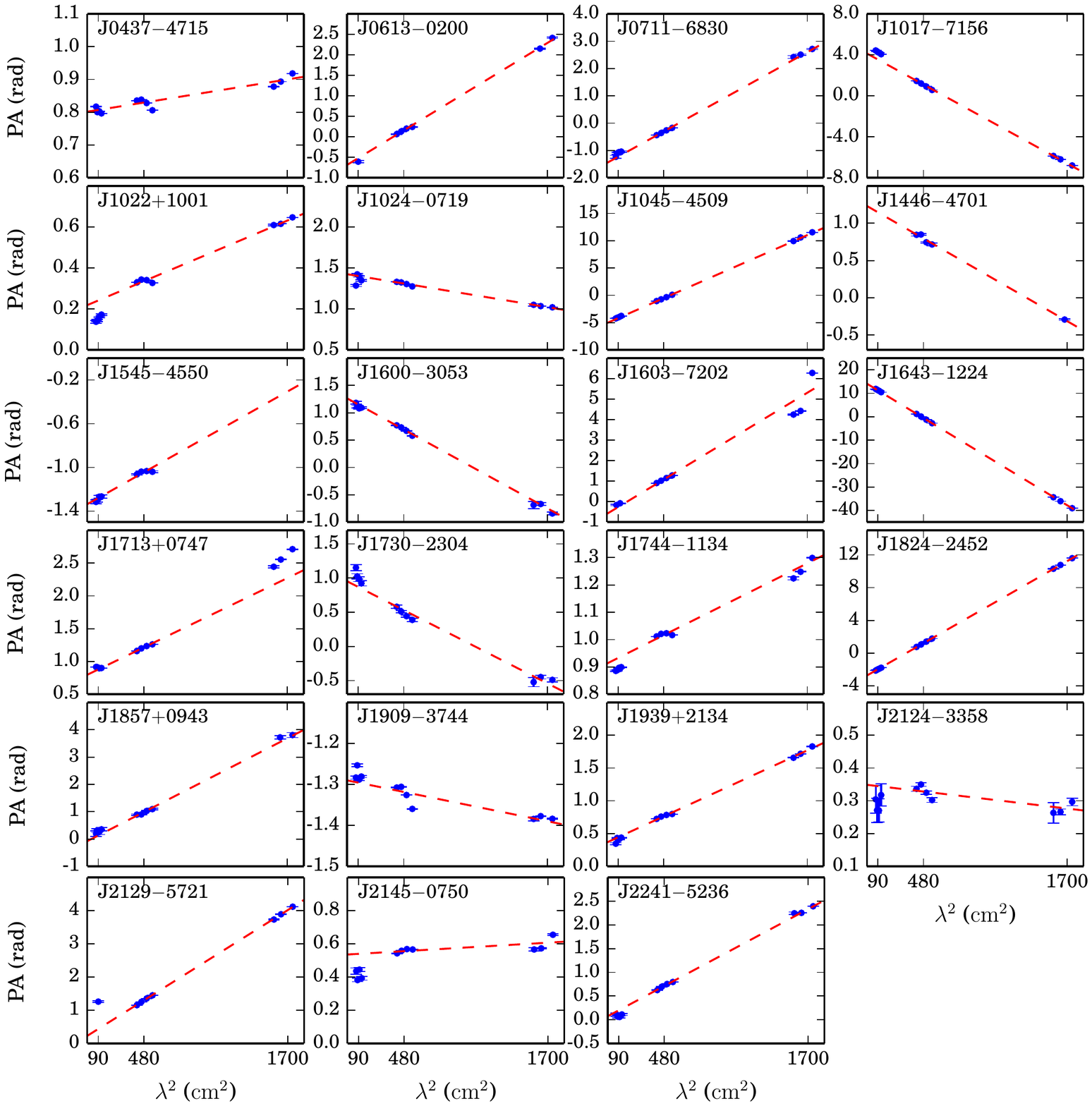}
\caption{Position angles as a function of $\lambda^2$ for $23$ MSPs. The fitted RMs 
are indicated with red dashed lines. PSR J1832$-$0836 is not included here because of the 
low S/N in the 10\,cm and 50\,cm bands for this pulsar.
} 
\label{rmFreq}
\end{center}
\end{figure*}

Our results are summarised in Table~\ref{rm}. Previously published results, obtained 
from the 20\,cm band alone, are shown in the second column. In columns 3, 4 and 5 we 
present our results determined across two bands (10-20, 10-50 and 20-50 respectively).  
In column 6 we present the RM value obtained by fitting across all three bands.  
In Fig.~\ref{rmFreq}, the mean PAs in the stable regions for each pulsar are plotted 
a function of $\lambda^2$. The best fitted RMs are indicated with red dashed lines. 

For some pulsars, our RMs are significantly different from previously published results. 
These are explained as follows.
First, previous measurements were obtained using only the 20\,cm band. 
In Fig.~\ref{rmFreq} it is clear that for pulsars such as J0437$-$4715,
J1022$+$1001 and J1744$-$1134, the PAs in the 20\,cm band deviate from the 
best fitted lines obtained using the wider band.
Second, previous measurements used PAs averaged over the pulse longitude 
while we only averaged PAs within phase ranges that PAs are stable. Therefore, 
the variation of RM across the pulse longitude would introduce deviations.

Fig.~\ref{rmFreq} shows that, for some pulsars, the PAs generally obey the 
$\lambda^2$ fit across a wide range of frequency (e.g., PSRs J0613$-$0200, 
J0711$-$6830, J1045$-$4509, J1643$-$1224, J1824$-$2452A). 
However, for other pulsars, the PAs can significantly deviate from the $\lambda^2$ fit
across bands (e.g., PSRs J1017$-$7156, J1713$+$0747) and show different trends within bands 
(e.g., PSRs J0437$-$4715, J1022$+$1001, J1730$-$2304, J1744$-$1134, J1909$-$3744, J2124$-$3358, 
J2145$-$0750).  
For PSRs J2124$-$3358 and J2129$-$5721, the deviation of PA in the 10\,cm band 
from the best fitted result is likely caused by the low S/N of the profile.
For PSRs J1603$-$7202 and J2145$-$0750, the PA curves vary dramatically within 
bands and cause the deviation of PAs from the $\lambda^2$ fit.

\begin{table*}
\centering
\caption{Interstellar RMs for $23$ MSPs in units of $\rm{rad\ m^{-2}}$. Previoulsy published results without footnotes are from~\citet{Yan11}.}
\label{rm}
\begin{tabular}{lcccccc}
\hline
PSR          &    Previously published &  Phase ranges        &    \multicolumn{4}{c}{Measured from mean profile}                   \\  
             &    20\,cm               &                      &    10\,cm - 20\,cm  &  10\,cm - 50\,cm   &  20\,cm - 50\,cm   &    fitting      \\   
\hline
J0437$-$4715            & $0.0   $ $\pm$ 0.4      &  $(0.01, 0.03)   $ & $0.60   $$\pm$ 0.01  & $0.618   $$\pm$ 0.004 &  $0.624   $$\pm$ 0.001 &  $0.58   $$\pm$ 0.09   \\  
J0613$-$0200$^\star$    & $9.7   $ $\pm$ 1.1      &  $(0.01, 0.03)   $ & $19.8   $$\pm$ 0.7   & $17.8    $$\pm$ 0.2   &  $17.20   $$\pm$ 0.08  &  $17.5   $$\pm$ 0.3   \\  
J0711$-$6830            & $21.6  $ $\pm$ 3.1      &  $(-0.28, -0.26) $ & $22.1   $$\pm$ 0.4   & $23.5    $$\pm$ 0.1   &  $23.89   $$\pm$ 0.05  &  $23.9   $$\pm$ 0.4   \\  
J1017$-$7156            & $-78   $ $\pm$ $3^a$    &  $(-0.025, 0.005)$ & $-82.1  $$\pm$ 0.2   & $-66.59  $$\pm$ 0.04  &  $-61.66  $$\pm$ 0.03  &  $-63    $$\pm$ 1    \\
J1022$+$1001            & $-0.6  $ $\pm$ 0.5      &  $(-0.01, 0.01)  $ & $4.68   $$\pm$ 0.06  & $2.95    $$\pm$ 0.01  &  $2.405   $$\pm$ 0.004 &  $2.4    $$\pm$ 0.1   \\  
                        &                         &                    &                      &                       &                        &                       \\
J1024$-$0719            & $-8.2  $ $\pm$ 0.8      &  $(-0.04, 0.03)  $ & $-1.88  $$\pm$ 0.09  & $-2.26   $$\pm$ 0.03  &  $-2.38   $$\pm$ 0.02  &  $-2.4   $$\pm$ 0.2    \\  
J1045$-$4509            & $92.0  $ $\pm$ 1.0      &  $(0.05, 0.07)   $ & $91.5   $$\pm$ 0.1   & $93.34   $$\pm$ 0.06  &  $93.91   $$\pm$ 0.07  &  $94.7   $$\pm$ 0.7    \\  
J1446$-$4701$^\ast$     & $-14   $ $\pm$ $3^a$    &  $(-0.05, 0.0)   $ &                      &                       &  $-8.98   $$\pm$ 0.11  &  $-9.1   $$\pm$ 0.2    \\
J1545$-$4550$^\dagger$  & $-0.6  $ $\pm$ $1.3^b$  &  $(-0.05, 0.0)   $ & $6.3    $$\pm$ 0.2   &                       &                        &  $6.1    $$\pm$ 0.5    \\
J1600$-$3053            & $-15.5 $ $\pm$ 1.0      &  $(0.01, 0.04)   $ & $-11.6  $$\pm$ 0.1   & $-11.77  $$\pm$ 0.09  &  $-11.8   $$\pm$ 0.1   &  $-11.8  $$\pm$ 0.3    \\  
                        &                         &                    &                      &                       &                        &                        \\
J1603$-$7202            & $27.7  $ $\pm$ 0.8      &  $(-0.01, 0.0)   $ & $31.2   $$\pm$ 0.4   & $28.91   $$\pm$ 0.09  &  $28.20   $$\pm$ 0.05  &  $35     $$\pm$ 2    \\  
J1643$-$1224            & $-308.1$ $\pm$ 1.0      &  $(-0.1, -0.05)  $ & $-306.8 $$\pm$ 0.2   & $-301.70 $$\pm$ 0.06  &  $-300.09 $$\pm$ 0.05  &  $-305.7 $$\pm$ 0.2    \\   
J1713$+$0747            & $8.4   $ $\pm$ 0.6      &  $(0.0, 0.01)    $ & $8.19   $$\pm$ 0.02  & $10.67   $$\pm$ 0.02  &  $11.45   $$\pm$ 0.03  &  $8.7    $$\pm$ 0.5    \\  
J1730$-$2304            & $-7.2  $ $\pm$ 2.2      &  $(-0.02, 0.0)   $ & $-13.4  $$\pm$ 0.2   & $-9.22   $$\pm$ 0.08  &  $-7.88   $$\pm$ 0.1   &  $-8.8   $$\pm$ 0.6    \\  
J1744$-$1134            & $-1.6  $ $\pm$ 0.7      &  $(0.0, 0.02)    $ & $3.24   $$\pm$ 0.02  & $2.34    $$\pm$ 0.01  &  $2.05    $$\pm$ 0.01  &  $2.2    $$\pm$ 0.2    \\  
                        &                         &                    &                      &                       &                        &                        \\
J1824$-$2452A           & $77.8  $ $\pm$ 0.6      &  $(-0.02, 0.04)  $ & $82.6   $$\pm$ 0.3   & $82.06   $$\pm$ 0.07  &  $81.91   $$\pm$ 0.04  &  $82.2   $$\pm$ 0.2    \\  
J1857$+$0943$^\ddagger$ & $16.4  $ $\pm$ 3.5      &  $(0.06, 0.062)  $ & $18.4   $$\pm$ 0.8   & $21.4    $$\pm$ 0.3   &  $22.4    $$\pm$ 0.3   &  $22.2   $$\pm$ 0.9    \\  
J1909$-$3744            & $-6.6  $ $\pm$ 0.8      &  $(-0.01, 0.01)  $ & $-0.38  $$\pm$ 0.02  & $-0.30   $$\pm$ 0.01  &  $-0.27   $$\pm$ 0.01  &  $-0.6   $$\pm$ 0.2    \\  
J1939$+$2134            & $6.7   $ $\pm$ 0.6      &  $(0.0, 0.04)    $ & $12.3   $$\pm$ 0.2   & $9.13    $$\pm$ 0.05  &  $8.11    $$\pm$ 0.01  &  $8.3    $$\pm$ 0.1    \\  
J2124$-$3358            & $-5.0  $ $\pm$ 0.9      &  $(-0.4, -0.38)  $ & $1.6    $$\pm$ 0.3   & $0.07    $$\pm$ 0.08  &  $-0.41   $$\pm$ 0.03  &  $-0.4   $$\pm$ 0.1    \\  
                        &                         &                    &                      &                       &                        &                        \\
J2129$-$5721$^\amalg$   & $23.5  $ $\pm$ 0.8      &  $(-0.02, 0.0)   $ &$0.00   $$\pm$ 0.06  & $16.61   $$\pm$ 0.02  &  $21.88   $$\pm$ 0.03  &  $22.3   $$\pm$ 0.3     \\  
J2145$-$0750            & $-1.3  $ $\pm$ 0.7      &  $(-0.03, -0.01) $ & $1.4    $$\pm$ 0.3   & $-0.31   $$\pm$ 0.09  &  $-0.85   $$\pm$ 0.04  &  $-0.8   $$\pm$ 0.1     \\  
J2241$-$5236            & $14    $ $\pm$ $6^c$    &  $(0.02, 0.04)   $ & $16.1   $$\pm$ 0.3   & $13.84   $$\pm$ 0.08  &  $13.14   $$\pm$ 0.04  &  $13.3   $$\pm$ 0.1    \\
\hline
\end{tabular}
~\\
$^a$~\citet{Keith12}; $^b$~\citet{Burgay13}; $^c$~\citet{Keith11}.
~\\
$^\star$ J0613$-$0200: 50\,cm, two subbands; 10\,cm, one subband. 
~\\
$^\ast$ J1446$-$4701: 50\,cm, one subband; 10\,cm, not used. 
~\\
$^\dagger$ J1545$-$4550: 50\,cm, not used. 
~\\
$^\ddagger$ J1857$+$0943: 50\,cm, two subband. 
~\\
$^\amalg$ J2129$-$5721: 10\,cm, one subband..
\end{table*}
 

The bottom parts of the right-side panels of Fig. \ref{0437} to \ref{2241} show measurements of 
apparent RM measured at specific phases for each MSP.
Since only phase bins whose linear polarization exceeds five times the baseline 
rms noise were used, and we only plot RMs whose uncertainty is smaller than $3\ \rm{rad\ m^{-2}}$,
the phase-resolved RMs only cover pulse phases where the linear polarization is 
strong and PAs generally obey the $\lambda^2$ fit.
For most pulsars, we can see systematic RM variations across the pulse longitude following the 
structure of the mean profile.
For instance, in PSR~J0437$-$4715 the RM shows complex variations from approximately $-8\ \rm{rad\ m^{-2}}$ 
to $8\ \rm{rad\ m^{-2}}$. For PSR~J1643$-$1224, one linear polarization component has 
a RM of approximately $-306\ \rm{rad\ m^{-2}}$ and the other $-300\ \rm{rad\ m^{-2}}$. 
We find that in some cases significant RM variations are associated with orthogonal or non-orthogonal 
mode transitions in PA (e.g., PSRs J1022$+$1001, J1600$-$3053, J1643$-$1224, J1713$+$0747).
For PSR J1744$-$1134, whose PA curve is smooth across the main pulse, the RMs show minor 
variations.
This is consistent with previous phase-resolved RM study of normal pulsars, which also shows 
that the greatest RM fluctuations seem coincident with the steepest gradients of the PA 
curve, whereas pulsars with flat PA curve show little RM variation~\citep{Noutsos09}.

\section{Summary of results and conclusions}

Our results indicate that:
\begin{itemize}

\item Most MSPs in our sample have very wide profiles with multiple components. This is not a surprise and 
	has been presented in numerous earlier publications. We have shown that 18 of the 24 MSPs exhibit emission
  over more than half of the pulse period and the overall pulse width is relatively constant for
  pulsars that have high S/N profiles in all three bands. The MSPs in our sample do not show the frequency
  evolution of the component separations~\citep{Kramer99} that has been observed in normal pulsars~\citep[e.g.,][]{Cordes78,Thorsett91,Mitra02,Mitra04,Chen14}.
	
\item The spectra for some of the pulsars in our sample significantly deviate from a single power-law across 
	the different observing bands. We have observed the spectral steepening at high frequencies and, for 
	some pulsars, we have shown positive spectral indices in the 50\,cm band. We have also observed the spectral 
	flattening within bands at high frequencies for PSRs J1022$+$1001 and J2241$-$5236. The spectral  
	steepening and turnover have been identified in normal pulsars~\citep[e.g.,][]{Maron00,Kijak11}. The flattening 
	or turn-up of the spectrum has been previously observed at extremely high frequencies ($\sim30$\,GHz)~\citep{Kramer96}, 
	and has been explained by refraction effects~\citep{Petrova02}. 
	However, such spectral features have not been observed in MSPs before, and previous measurements of MSP flux 
	densities over a wide frequency range did not show spectral turnovers or breaks~\citep{Kramer99,Kuzmin01}. 

\item For almost all of the MSPs in our sample, the observed three-band PA variations across the profile 
	are extremely complicated and cannot be fitted using the RVM. We show complex details of 
	the PA variation for several MSPs, which were previously thought to have relatively flat or smooth PA 
	profiles (e.g., PSRs J1024$-$0719, J1600$-$3053, J1744$-$1134, J2124$-$3358). Across bands, the 
	PA profiles can evolve significantly (e.g., PSRs J0437$-$4715, J0711$-$6830, J1603$-$7202, J1730$-$2304).

	One exception is PSR J1022$+$1001, whose PA profile is relatively smooth in all three bands 
	except for a discontinuity close to phase zero. At 10\,cm, the PA variation fits the RVM very 
	well. The PA variation departs from the RVM progressively with decreasing frequency. One model to explain 
	this would be that at higher frequencies and lower emission heights, the magnetic field is 
	closer to a simple dipolar field. As the frequency decreases, the magnetic field departs from 
	this simple dipolar form. It is worth noting that PSR J1022$+$1001 has the longest pulse period of 
	the pulsars in our sample.

\item We have observed systematic variations of apparent RM across the pulse longitude following the 
	structure of the mean profile, indicating that such variations are likely to arise from the 
	pulsar magnetosphere. We have also shown that the PA of some pulsars does not follow the $\lambda^2$ 
	relation. As discussed in~\citet{Noutsos09}, possible explanations of these phenomena includes Faraday 
	rotation in the pulsar magnetosphere~\citep{Kennett98,Wang11}, the superposition and frequency 
	dependence of quasi-orthogonal polarization modes~\citep{Ramach04} and interstellar scattering~\citep{Kara09}.

\item	Different pulse components usually have differing spectral indices, apparent RMs and fractional 
	polarizations. Measurements of flux density as a function of frequency for individual components can 
	significantly differ from that obtained by averaging over the entire profile. The spectral shape also 
	often	deviates from a single power-law. In some cases, the peaks of pulse components coincide with 
	the local maximum or minimum of the phase-resolved spectral index. The fractional polarization increases 
	with increasing frequency for some components, but decreases for other components. These results suggest 
	that there are multiple emission regions or structures within the pulsar magnetosphere and that pulse 
	components originate in different locations within the magnetosphere~\citep[e.g.,][]{Dyks10}. 

\end{itemize}

The main goal of this paper has been to inspire and promote our studies and understanding 
of the MSP emission mechanism by publishing high quality, multi-frequency 
polarization profiles.
All the raw data and resulting averaged profiles are available for 
public access online.

Producing a model to describe all these observations will be extremely 
challenging and made more-so by the gaps in the frequency coverage that we 
currently have available at the Parkes telescope.  
In order to mitigate this problem, we are developing a new ultra-wideband 
receiver system that will provide simultaneous observations from approximately 
$0.7$ to $4$\,GHz.  
As our telescope sensitivity continues to improve, millisecond pulsar profiles seem to 
become more and more complicated. However, it is still likely that even more low-level 
components exist in these pulsars. A full understanding of the pulse profiles will 
only be possible with the sensitivity provided by future telescopes such as the 
five-hundred-metre-spherical telescope (FAST) and the Square Kilometre Array (SKA).

\section*{Acknowledgments}
\thanks{
The Parkes radio telescope is part of the Australia Telescope National 
Facility which is funded by the Commonwealth of Australia for operation as a 
National Facility managed by CSIRO. 
This work was supported by the Australian Research Council through grant 
DP140102578. SD is supported by China Scholarship Council
(CSC). GH is a recipient of a Future Fellowship from the Australian Research 
Council. VR is a recipient of a John Stocker postgraduate scholarship from the
Science and Industry Endowment Fund of Australia. LW acknowledges support from 
the Australian Research Council. RXX is supported by the National Basic Research 
Program of China (973 program, 2012CB821800), the National Natural Science Foundation 
of China (Grant No. 11225314) and XTP XDA04060604. We acknowledge the help and 
support of F. Jenet who supervised AM's contribution to this work.
This work made use of NASA’s ADS system.
}

\bibliography{prof}

\begin{appendix}

\section{Multi-frequency Polarization Profiles}

\begin{table*}
\centering
\caption{References, duty cycles and S/N for MSPs in our sample.}
\label{ref}
\begin{tabular}{lccccc}
\hline
PSR          &    References                       &  Duty cycle           &    \multicolumn{3}{c}{S/N}                   \\  
             &                                     &                       &    50\,cm    &  20\,cm      &  10\,cm        \\   
\hline
J0437$-$4715 & \citet{Johnston93,Manchester95_1}   & 0.05  &    14285.8 	 &   33512.4  	&     5445.6    \\  
             & \citet{Navarro97,Yan11}             &       &               &              &               \\  
J0613$-$0200 & \citet{Xilouris98,Stairs99}         & 0.2   &      812.3 	 &    1490.1  	&      396.5    \\  
             & \citet{Ord04,Yan11}                 &       &               &              &               \\  
J0711$-$6830 & \citet{Manchester04,Ord04,Yan11}    & 0.05  &     1194.0 	 &    3368.4  	&      488.8    \\  
J1017$-$7156 & \citet{Keith12}                     & 0.2   &      634.4 	 &    1057.0  	&      233.6    \\
J1022$+$1001 & \citet{Xilouris98,1022Kramer99}     & 0.2   &     4827.2 	 &    6979.8  	&     1577.0    \\  
             & \citet{Stairs99,Ord04,Yan11}        &       &               &              &               \\
             &                                     &       &               &              &               \\
J1024$-$0719 & \citet{Xilouris98,Ord04,Yan11}      & 0.05  &      613.0 	 &    1459.8  	&      273.6    \\  
J1045$-$4509 & \citet{Manchester04,Ord04,Yan11}    & 0.2   &     1097.4 	 &    2155.7  	&      456.8    \\  
J1446$-$4701 & \citet{Keith12}                     & 0.2   &       62.2 	 &     215.2  	&       31.6    \\
J1545$-$4550 & \citet{Burgay13}                    & 0.2   &             	 &     203.5  	&      157.8    \\
J1600$-$3053 & \citet{Ord04,Yan11}                 & 0.2   &      213.6 	 &    2158.3  	&     1045.2    \\  
             &                                     &       &               &              &               \\
J1603$-$7202 & \citet{Manchester04,Ord04,Yan11}    & 0.2   &     1603.6 	 &    3215.6  	&      446.8    \\  
J1643$-$1224 & \citet{Xilouris98,Stairs99}         & 0.2   &     1505.3 	 &    3097.7  	&     1073.4    \\   
             & \citet{Ord04,Yan11}                 &       &               &              &               \\  
J1713$+$0747 & \citet{Xilouris98,Stairs99}         & 0.2   &     1751.4 	 &   10294.1  	&     3894.0    \\  
             & \citet{Ord04,Yan11}                 &       &               &              &               \\  
J1730$-$2304 & \citet{Xilouris98,Kramer98}         & 0.2   &     1138.4 	 &    2645.2  	&     1665.1    \\  
             & \citet{Stairs99,Ord04,Yan11}        &       &               &              &               \\  
J1744$-$1134 & \citet{Xilouris98,Kramer98}         & 0.2   &     1808.5 	 &    4516.1  	&     1025.0    \\  
             & \citet{Stairs99,Ord04,Yan11}        &       &               &              &               \\  
             &                                     &       &               &              &               \\
J1824$-$2452A& \citet{Ord04,Yan11,Stairs99}        & 0.05  &      432.2 	 &     620.0  	&      136.3    \\  
J1832$-$0836 & \citet{Burgay13}                    & 0.2   &            	 &     100.8  	&       24.2    \\  
J1857$+$0943 & \citet{Thorsett90,Xilouris98}       & 0.2   &      376.7 	 &    1563.0  	&      498.9    \\  
             & \citet{Ord04,Yan11}                 &       &               &              &               \\  
J1909$-$3744 & \citet{Ord04,Yan11}                 & 0.2   &     1702.4 	 &    9413.7  	&     1971.3    \\  
J1939$+$2134 & \citet{Thorsett90,Xilouris98}       & 0.2   &     2066.5 	 &    1562.6  	&      565.0    \\  
             & \citet{Stairs99,Ord04,Yan11}        &       &               &              &               \\
             &                                     &       &               &              &               \\
J2124$-$3358 & \citet{Manchester04,Ord04,Yan11}    & 0.05  &      332.2 	 &     411.0  	&      135.4    \\  
J2129$-$5721 & \citet{Manchester04,Ord04,Yan11}    & 0.2   &      750.6 	 &    1829.3  	&       59.5    \\  
J2145$-$0750 & \citet{Xilouris98,Stairs99}         & 0.2   &     4051.6 	 &    8680.7  	&     1483.6    \\  
             & \citet{Manchester04,Ord04,Yan11}    &       &               &              &               \\
J2241$-$5236 & \citet{Keith11}                     & 0.2   &     4270.2 	 &    3549.0  	&      311.0    \\
\hline
\end{tabular}
\end{table*}

\begin{figure*}
\begin{center}
\includegraphics[width=6 in,trim=0 0.1cm 0 0.1cm]{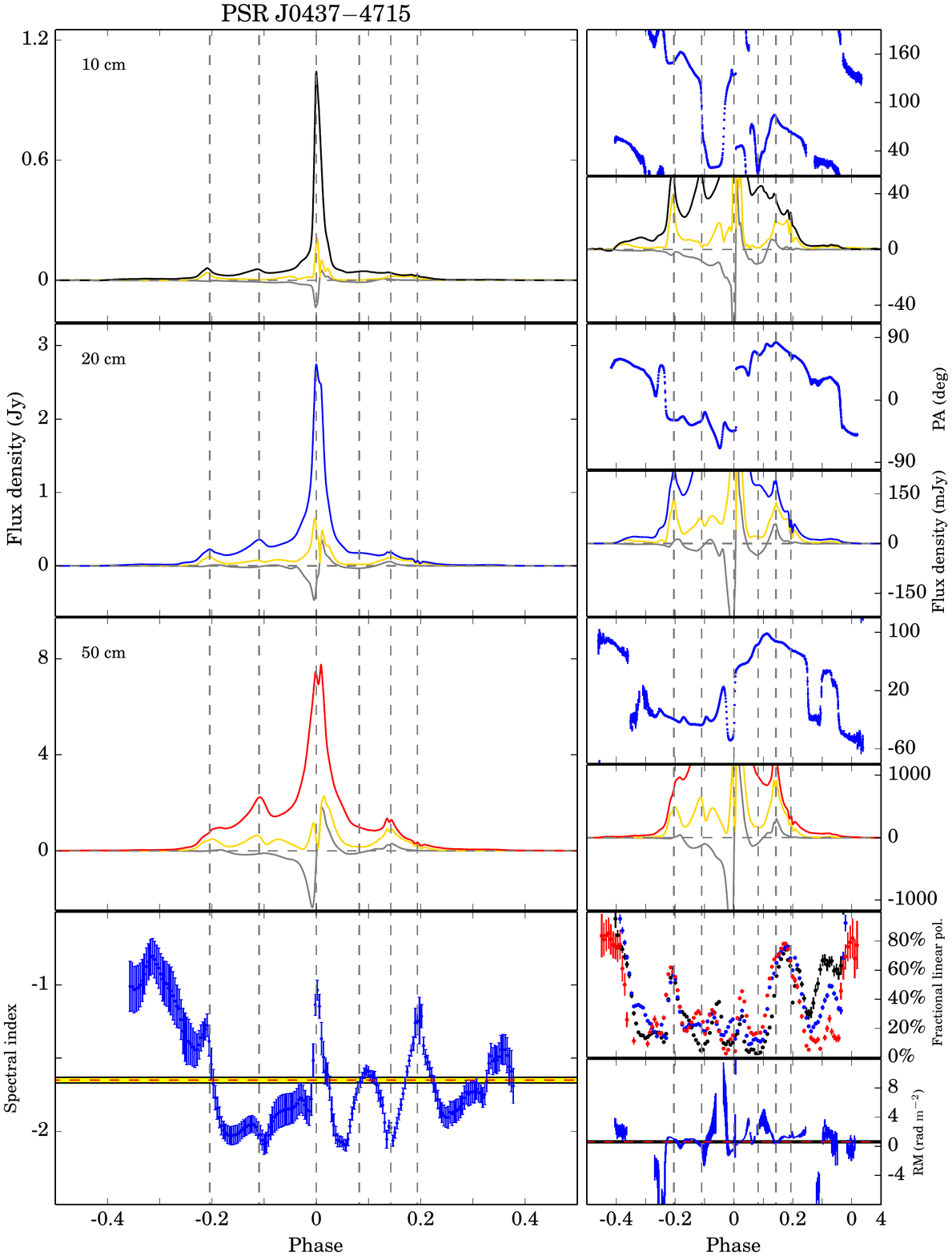}
\caption{Multi-frequency polarization pulse profiles and phase-resolved results for PSR J0437$-$4715. 
Multiple overlapping components and complex polarization 
variations are shown across the pulse profile. 
The overall pulse width is more than $300^{\circ}$ in all three bands.
The main peak has two components with the second component having a steeper 
spectrum and disappearing at high frequencies. The leading and trailing 
parts of the main peak have steeper spectral indices.
However, the outer edges of the profile have flat spectra. We have tried 
different baseline duty cycles from 0.1 to 0.05 in our processing and found 
that the results we present here are not significantly affected.
The profile features are consistent with the Murchison Widefield Array (MWA) 
observation at a frequency of 192\,MHz, which shows that at low frequencies 
the central bright component is flanked by multiple outer components~\citep{Bhat14}.
The PA curves vary dramatically across observing bands. While the orthogonal 
transition close to the main profile peak exists in all three bands, the previously
reported non-orthogonal transition at 20\,cm around phase $-0.23$ disappears in 
the other two bands~\citep{Yan11}.
We observed a probable orthogonal transition at 50\,cm around phase 0.25, and 
non-orthogonal transitions at 50\,cm around phase 0.3 and $-0.02$ and at 10\,cm 
around phase 0.05.
The phase-resolved spectral indices, fractional linear polarizations and apparent 
RMs vary dramatically across the profile.
Close to phase zero, we observed step changes in the phase-resolved spectral index and 
apparent RM, which coincide with the change in the distribution of phase-resolved electric 
field magnitude observed by~\citet{Oslowski14}.
}
\label{0437}
\end{center}
\end{figure*}

\begin{figure*}
\begin{center}
\includegraphics[width=6 in]{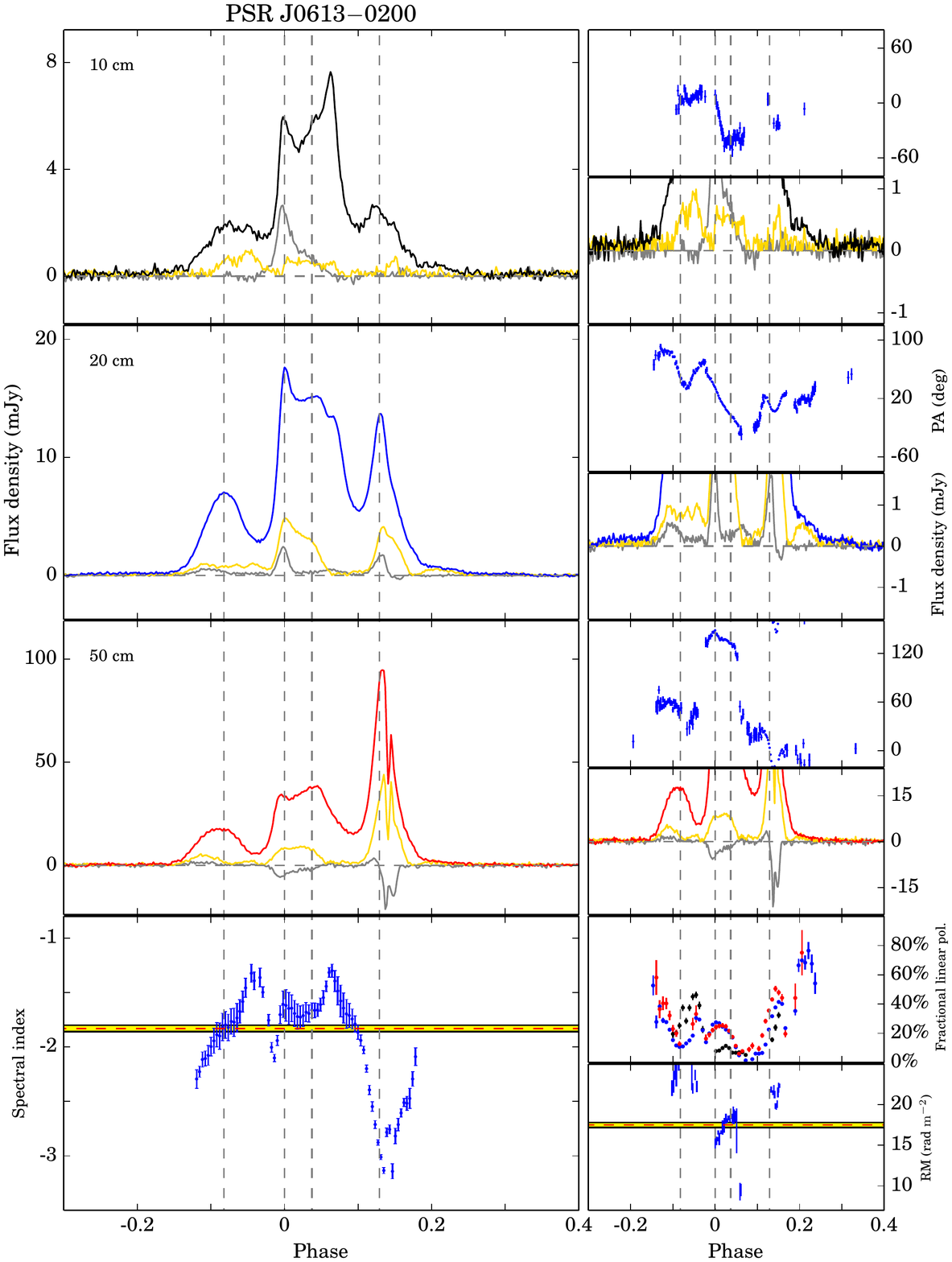}
\caption{Multi-frequency polarization pulse profiles and phase-resolved results for PSR J0613$-$0200. Our high S/N 
profiles provide more details in the PA curve compared to previous observations, and we show that 
the PA curves are complex and very different in three bands. 
At 20\,cm, the discontinuous PA at the leading edge of the trailing 
component reported by \citet{Yan11} is not observed, and the PA curve seems 
to be continuous.
The main pulse of the profile shows clear frequency evolution, and most 
significantly, the trailing peak has a very steep spectrum. The 
trailing peak splits into two peaks at low frequencies as previously 
observed by \citet{Stairs99}.
From the high frequencies to low frequencies, the fractional linear 
polarization increases, and the trailing component becomes highly linearly 
polarized. 
At 50\,cm, the circular polarization swaps sign compared to higher 
frequencies.
The three main pulse components of the profile clearly have different 
apparent RMs.}
\label{0613}
\end{center}
\end{figure*}

\begin{figure*}
\begin{center}
\includegraphics[width=6 in]{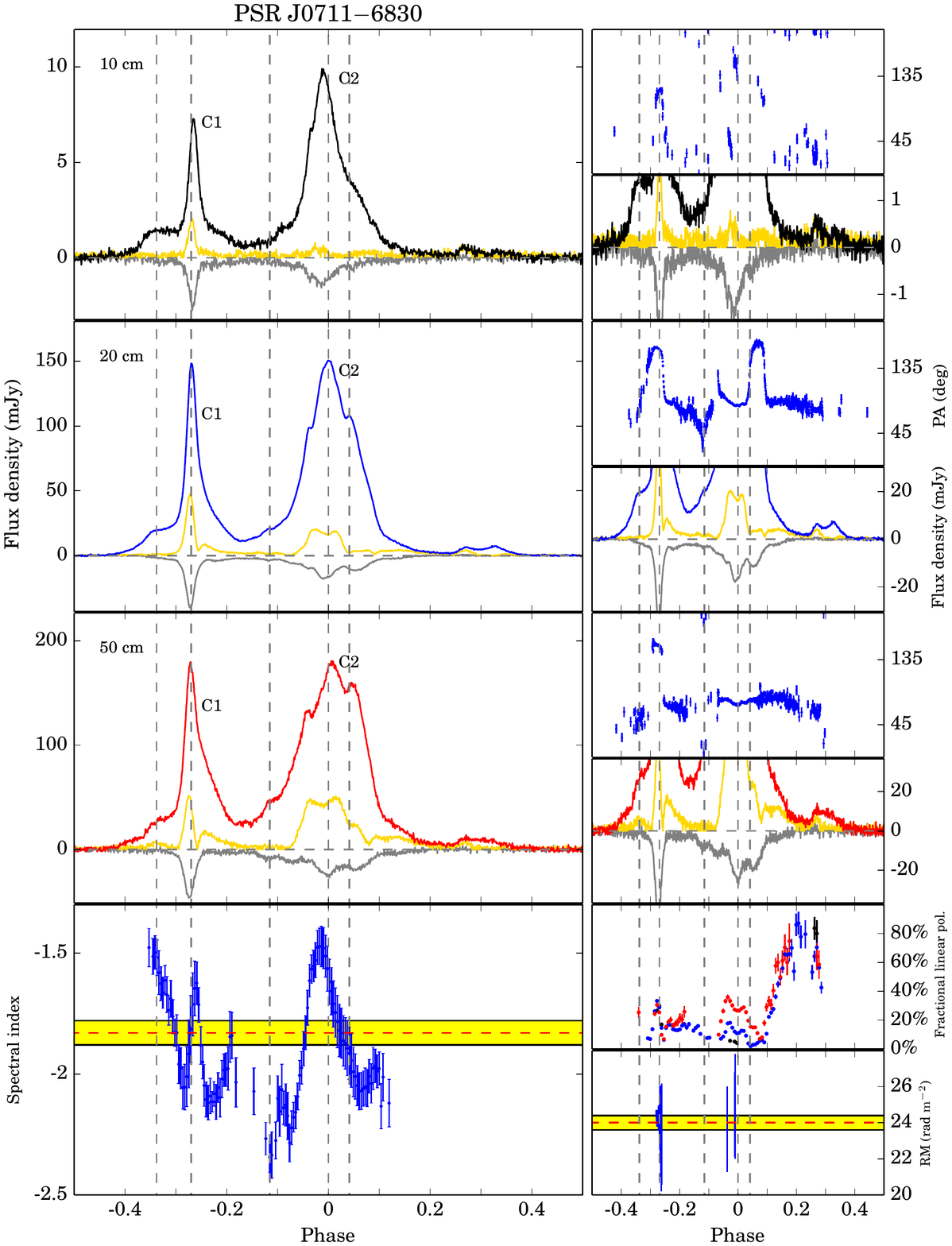}
\caption{Multi-frequency polarization pulse profiles and phase-resolved results for PSR J0711$-$6830. 
The double-peaked weak component following the second peak is clear.
The orthogonal mode transition after the peak of the leading component 
seen by~\citet{Yan11} is confirmed at 20\,cm and is seen at 50\,cm.
However the orthogonal mode transition near the trailing edge 
of the main peak is not present at 50\,cm.
The leading component has a slightly steeper spectrum than the main pulse. 
The fractional linear polarization of the main peak decreases significantly 
with increasing frequency. 
}
\label{0711}
\end{center}
\end{figure*}

\begin{figure*}
\begin{center}
\includegraphics[width=6 in]{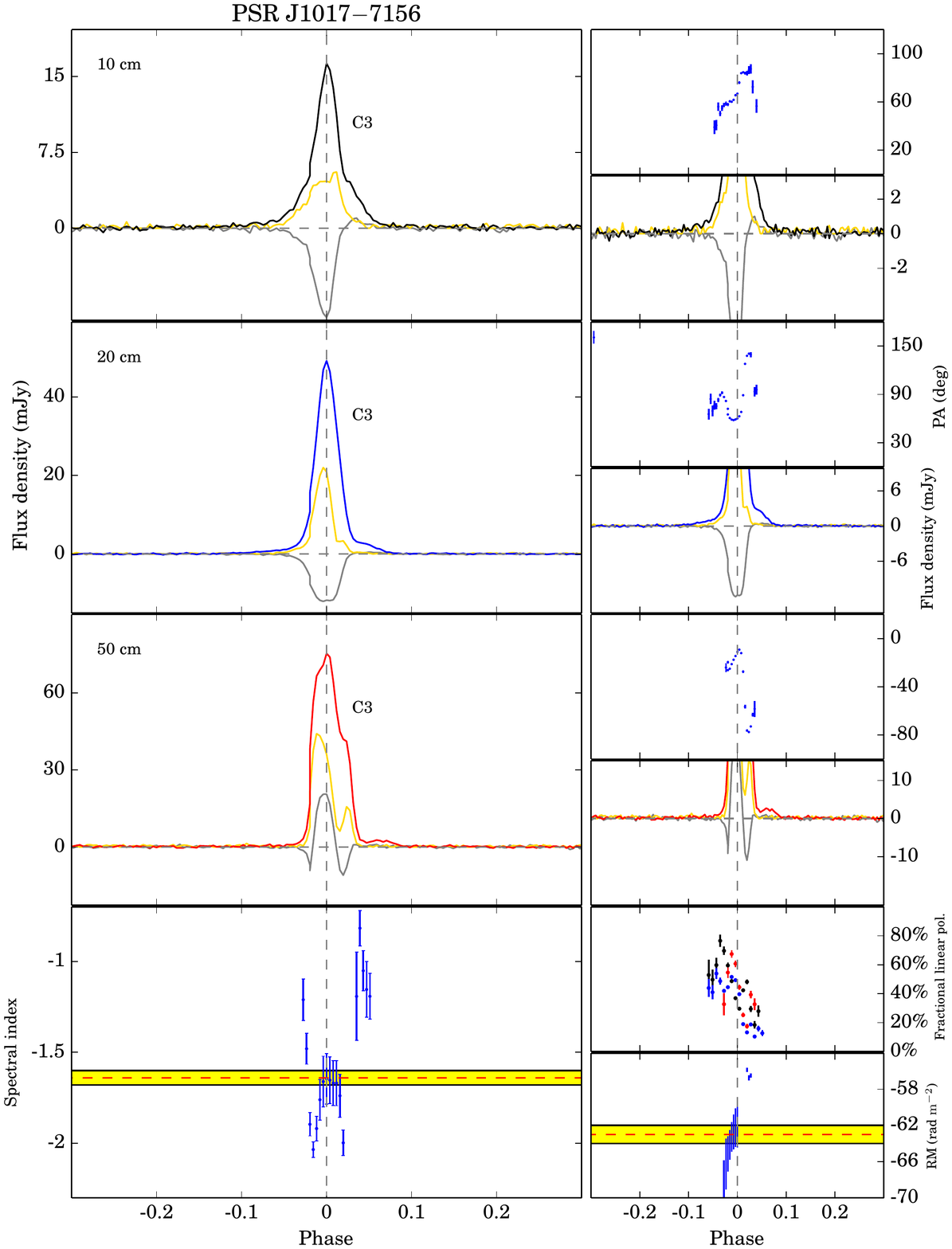}
\caption{Multi-frequency polarization pulse profiles and phase-resolved results for PSR J1017$-$7156. 
We show that the PA variations are more complex than was observed in 
previous work.
While the leading and trailing edges of the main pulse have a steeper spectrum 
compared with the central peak, the trailing component around phase 0.04 has
a much flatter spectrum.
Both the linear and circular polarization have multiple components and show 
significant evolution with frequency. Especially, the circular polarization of 
the main pulse seems to consist of two components of opposite sign with 
the left-circular component having narrower width and much steeper spectrum.
}
\label{1017}
\end{center}
\end{figure*}

\begin{figure*}
\begin{center}
\includegraphics[width=6 in]{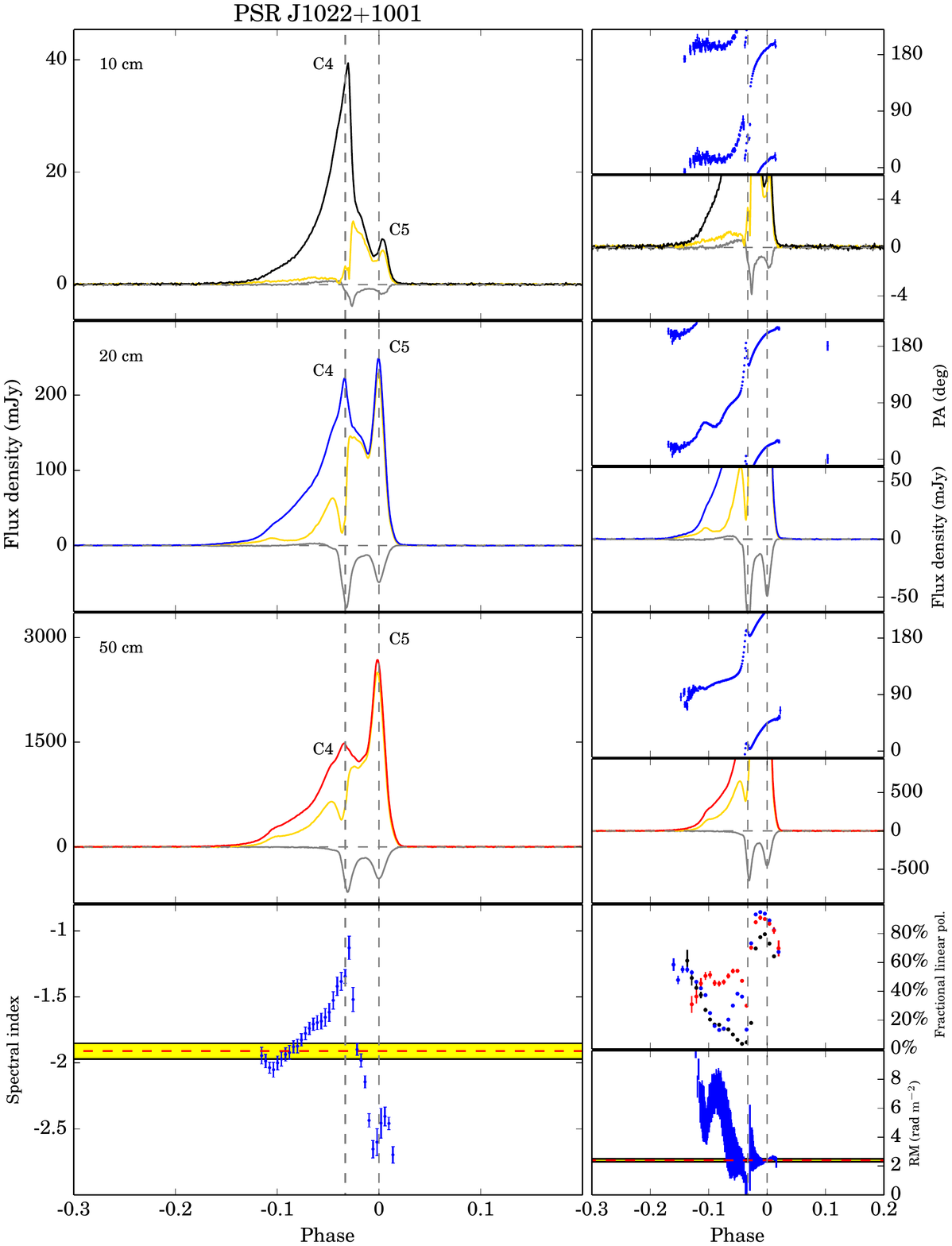}
\caption{Multi-frequency polarization pulse profiles and phase-resolved results for PSR J1022$+$1001. 
At 10\,cm, except for the discontinuity close to phase zero, the PA variation 
fits the RVM very well. As the frequency decreases, the PA variation 
departs from the RVM progressively.
The spectral indices of two main peaks are significantly different so that 
the relative strength of the two main peaks evolves dramatically with 
frequency.
While the second peak remains highly linearly polarized, the first peak 
depolarizes rapidly with increasing frequency.
We also see systematic variation of the phase-resolved apparent RMs.
}
\label{1022}
\end{center}
\end{figure*}

\begin{figure*}
\begin{center}
\includegraphics[width=6 in]{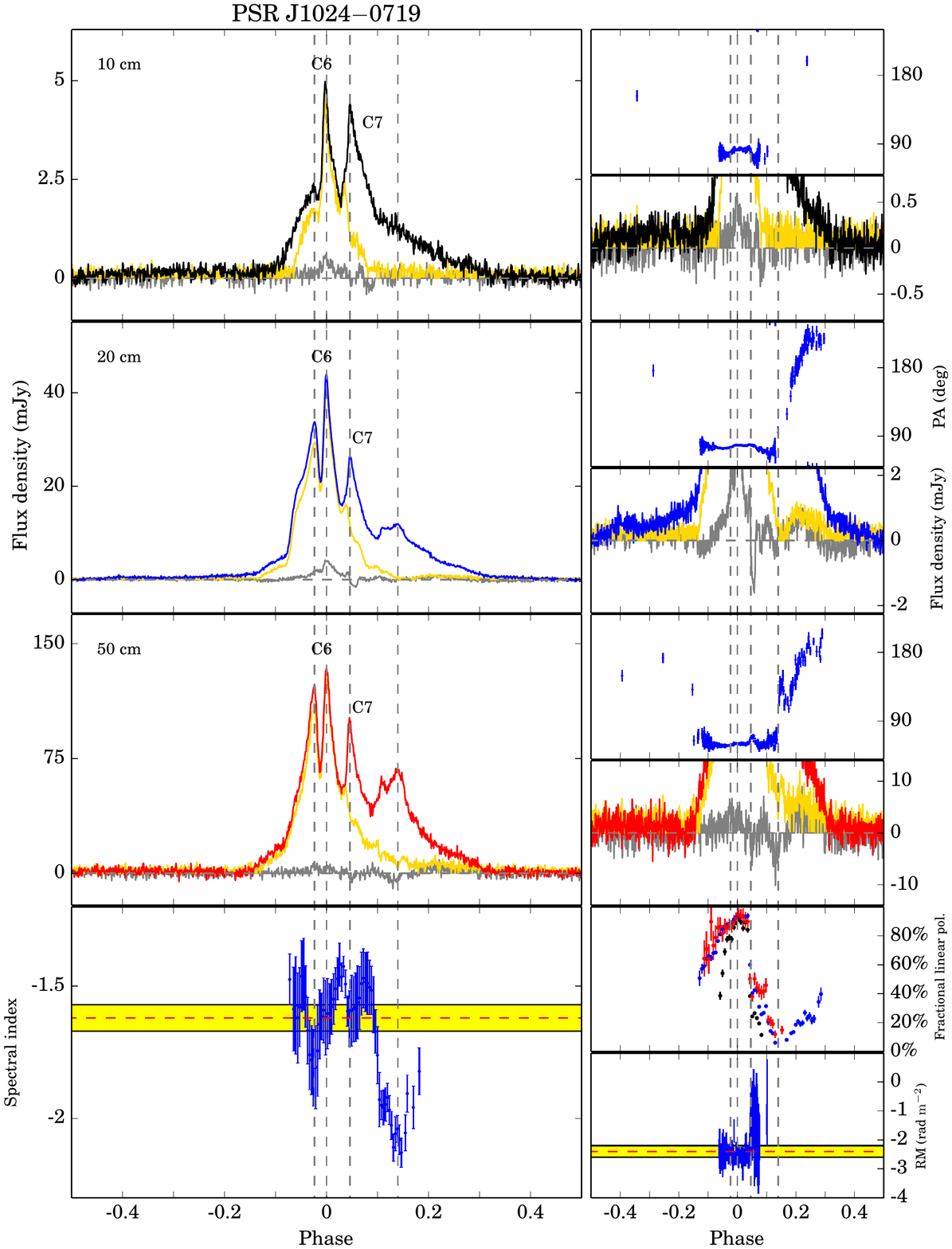}
\caption{Multi-frequency polarization pulse profiles and phase-resolved results for PSR J1024$-$0719. 
Besides the flat PA curve across the main part of the profile as 
previously reported, we also show the PAs of the trailing component which 
increase with phase at 20\,cm and 50\,cm.
The leading component and the trailing component of the profile have much 
steeper spectra compared with the central peaks. 
The leading part of the profile is highly linearly polarized and has stable  
phase-resolved RMs. As the fractional linear polarization drops down at the 
trailing part, the RMs show some variations.
}
\label{1024}
\end{center}
\end{figure*}

\begin{figure*}
\begin{center}
\includegraphics[width=6 in]{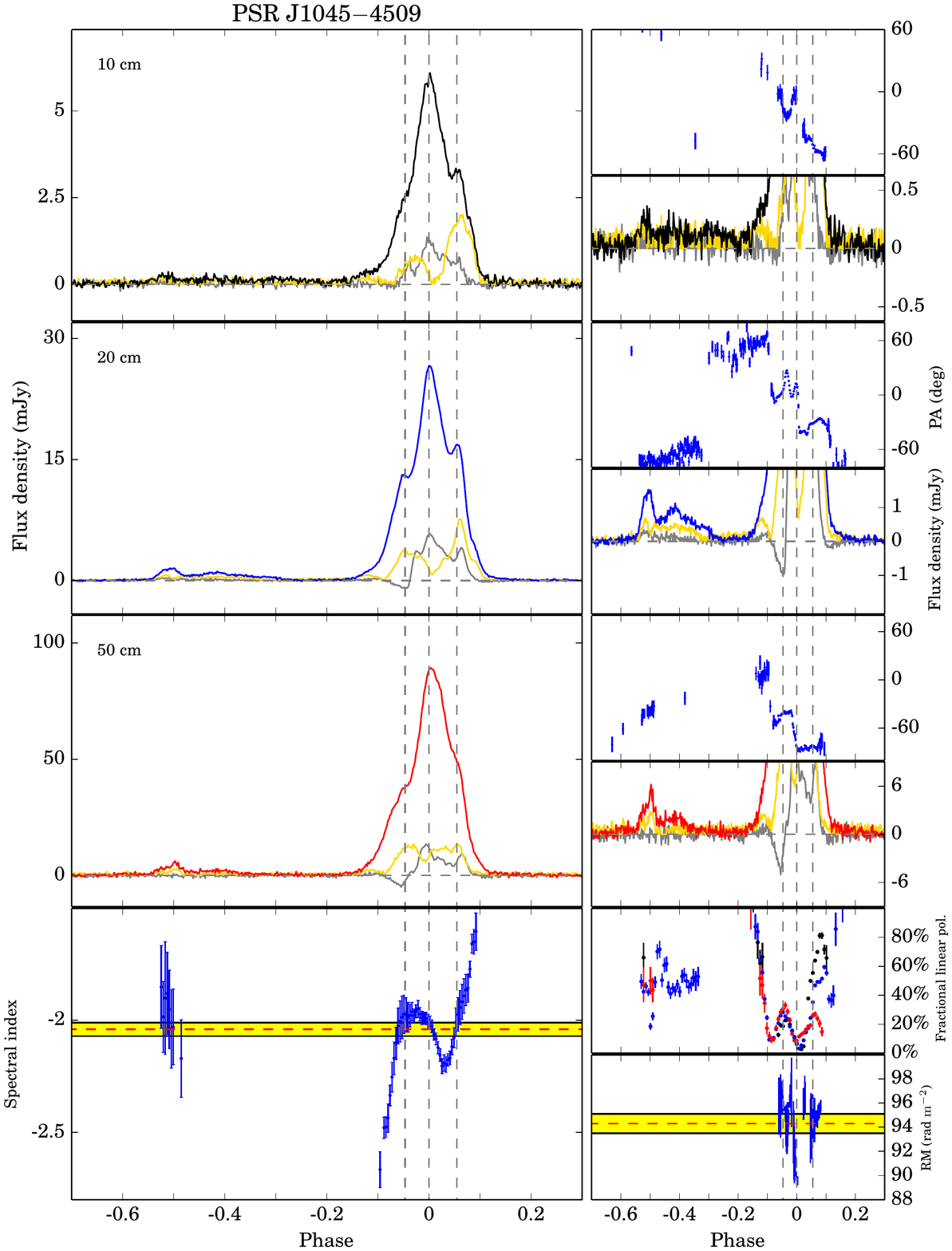}
\caption{Multi-frequency polarization pulse profiles and phase-resolved results for PSR J1045$-$4509. 
We confirm that the leading emission is joined to the main pulse by a low-level bridge of emission as 
seen by~\citet{Yan11}.
We show the complex PA curve with more detail, and determine the PA of the 
low-level bridge connecting the leading emission and the main pulse.
At the leading edge of the main pulse, there is a non-orthogonal transition 
rather than an orthogonal transition as suggested by \citet{Yan11}.
Around phase zero, a non-orthogonal transition can be seen in all 
three bands.
The PA of the low-level bridge emission seems to be discontinuous with 
the other PA variations and could be an orthogonal mode.
To calculate the phase-resolved spectral index of the leading emission, 
we averaged the profile in frequency in the 10\,cm band and only 
divided the 50\,cm band into two subbands.
We show that the leading edge of the main pulse has a steeper spectrum 
than that of the trailing edge.
We also see an increase in the linear polarization of the trailing 
edge of the main pulse at high frequencies.
}
\label{1045}
\end{center}
\end{figure*}

\begin{figure*}
\begin{center}
\includegraphics[width=6 in]{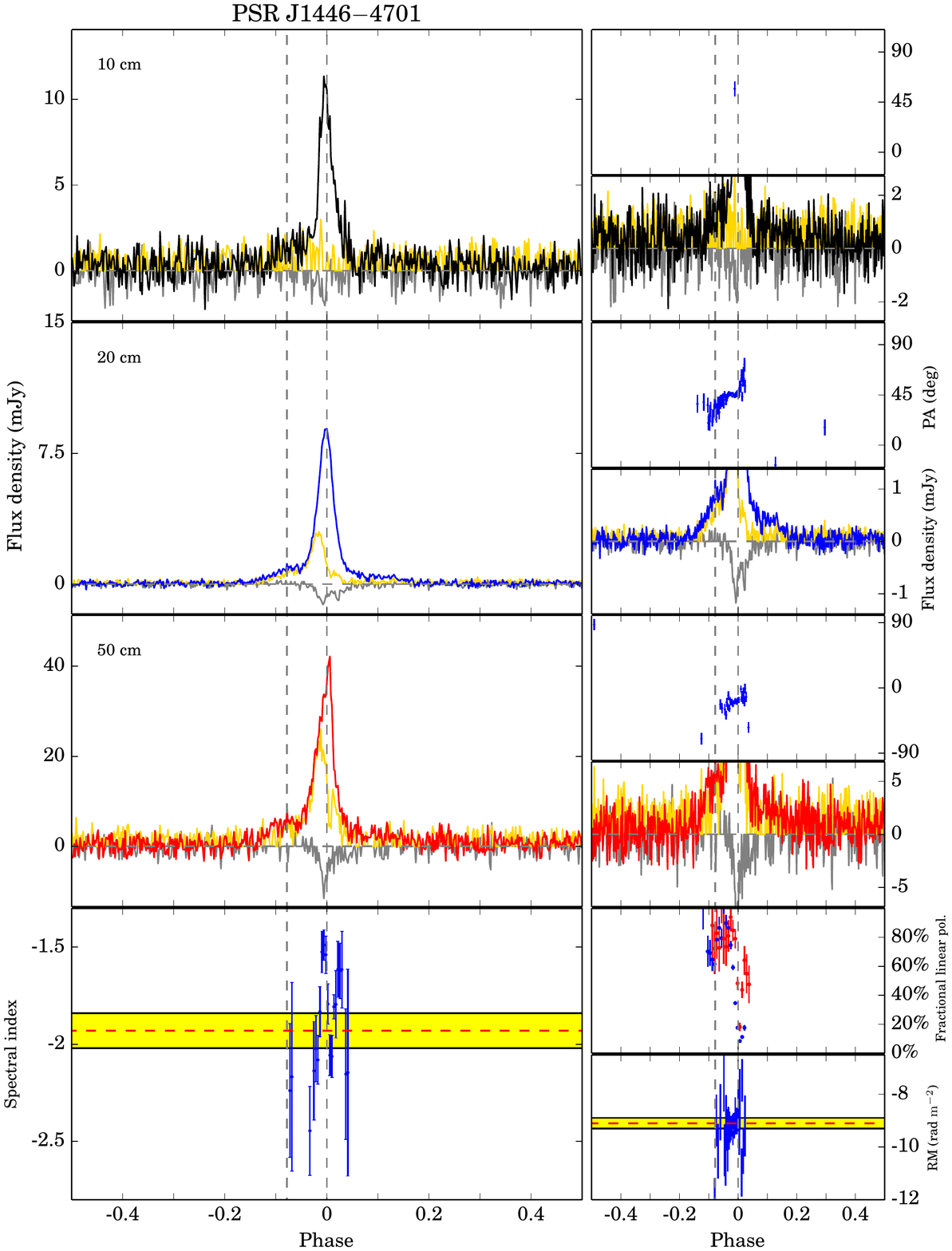}
\caption{Multi-frequency polarization pulse profiles and phase-resolved results for PSR J1446$-$4701. 
The PAs are flat over the main pulse, but show variations over the leading and 
trailing parts.
To calculate the phase-resolved spectral index, we averaged the profile 
in frequency in the 10\,cm band and only divided the 50\,cm band 
into two subbands.
The linear polarization is much stronger at low frequencies.
}
\label{1446}
\end{center}
\end{figure*}

\begin{figure*}
\begin{center}
\includegraphics[width=6 in]{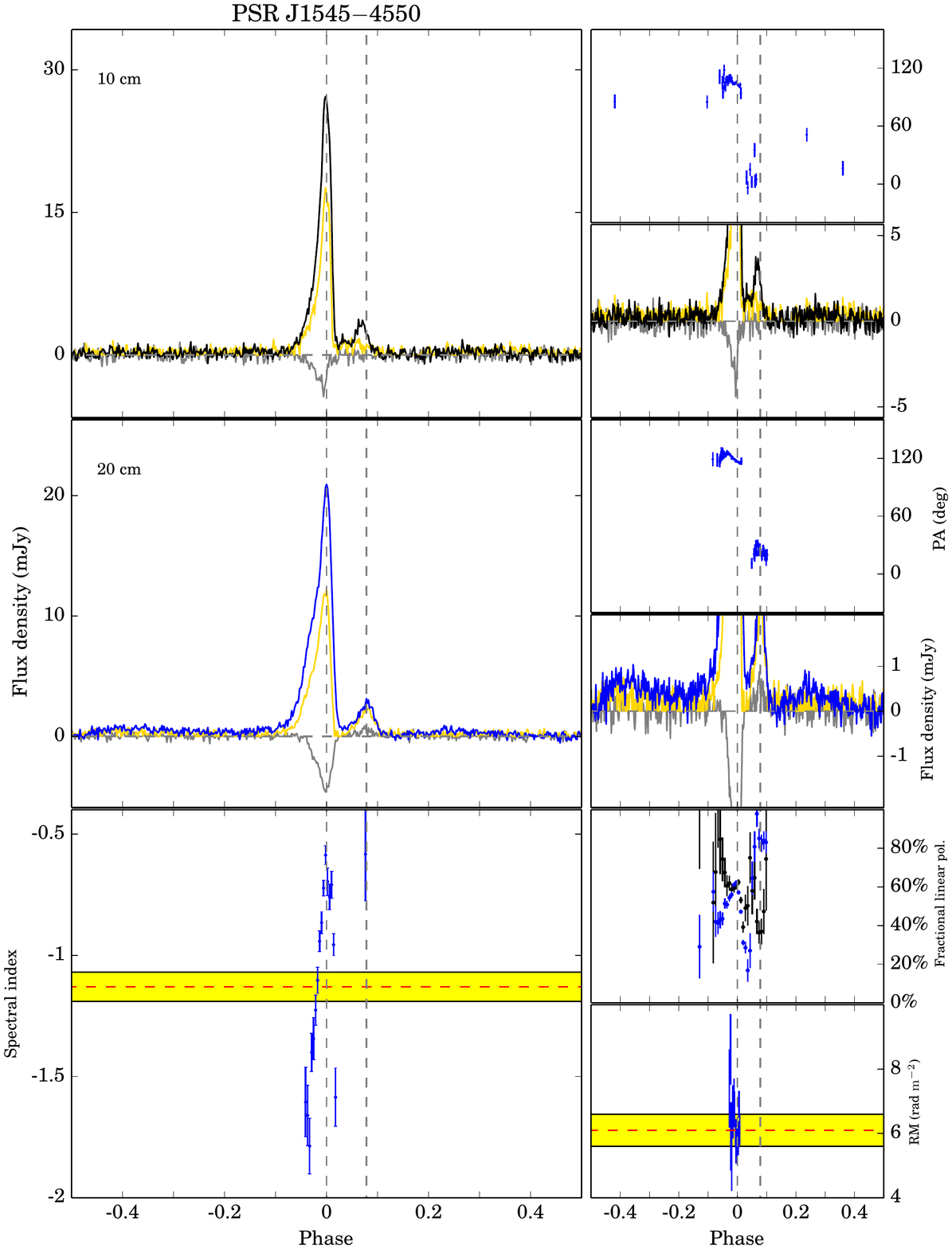}
\caption{Multi-frequency polarization pulse profiles and phase-resolved results for PSR J1545$-$4550. 
At 20\,cm, \citet{Burgay13} show a component around phase $0.35$ that 
we do not see in our analysis. We have confirmed with the High Time Resolution 
Universe (HTRU) collaboration that this extra component resulted from an error 
in their analysis.  
At 10\,cm, our results are consistent with those in~\citet{Burgay13}.
At 50\,cm, we only have a few observations and the S/N is low, therefore 
we do not present the polarization profile here.
We also show that low-level emission extends over at least $80$ 
per cent of the pulse period.
There is an orthogonal transition between the main pulse and 
the trailing component.
}
\label{1545}
\end{center}
\end{figure*}

\begin{figure*}
\begin{center}
\includegraphics[width=6 in]{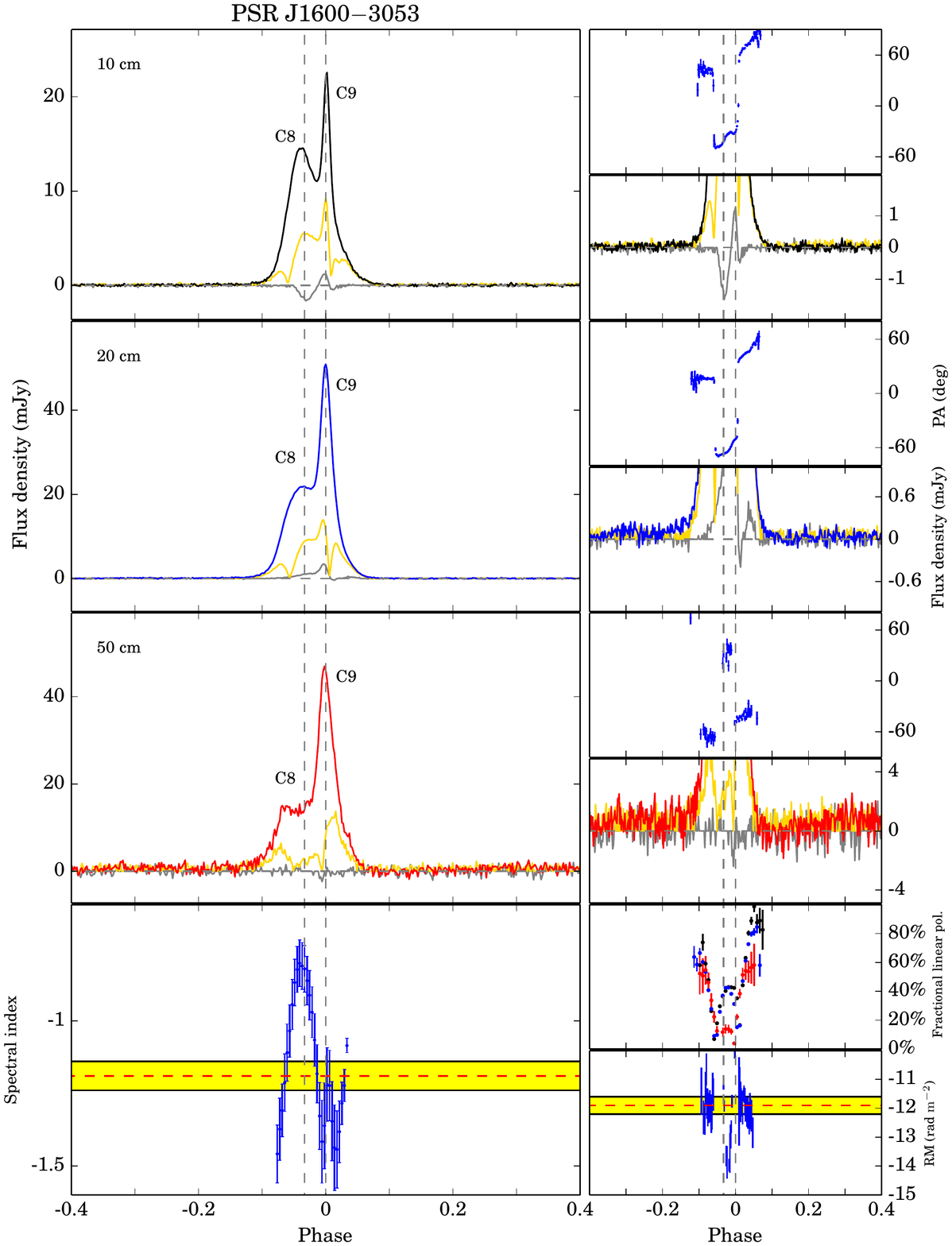}
\caption{Multi-frequency polarization pulse profiles and phase-resolved results for PSR J1600$-$3053. 
We show orthogonal transitions near the peaks of the two main 
components, which is consistent with previously published results.
The leading component of the main pulse has a flatter spectrum compared with 
the main component.
The central part of the pulse profile depolarizes rapidly with decreasing 
frequency. 
We see a sign swap of the circular polarization of the leading component 
from 20\,cm to 10\,cm, and at 50\,cm, the circular polarization becomes almost 
zero across the whole profile.
}
\label{1600}
\end{center}
\end{figure*}

\begin{figure*}
\begin{center}
\includegraphics[width=6 in]{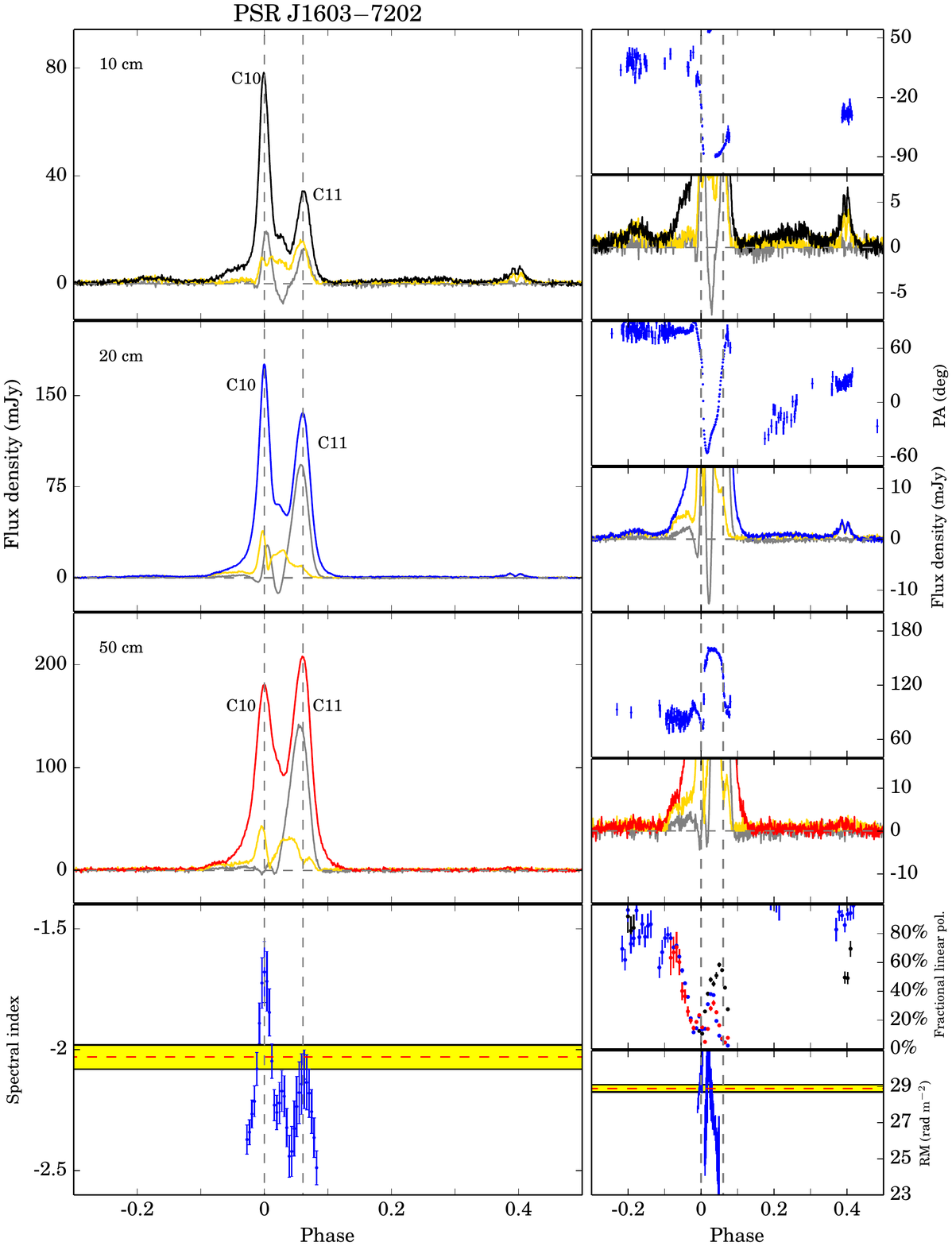}
\caption{Multi-frequency polarization pulse profiles and phase-resolved results for PSR J1603$-$7202. 
The broad low-level feature preceding the main pulse and the double-peak trailing 
pulse can be clearly identified.
We discovered new low-level emission connecting the main pulse and the 
double-peak trailing pulse, and it becomes stronger at 10\,cm.
The relative strength of the two main peaks evolves significantly with frequency.
As frequency goes down, the second main peak becomes highly circularly polarized.
}
\label{1603}
\end{center}
\end{figure*}

\begin{figure*}
\begin{center}
\includegraphics[width=6 in]{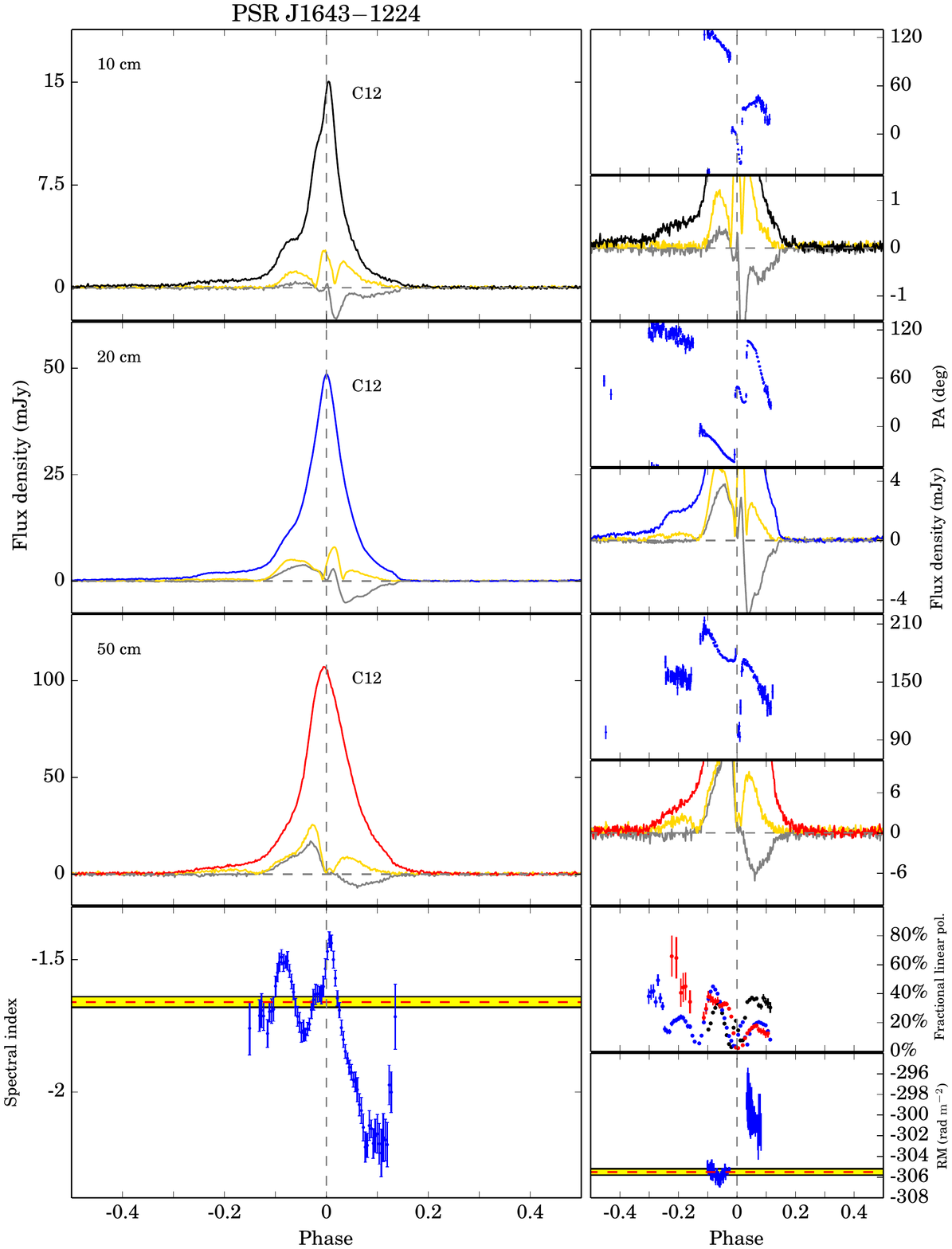}
\caption{Multi-frequency polarization pulse profiles and phase-resolved results for PSR J1643$-$1224. 
At 10\,cm and 20\,cm, the PA of the broad feature preceding the main pulse is determined and 
found to be discontinuous with the rest of the PA variation, showing 
a new orthogonal transition.
The main pulse clearly has multiple components and the trailing part has much 
steeper spectrum than other parts. The leading and trailing parts of the pulse 
have different apparent RMs.
}
\label{1643}
\end{center}
\end{figure*}

\begin{figure*}
\begin{center}
\includegraphics[width=6 in]{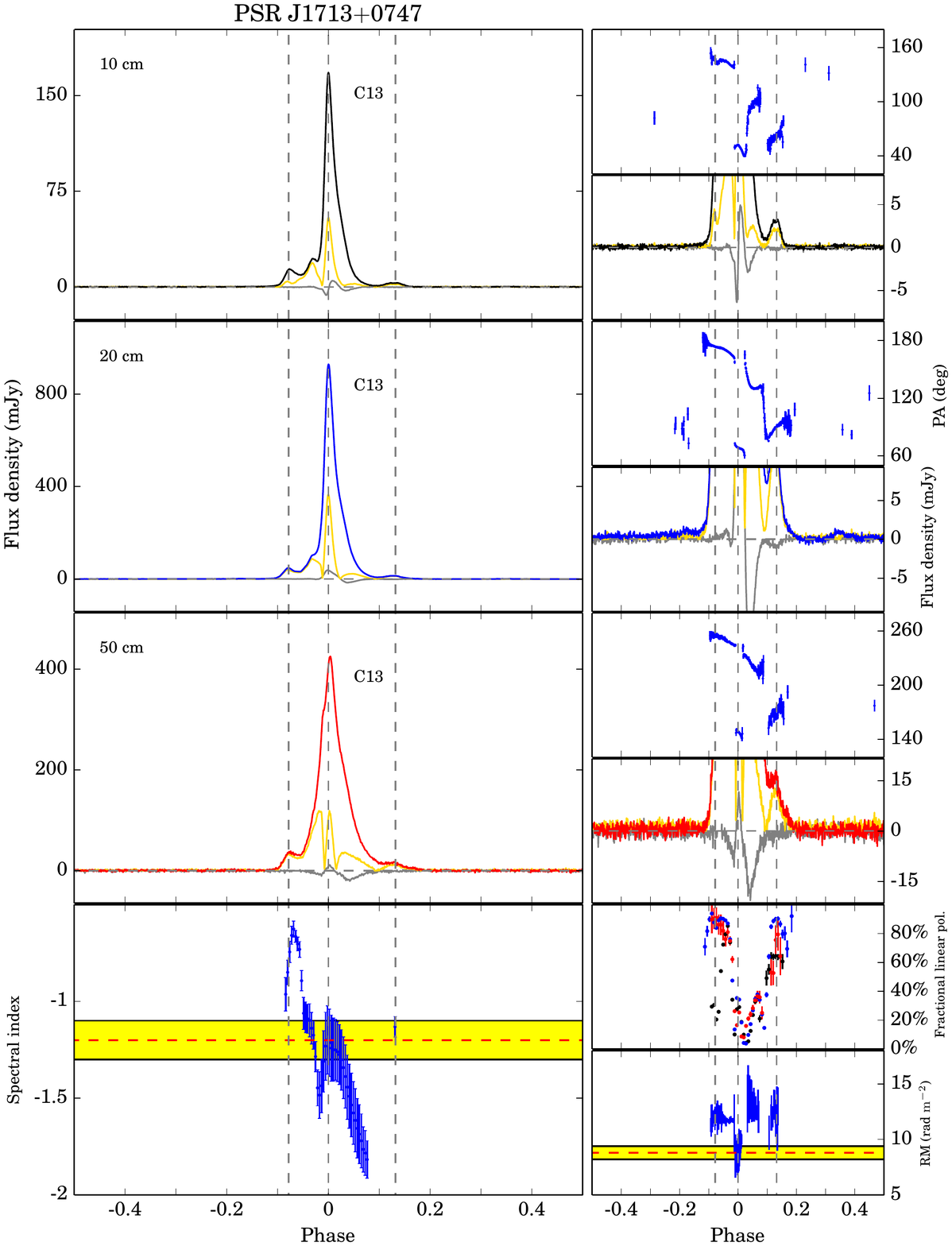}
\caption{Multi-frequency polarization pulse profiles and phase-resolved results for PSR J1713$+$0747. 
At 20\,cm and 50\,cm, we show the almost complete linearly polarized leading and trailing components, 
which is consistent with previously published results.
We detected weak emission around phase $-0.2$ and 0.35 at 20\,cm, which 
increases the overall width from $104^{\circ}$ (as previously thought) to 
$199^{\circ}$. 
The non-orthogonal transition preceding the trailing pulse component reported 
by \citet{Yan11} is observed at 10\,cm and 50\,cm, but at 20\,cm 
the PA transition is continuous.
The linear polarization of the leading and trailing components become stronger 
at low frequencies relative to the rest of the profile.
The main peak of the total intensity profile clearly has multiple components, and different 
components have different apparent RMs.
}
\label{1713}
\end{center}
\end{figure*}

\begin{figure*}
\begin{center}
\includegraphics[width=6 in]{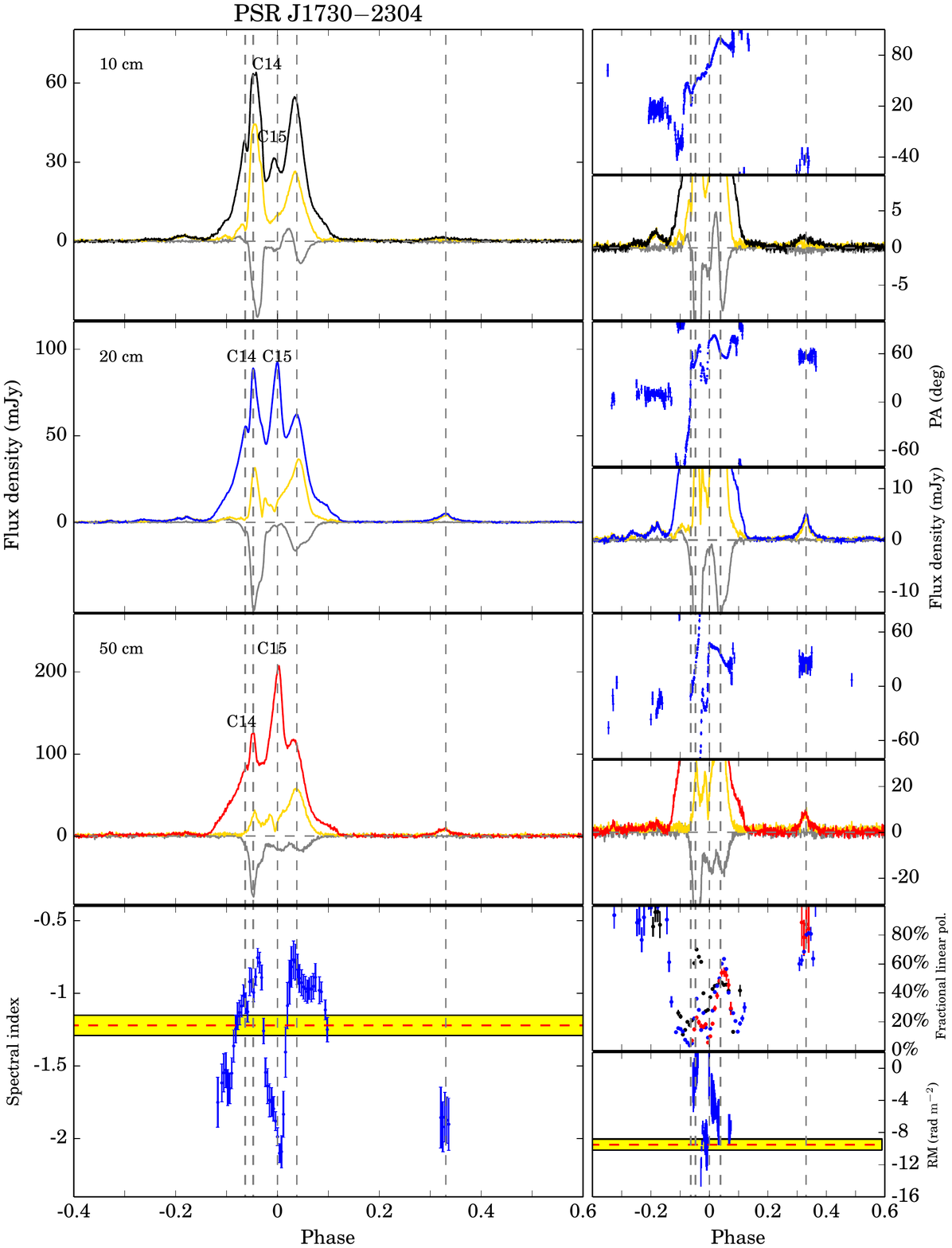}
\caption{Multi-frequency polarization pulse profiles and phase-resolved results for PSR J1730$-$2304. 
We clearly show the weak leading and trailing components already reported, 
and at 20\,cm we detect a weaker leading component not discovered before (around phase $-0.32$). 
This increases overall width of the pulse from $232^{\circ}$ to $248^{\circ}$.
The pulse profile is very complex, with four clear peaks across the main pulse.
To calculate the phase-resolved spectral index of the trailing component
around phase 0.33, we averaged the profile in frequency in the 10\,cm 
band and only divided the 50\,cm band into two subbands.
The central peak at 20\,cm band has a steeper spectrum compared with other 
components.
As the frequency goes down, the second peak depolarizes rapidly.
The PA variations are very complex and are different in the three bands, 
leading to apparent RM variations across the profile.
}
\label{1730}
\end{center}
\end{figure*}

\begin{figure*}
\begin{center}
\includegraphics[width=6 in]{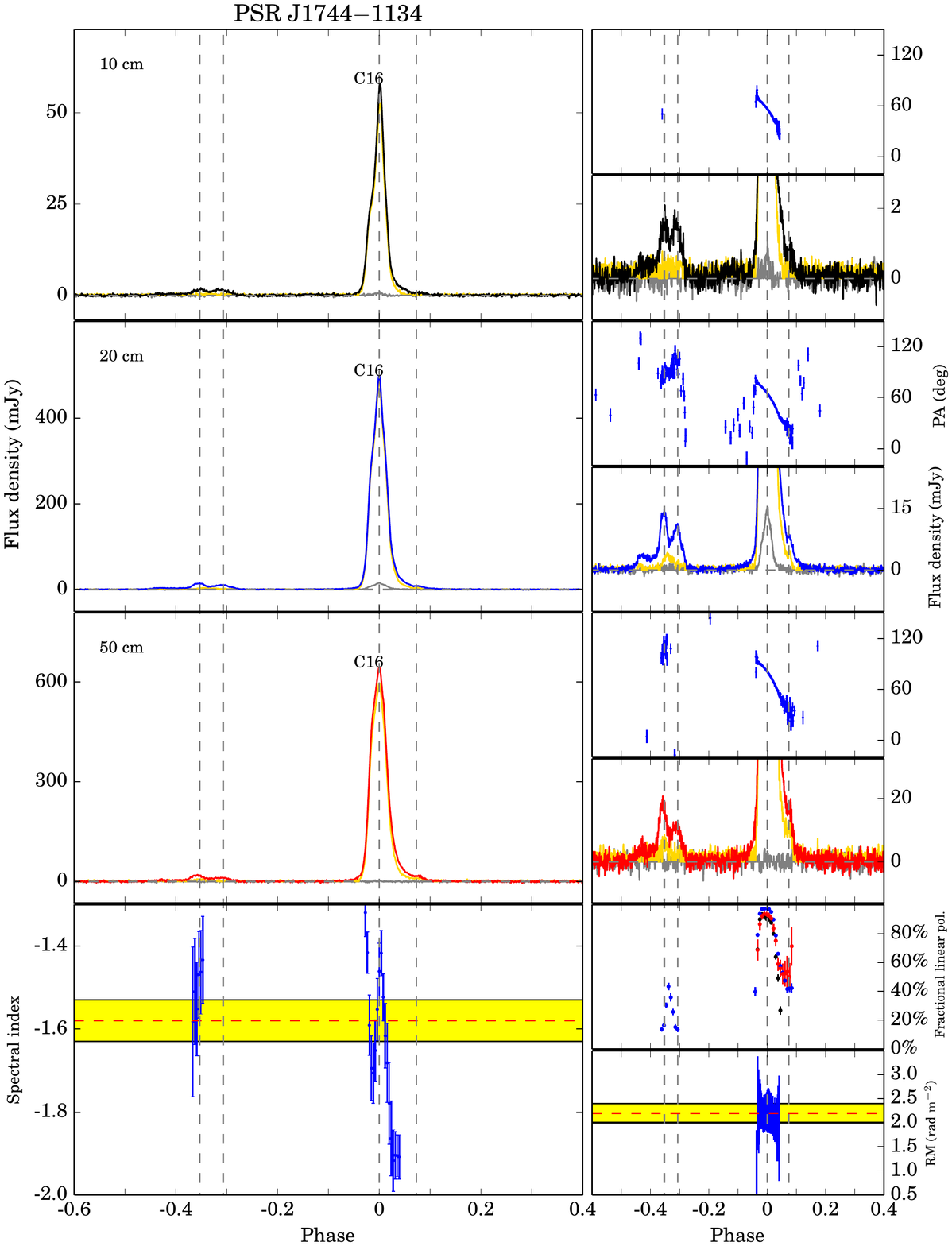}
\caption{Multi-frequency polarization pulse profiles and phase-resolved results for PSR J1744$-$1134. 
The multiple-component precursor is clearly identified and no significant 
post-cursor component is observed.
While the PAs of the main pulse show a smooth decrease, those of the precursor 
have clear structures and do not simply connect with the rest of the PA 
variations.
The shape of the PA curves are similar in the three bands and the phase-resolved 
RMs are almost constant.
The main pulse is highly linearly polarized from 10\,cm to 50\,cm. 
The circular polarizaion of main pulse grows stronger from 10\,cm to 20\,cm, 
but is weaker at 50\,cm.
To calculate the phase-resolved spectral index of the leading component
around phase $-0.35$, we averaged the profile in frequency in the 10\,cm 
band and only divided the 50\,cm band into two subbands.
}
\label{1744}
\end{center}
\end{figure*}

\begin{figure*}
\begin{center}
\includegraphics[width=6 in]{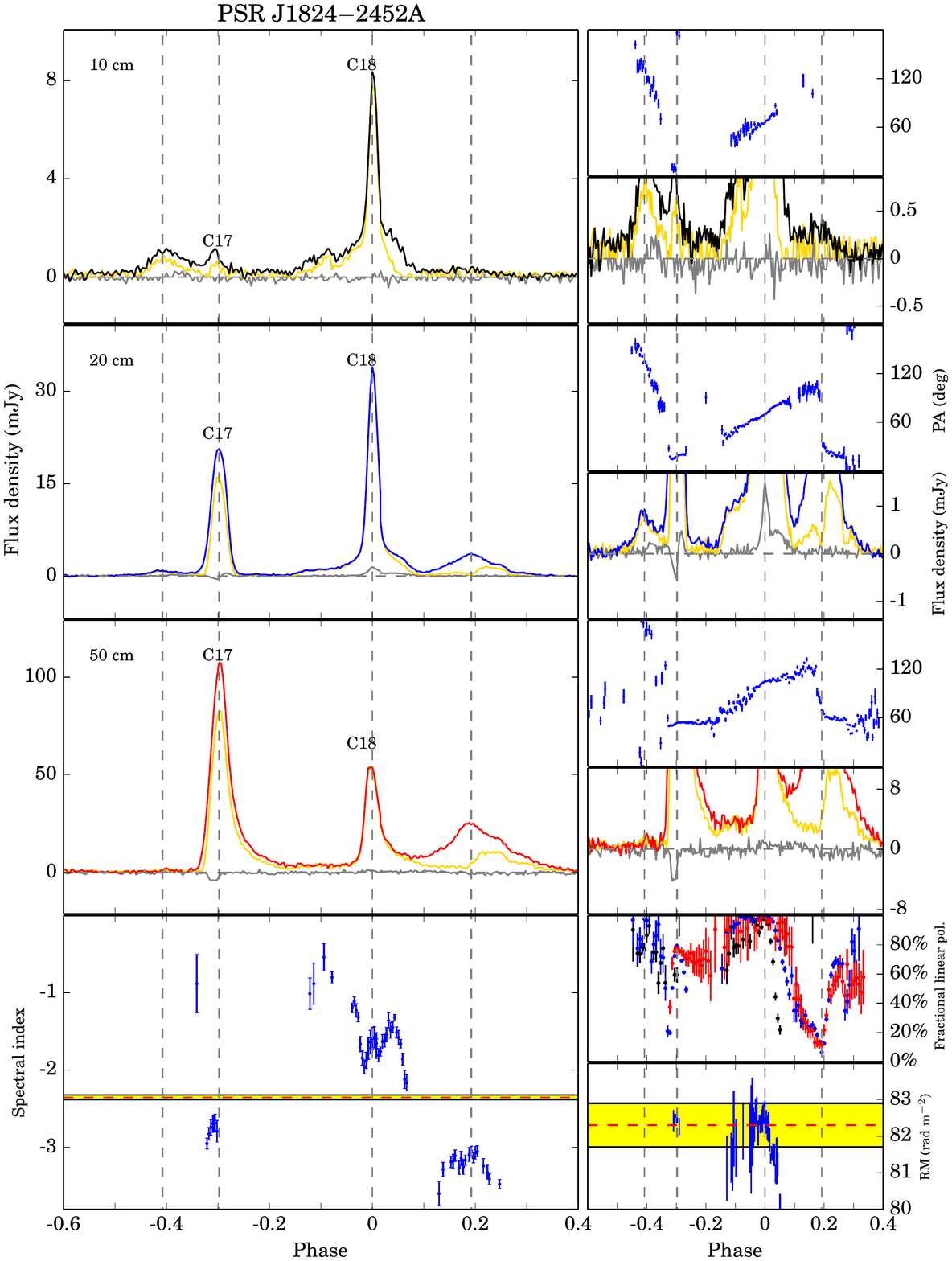}
\caption{Multi-frequency polarization pulse profiles and phase-resolved results for PSR J1824$-$2452A. 
The weak component around phase $-0.4$ is clearly shown at 10\,cm and 20\,cm and is highly 
linearly polarized with a flat spectrum. 
We also show that there is low-level bridge emission 
connecting the two main components of the pulse profile.
The PAs of preceding components are continuous themselves, but are 
discontinuous with the rest of the PA variations.
To calculate the phase-resolved spectral index of the trailing component
around phase 0.2, we averaged the profile in frequency in the 10\,cm 
band.
The frequency evolution of the total intensity is significant and 
our results are consistent with previous low frequency observations~\citep[e.g.,][]{Stairs99}.
The phase-resolved spectral indices show huge variations related to the 
different components. 
}
\label{1824}
\end{center}
\end{figure*}

\begin{figure*}
\begin{center}
\includegraphics[width=6 in]{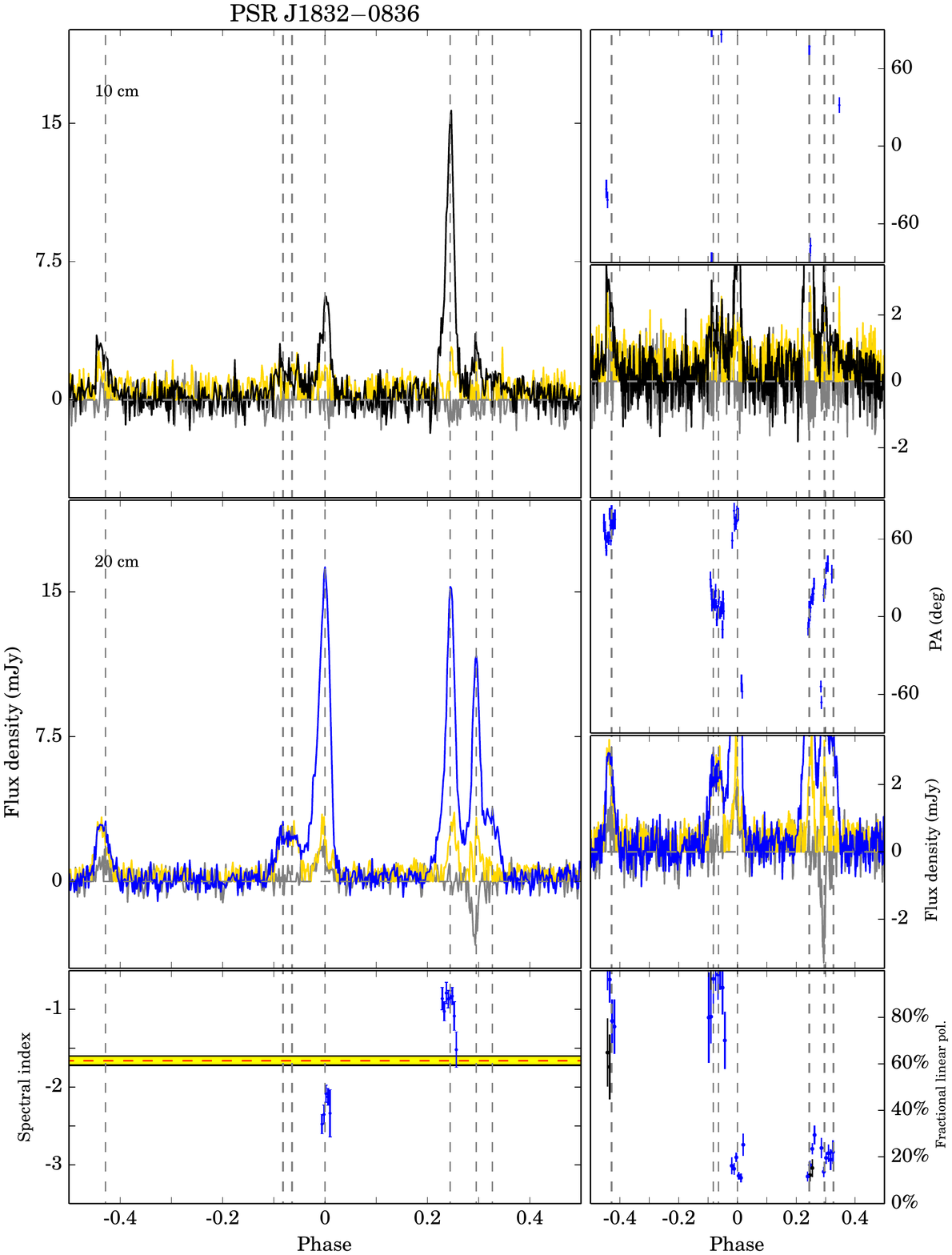}
\caption{Multi-frequency polarization pulse profiles and phase-resolved results for PSR J1832$-$0836. 
At 50\,cm, we only have a few observations and the S/N are low, therefore 
we do not present the polarization profile here.
The components around phase $-0.45$ and $-0.08$ are highly linearly polarized 
and have relatively flat spectrum. 
The PAs around phase $-0.05$ and $0.3$ seem to be discontinuous, but is hard 
to confirm because of the low S/N.
}
\label{1832}
\end{center}
\end{figure*}

\begin{figure*}
\begin{center}
\includegraphics[width=6 in]{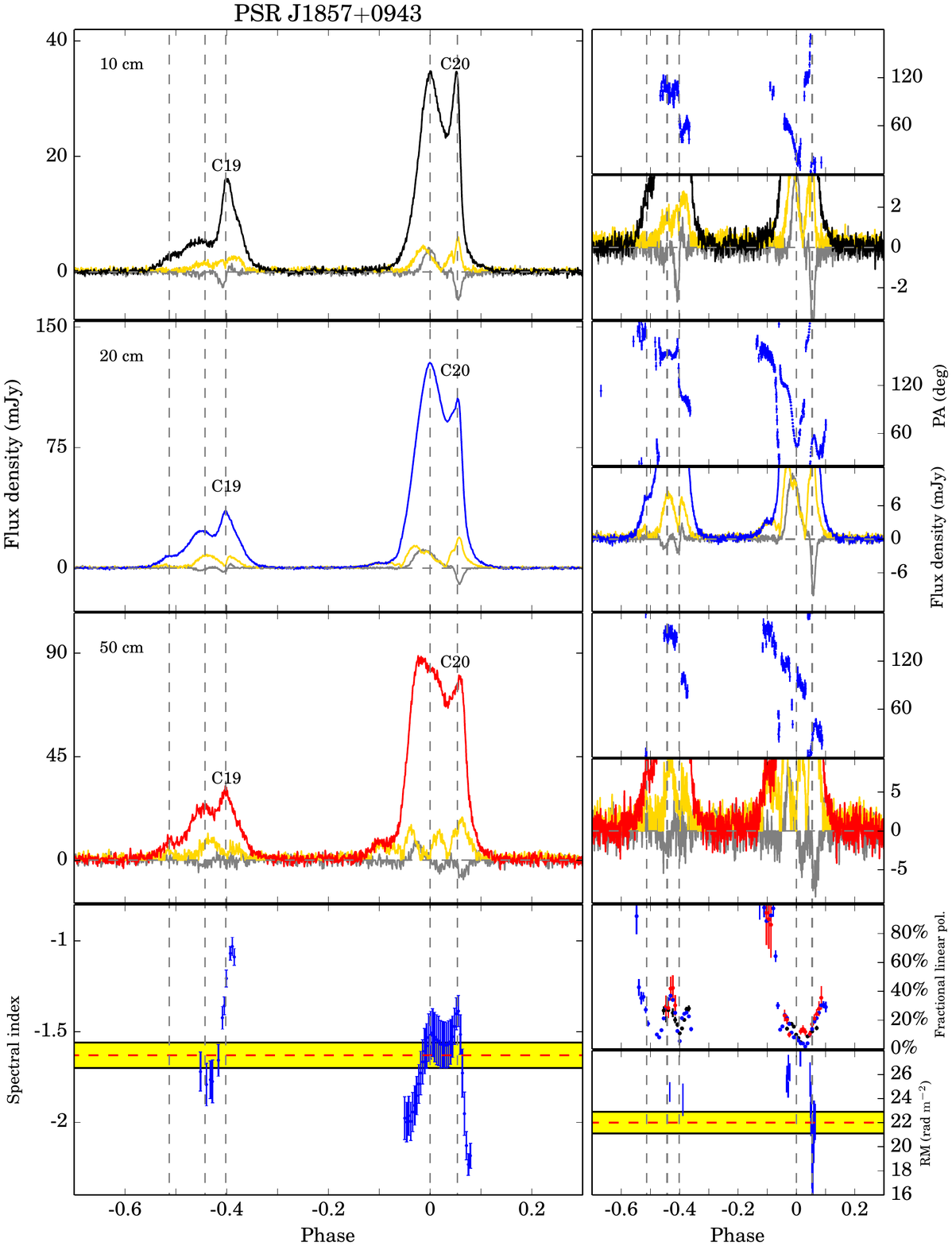}
\caption{Multi-frequency polarization pulse profiles and phase-resolved results for PSR J1857$+$0943. 
We show more details of the PA variation, which is very complex and inconsistent
with the RVM.
At the leading edge of the main pulse, the PA decreases rapidly followed by an 
orthogonal mode transition. 
Around phase $0.05$, there is evidence of a non-orthogonal transition.
Close to the peak of the interpulse, the PA shows a discontinuity at 20\,cm, 
but becomes continuous at 10\,cm.
Both the main pulse and the interpulse have multiple components. The main peak of 
the interpulse has a relatively flat spectrum. The spectrum of the second peak of 
the main pulse is flatter than that of the first peak. The frequency development we 
observe is consistent with previous published results~\citep[][]{Thorsett90}. 
At 50\,cm there is a new linear polarization component appearing close to 
the center of the main pulse.
}
\label{1857}
\end{center}
\end{figure*}

\begin{figure*}
\begin{center}
\includegraphics[width=6 in]{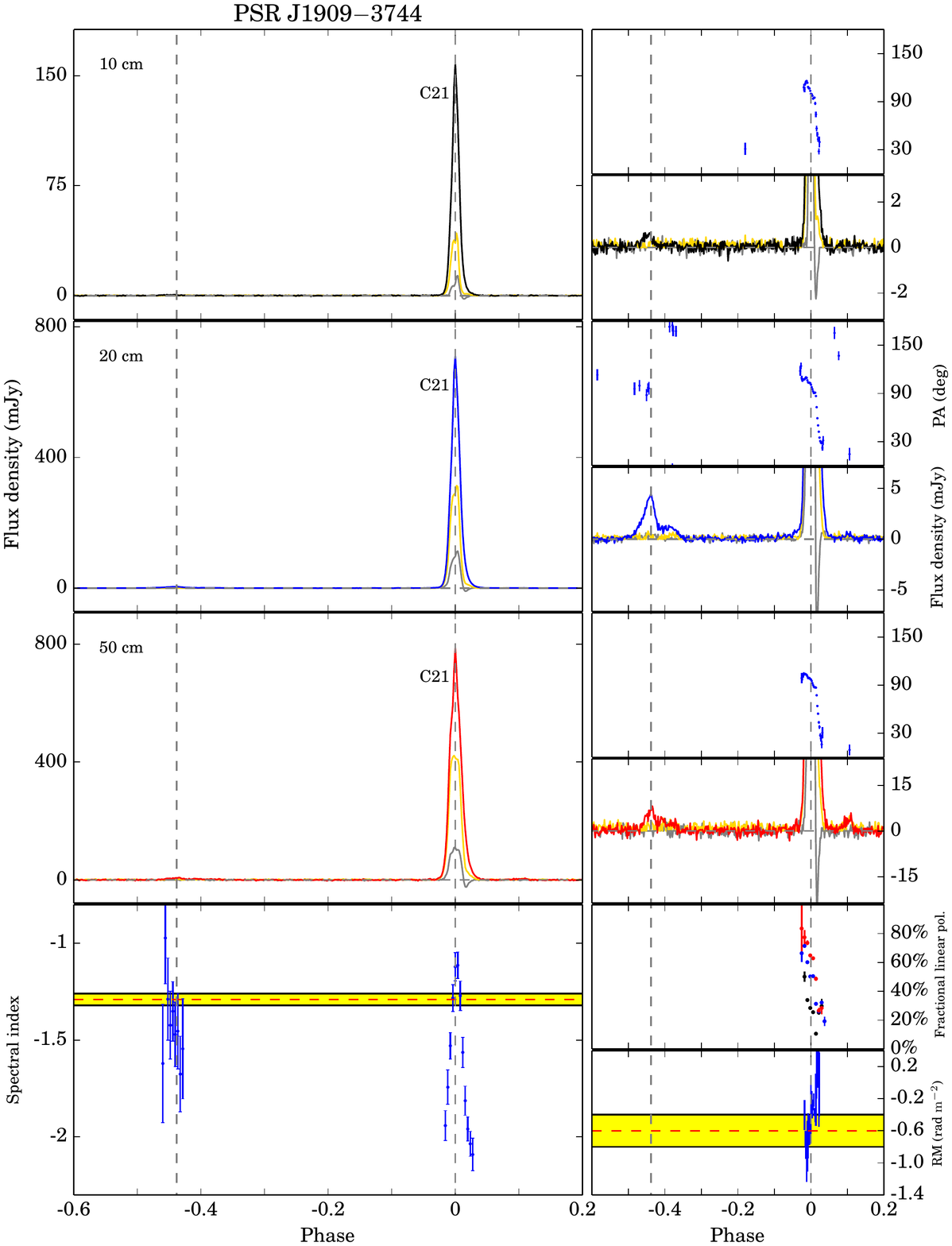}
\caption{Multi-frequency polarization pulse profiles and phase-resolved results for PSR J1909$-$3744. 
We show a narrow main pulse and a weak feature preceding the main pulse by 
approximately $0.45$ in phase.
There is little frequency evolution of the pulse profile, however, the fractional 
linear polarization increases as the frequency decreases.
To calculate the phase-resolved spectral index of the leading component
around phase $-0.45$, we averaged the profile in frequency in the 10\,cm 
and 50\,cm band.
}
\label{1909}
\end{center}
\end{figure*}

\begin{figure*}
\begin{center}
\includegraphics[width=6 in]{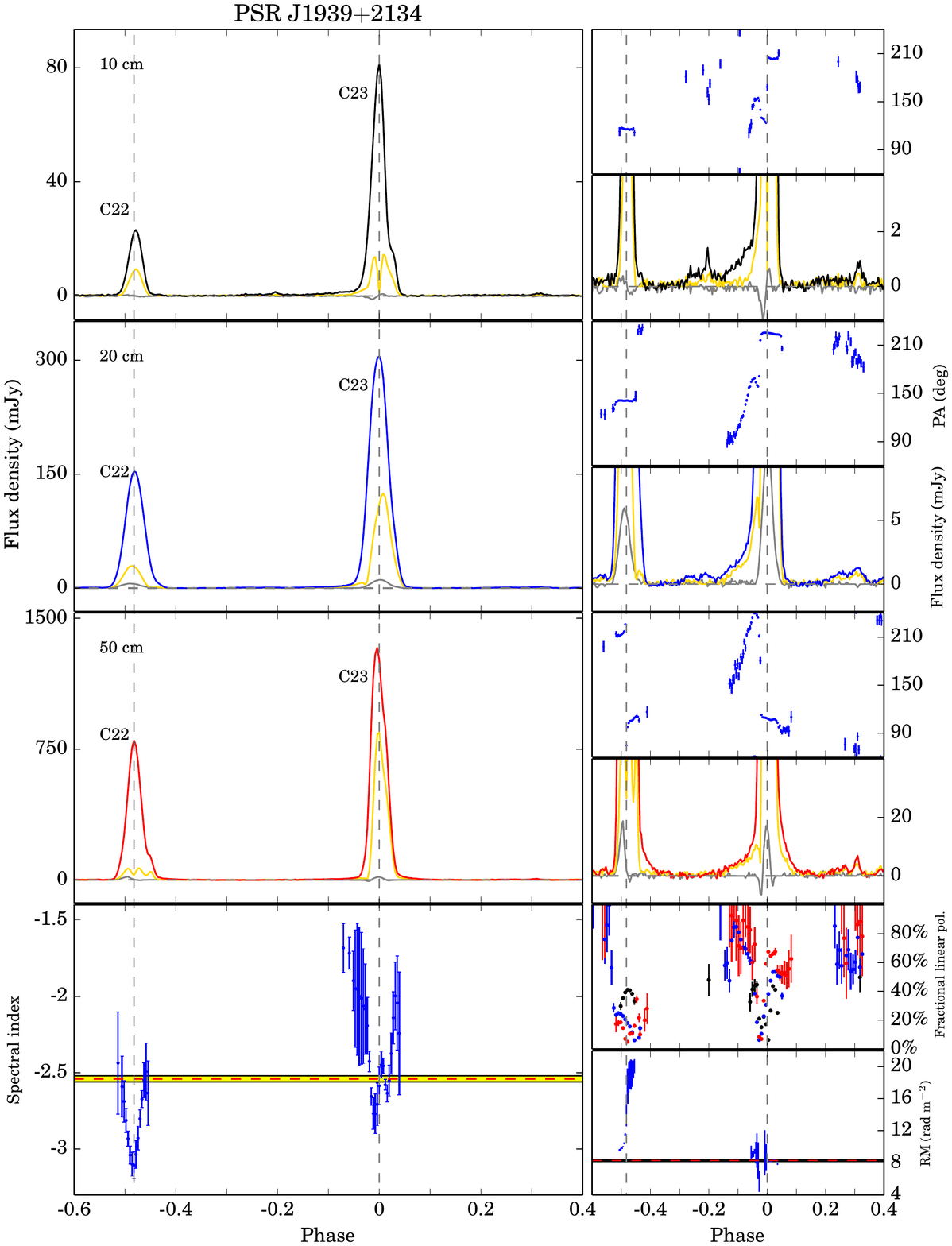}
\caption{Multi-frequency polarization pulse profiles and phase-resolved results for PSR J1939$+$2134. 
Because of the high $\rm{DM}/P$, our observations 
are significantly affected by DM smearing, and we do not see the secondary 
maxima at the trailing edges of both the main pulses and interpulse
~\citep{Thorsett90,Stairs99,Ord04}.
We confirm the existence of weak components preceding both the main pulse 
and interpulse seen by~\citet{Yan11}  and show that they are highly linearly 
polarized and stronger at 10\,cm. 
Our results show stronger left-circular emission in the main pulse compared 
to \citet{Yan11}.
The interpulse has a steeper spectrum compared with the main pulse, and has 
a significantly different RM.
The fractional linear polarization of the main pulse increases significantly 
as frequency decreases while that of the interpulse decreases.
}
\label{1939}
\end{center}
\end{figure*}

\begin{figure*}
\begin{center}
\includegraphics[width=6 in]{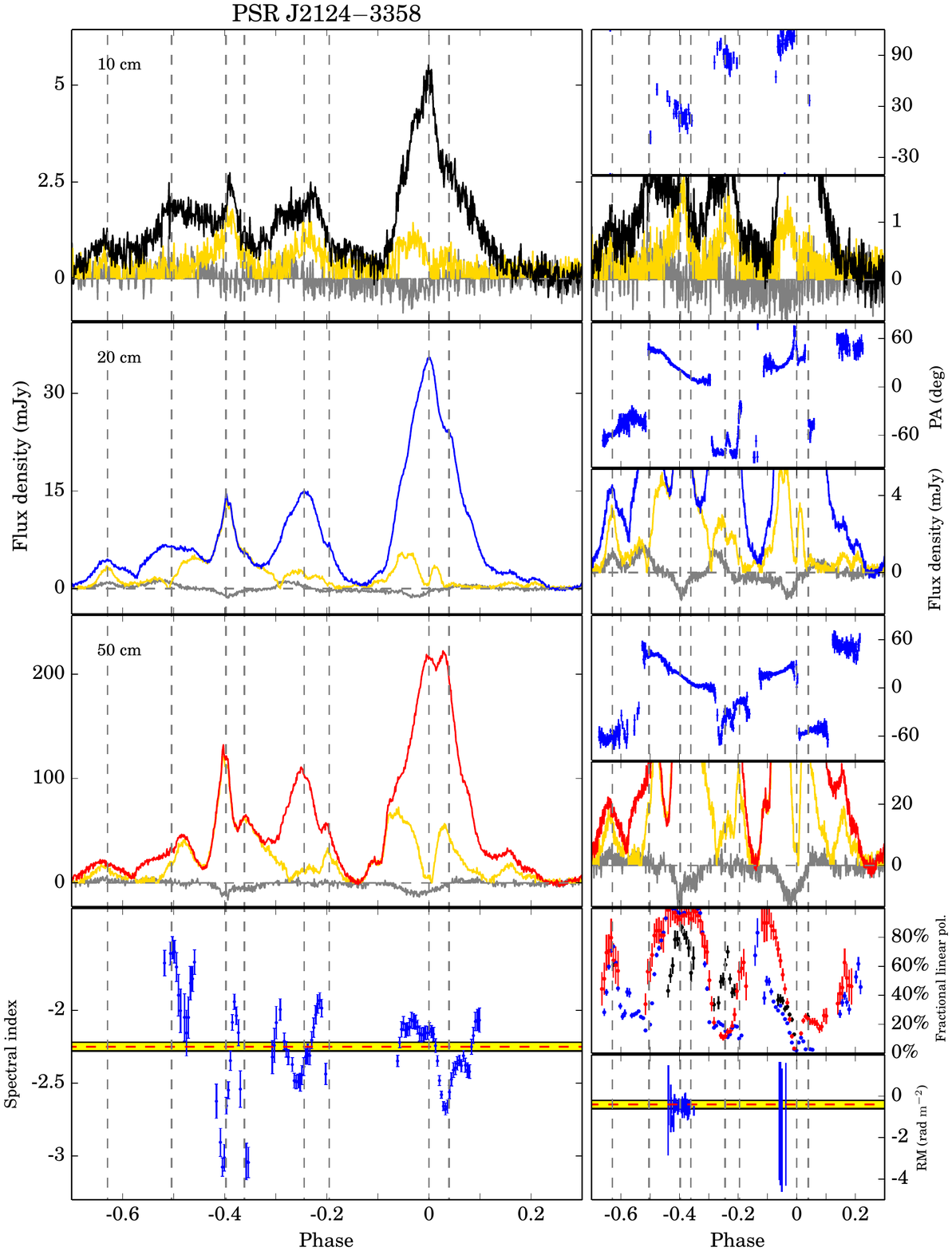}
\caption{Multi-frequency polarization pulse profiles and phase-resolved results for PSR J2124$-$3358. 
We are able to provide more details of the PA variation and show that it has 
complex structures.
At 20\,cm, around phase $0.03$ and $-0.5$, there is evidence of two 
orthogonal mode transitions.
At 50\,cm, around phase $0.1$, there is a non-orthogonal transition of 
$\sim110^{\circ}$.
Because of the complexity of the profile, profile evolution is hard to 
describe. There are large variations of spectral index across the pulse 
longitude and these seem to be related to the different components, but 
this is complicated by the overlap of different components.
We have tested that the phase-resolved spectral indices we present here are 
not significantly affected by the choice of baseline duty cycle and are 
generally consistent with those of~\citet{Manchester04}.
The fractional linear polarization of the main pulse increases at lower 
frequencies.
}
\label{2124}
\end{center}
\end{figure*}

\begin{figure*}
\begin{center}
\includegraphics[width=6 in]{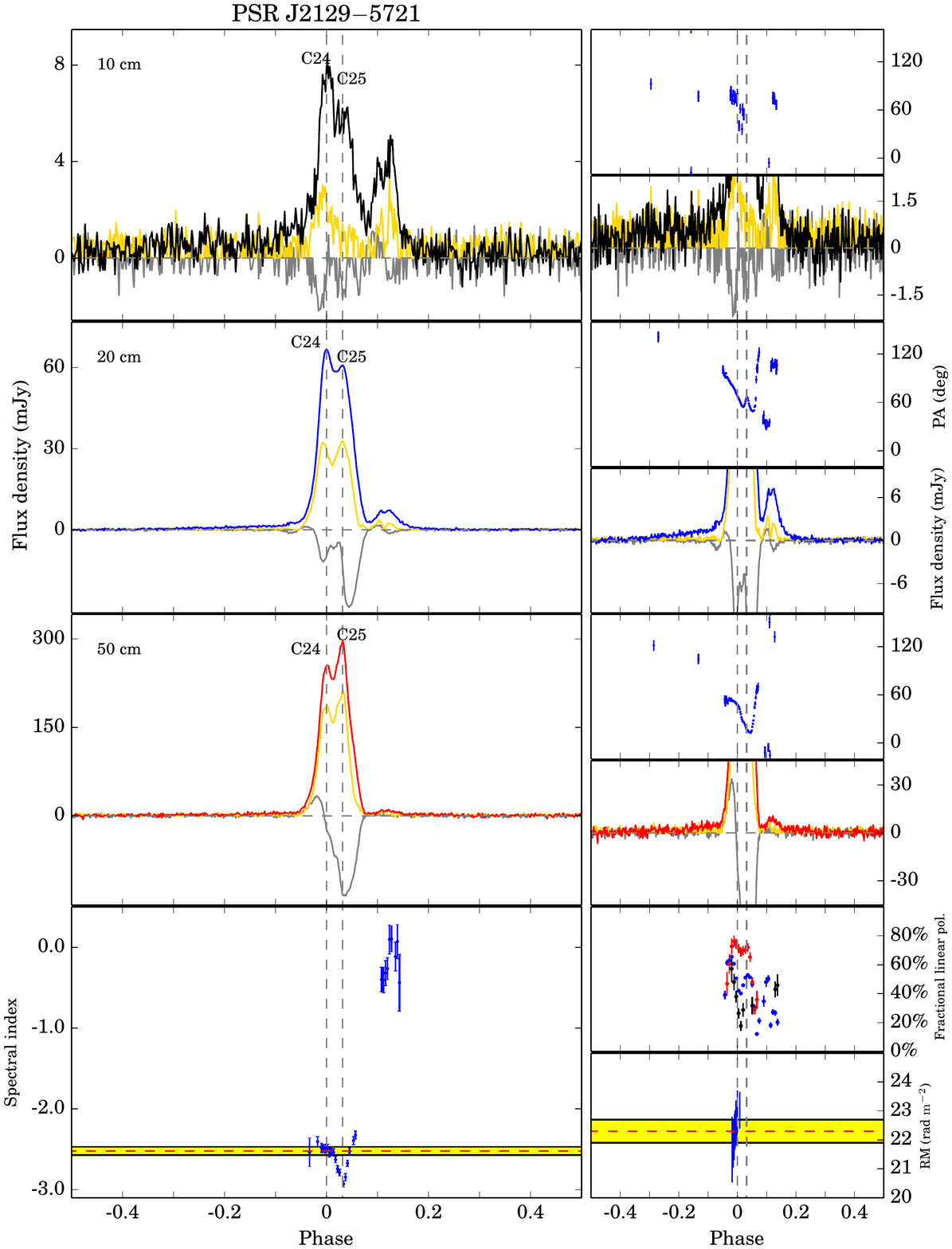}
\caption{Multi-frequency polarization pulse profiles and phase-resolved results for PSR J2129$-$5721. 
At 20\,cm, the weaker leading shelf of emission seen by~\citet{Yan11} extends to at least 
phase of $-0.4$, and the post-cursor clearly has multiple components.
We show more details of PA in the trailing edge of the main pulse.  
The PA decreases across the main pulse, and then increases quickly followed 
by an orthogonal mode transition at 20\,cm.
The post-cursor of the main pulse has much flatter spectrum.
The fractional linear polarization of the main pulse increases as frequency 
decreases.
To calculate the phase-resolved spectral index of the trailing component
around phase 0.1, we averaged the profile in frequency in the 10\,cm 
and 50\,cm band.
We show that the trailing component has a very flat spectrum.
}
\label{2129}
\end{center}
\end{figure*}

\begin{figure*}
\begin{center}
\includegraphics[width=6 in]{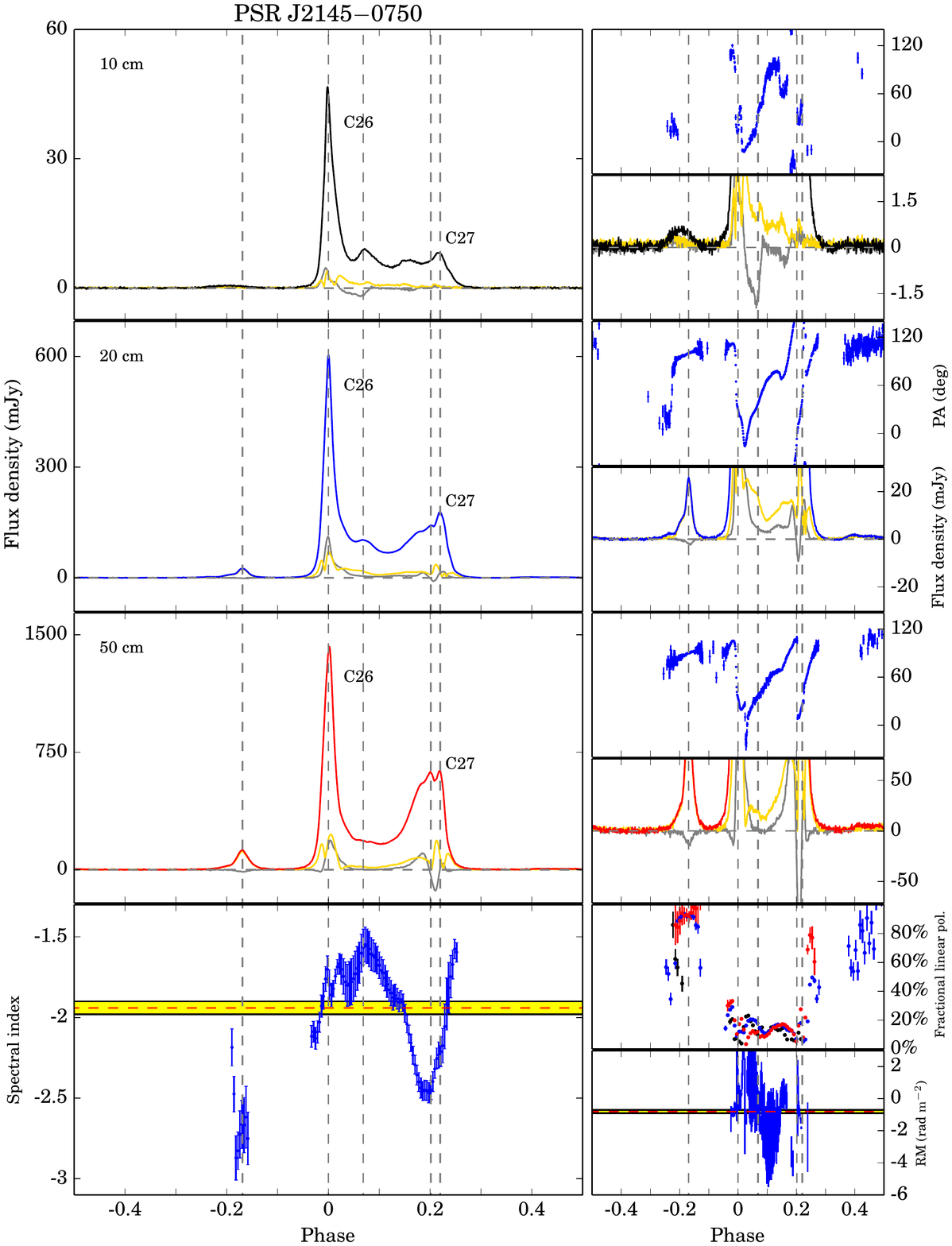}
\caption{Multi-frequency polarization pulse profiles and phase-resolved results for PSR J2145$-$0750. 
At 20\,cm, around phase $0.4$, there is evidence of new low-level emission which 
significantly extends the overall width of this MSP from $187^{\circ}$ to 
$277^{\circ}$.
To calculate the phase-resolved spectral index of the leading component
around phase $-0.18$, we averaged the profile in frequency in the 10\,cm 
band.
The trailing emission and the weak leading component have steeper spectra 
compared with other components.
}
\label{2145}
\end{center}
\end{figure*}

\begin{figure*}
\begin{center}
\includegraphics[width=6 in]{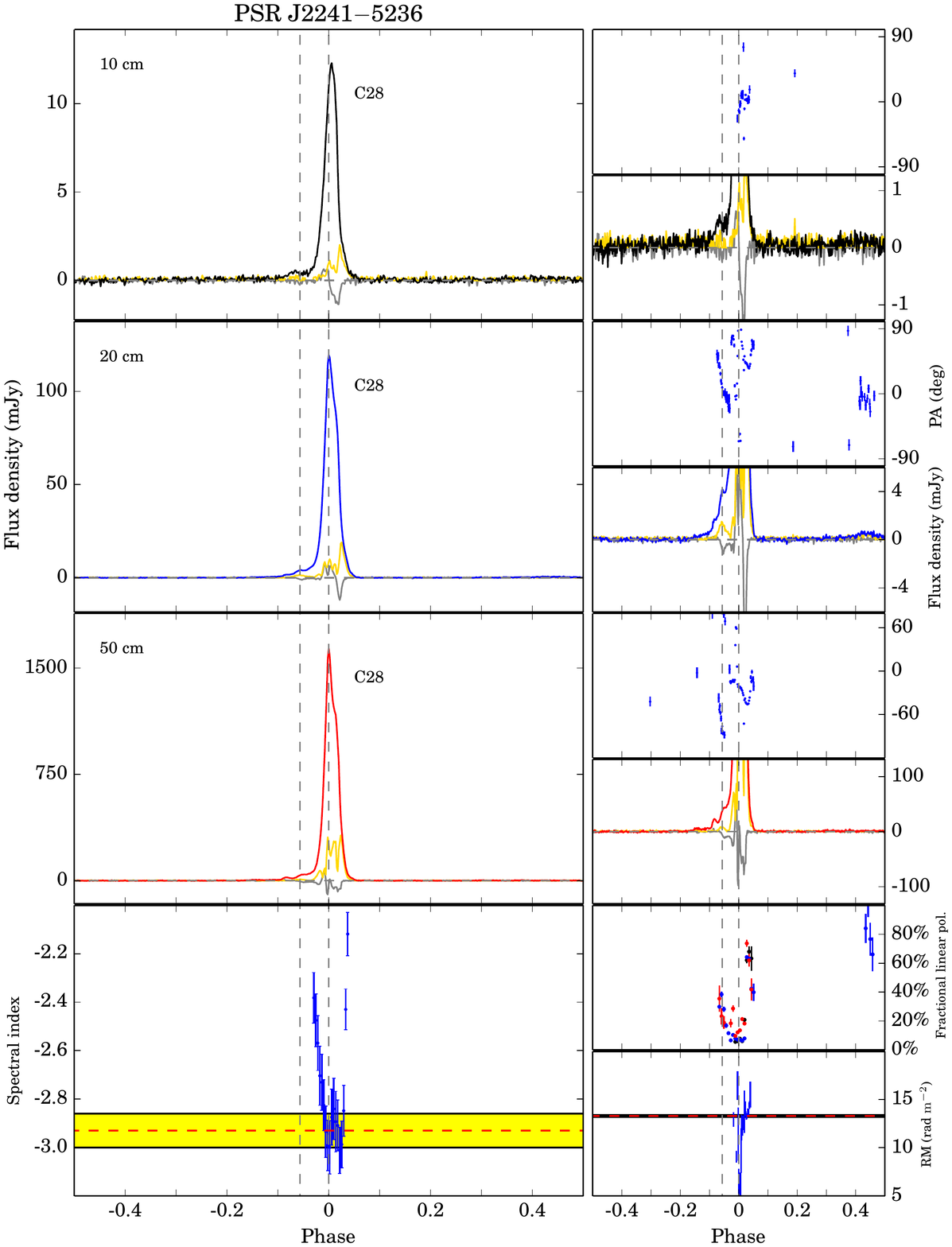}
\caption{Multi-frequency polarization pulse profiles and phase-resolved results for PSR J2241$-$5236. 
At 20\,cm, we show a new low-level component around phase $0.4$ with a width of 
approximately 0.2.
We also show more details of the complex PA variations and there is evidence 
for two orthogonal mode transitions close to the peak.
The frequency evolution of the pulse profile is hard to see, but the fractional 
linear polarization increases at lower frequencies.
}
\label{2241}
\end{center}
\end{figure*}

In this section, we present the multi-frequency polarization pulse profiles and phase-resolved results 
for each MSP in Figures~\ref{0437} to~\ref{2241}. 
The left-hand panels show the pulse profile in the 10\,cm (top), 20\,cm 
(second panel) and 50\,cm (third panel) observing bands. The black, blue and red 
lines in these panels respectively indicate the total intensity, Stokes I, profile in the three bands. 
The yellow line indicates linear polarization and the grey line shows circular polarization. 
The bottom panel on the left-hand side presents the phase-resolved spectral index.   
The red dashed line and yellow highlighted region represent the measured spectral 
index and its uncertainty as presented in Table~\ref{tableFlux}.
In the right-hand panels we have two panels for each of the 10\,cm, 20\,cm and 50\,cm bands. 
The upper panel shows the position angle of the linear polarization at the 
band central frequency (in degrees).
The lower panels shows a zoom-in around the profile baseline to show weak profile 
features. The colour scheme is the same as in the left-hand panels.   
The bottom two panels on the right-hand side show the phase-resolved fractional 
linear polarization for the three observing bands using the same colour scheme as above, 
and the phase-resolved RM. The red dashed line and yellow highlighted region represent the 
measured RM value and its uncertainty. 
In all panels, vertical dashed lines show the positions of peaks in the 20\,cm total intensity profile.

In Table~\ref{ref}, we give the references, baseline duty cycle and S/N for the pulse profile in each 
band for each pulsar. Specific comments for each individual pulsar and on the comparison with 
previous work are given in the caption of each figure. 

\newpage

\end{appendix}

\end{document}